%% file: ms_astroph.tex
\documentclass[numberedappendix]{emulateapj}
\usepackage{natbib}
\usepackage{lscape}
\begin{document}
\journalinfo{Accepted for publication in \textit{The Astronomical Journal}}
\title{Clean Kinematic Samples in Dwarf Spheroidals: An Algorithm for Evaluating Membership and Estimating Distribution Parameters When Contamination is Present}
\shorttitle{EM Algorithm}
\author{Matthew G. Walker\altaffilmark{1,2}, Mario Mateo\altaffilmark{2}, Edward W. Olszewski\altaffilmark{3}, Bodhisattva Sen\altaffilmark{4}, and Michael Woodroofe\altaffilmark{4}}
\email{walker@ast.cam.ac.uk}
\altaffiltext{1}{Institute of Astronomy, University of Cambridge, UK}
\altaffiltext{2}{Department of Astronomy, University of Michigan, Ann Arbor, MI}
\altaffiltext{3}{Steward Observatory, The University of Arizona, Tucson, AZ}
\altaffiltext{4}{Department of Statistics, University of Michigan, Ann Arbor, MI}

\begin{abstract} 

We develop an algorithm for estimating parameters of a distribution sampled with contamination.  We employ a statistical technique known as ``expectation maximization'' (EM).  Given models for both member and contaminant populations, the EM algorithm iteratively evaluates the membership probability of each discrete data point, then uses those probabilities to update parameter estimates for member and contaminant distributions.  The EM approach has wide applicability to the analysis of astronomical data.  Here we tailor an EM algorithm to operate on spectroscopic samples obtained with the Michigan-MIKE Fiber System (MMFS) as part of our Magellan survey of stellar radial velocities in nearby dwarf spheroidal (dSph) galaxies.  These samples, to be presented in a companion paper, contain discrete measurements of line-of-sight velocity, projected position, and pseudo-equivalent width of the Mg-triplet feature, for $\sim 1000 - 2500$ stars per dSph, including some fraction of contamination by foreground Milky Way stars.  The EM algorithm uses all of the available data to quantify dSph and contaminant distributions.  For distributions (e.g., velocity and Mg-index of dSph stars) assumed to be Gaussian, the EM algorithm returns maximum-likelihood estimates of the mean and variance, as well as the probability that each star is a dSph member.  These probabilities can serve as weights in subsequent analyses.  Applied to our MMFS data, the EM algorithm identifies more than $5000$ stars as probable dSph members.  We test the performance of the EM algorithm on simulated data sets that represent a range of sample size, level of contamination, and amount of overlap between dSph and contaminant velocity distributions.  The simulations establish that for samples ranging from large ($N \sim 3000$, characteristic of the MMFS samples) to small ($N\sim 30$, resembling new samples for extremely faint dSphs), the EM algorithm distinguishes members from contaminants and returns accurate parameter estimates much more reliably than conventional methods of contaminant removal (e.g., sigma clipping).  

\end{abstract}

\section{Introduction}

Most astronomical data sets are polluted to some extent by foreground/background objects (``contaminants'') that can be difficult to distinguish from objects of interest (``members'').  Contaminants may have the same apparent magnitudes, colors, and even velocities as members, and so satisfy many of the criteria used to identify members prior to and/or after observation.  Obviously, all else being equal, analyses in which those contaminants are properly identified are superior to analyses in which they are not.  Typically, one attempts to remove contaminants prior to analysis by assuming some reasonable distribution (e.g., Gaussian) for the member population, and then rejecting outliers iteratively until the number of outliers and the parameters (e.g., mean and variance) of the assumed distribution stabilize.  

This conventional approach is problematic in that the results depend on the arbitrary definition of some threshold (e.g., $3\sigma$) separating members from outliers.  Further, as they do not incorporate any explicit consideration of the contaminant distribution, conventional methods provide no means of identifying likely contaminants that lie near the center of the member distribution.  The resulting samples are thus likely to retain contaminants that will then receive equal weight as true members in subsequent analyses.

In this paper we promote a different approach to the problem of sample contamination, drawing upon a technique known in the statistics literature as ``expectation maximization'' (EM; \citealt{dempster77}).  We introduce an iterative EM algorithm for estimating the mean and variance of a distribution sampled with contamination.  The EM method differs from a conventional, sigma-clipping method in two key respects.  First, whereas the conventional method answers the question of membership with a yes or no, the EM method yields a probability of membership.  This probability serves as a weight during subsequent analysis.  No data points are discarded; rather, likely contaminants receive appropriately little weight.  Second, in evaluating membership probability, the EM method explicitly considers the distributions of both members \textit{and} contaminants.  Thus likely contaminants that happen to lie near the center of the member distribution can be identified as such and weighted accordingly.  EM therefore allows a full maximum-likelihood treatment of the entire data set.  The EM approach has potentially wide applicability to many types of data.  In previous work, \citet{sen07} briefly discuss the application of an EM algorithm to dSph data as a means for separating members from contaminants.  Here we exapand upon this work and develop an EM algorithm in order to estimate, given contaminated spectroscopic samples, mean velocities and velocity dispersions of pressure-supported galaxies.

In companion papers \citep[Papers I and II, respectively]{walker07a,walker08}, we present a spectroscopic sample from our Michigan/MIKE Fiber System (MMFS) survey of stellar radial velocities in nearby dwarf spheroidal (dSph) galaxies.  For each star we measure the line-of-sight velocity and the pseudo-equivalent width of the Mg-triplet absorption feature.  As of 2008 August, the MMFS data set contains measurements of 7103 red giant candidates in the dSphs Carina, Fornax, Sculptor and Sextans, each a satellite of the Milky Way.  Along the line of sight to each of the targeted dSphs, the Milky Way contributes stars with magnitudes and colors satisfying our CMD-based target selection criteria (Figure 1 of Paper I).  Therefore we expect the MMFS sample to carry a degree of contamination that varies with the density contrast between dSph and foreground.  

\begin{figure*}
  \plotone{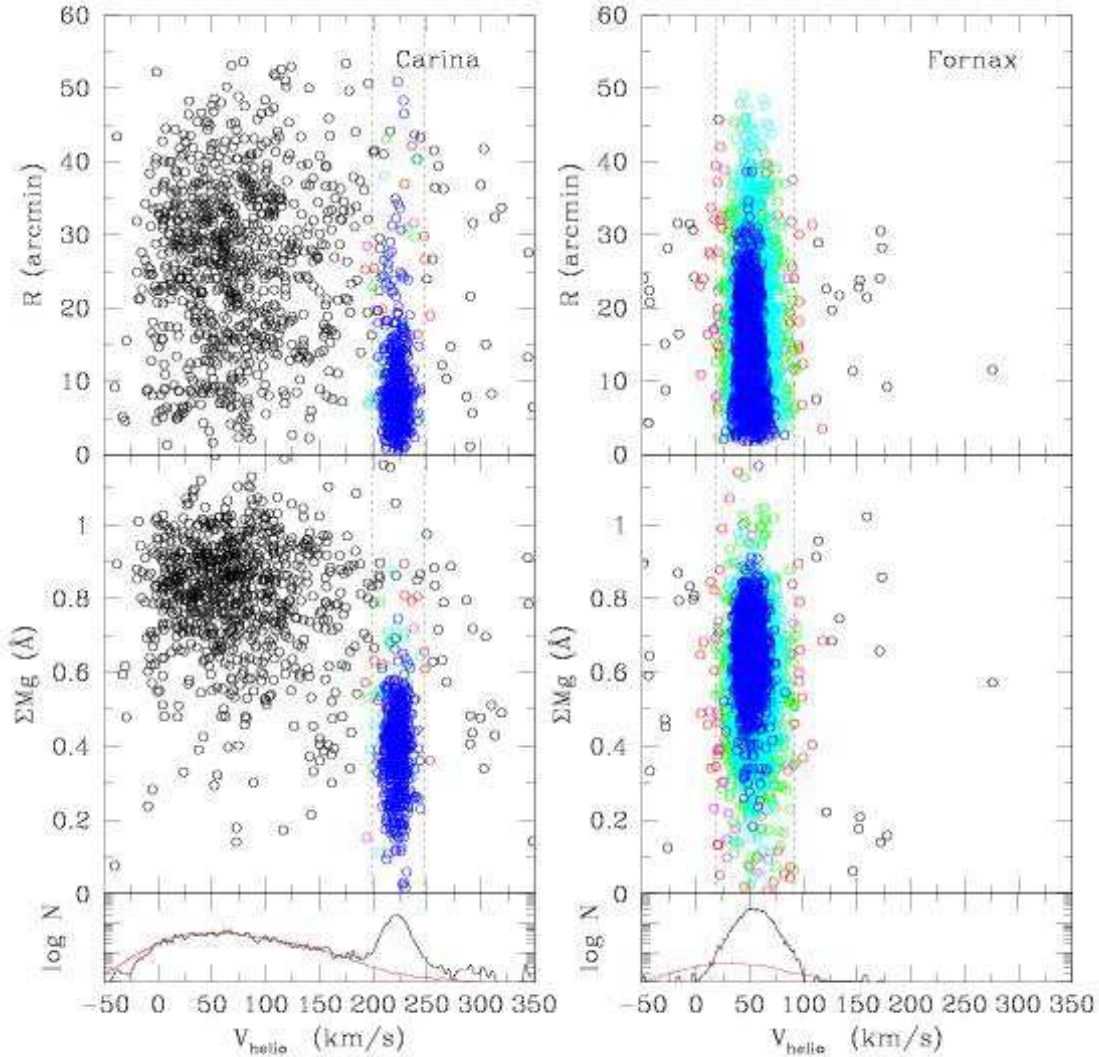}
  \caption{\scriptsize Projected distance from the dSph center (top panels) and magnesium index (bottom panels) versus velocity, Carina and Fornax dSphs.  Marker color indicates probability that the star belongs to the dSph population, evaluated using the EM algorithm.  Black (red; magenta; green; cyan; blue) markers signify $\hat{P}_{M} \leq 0.01$ ($\hat{P}_{M} > 0.01; > 0.50; > 0.68; > 0.95; > 0.99$).  In each main panel a pair of vertical, dotted lines encloses stars that would pass conventional membership tests based on an iterative, $3\sigma$ velocity threshold.  Sub-panels at bottom give the observed velocity distribution.  The red curve in the bottom is the expected velocity distribution of foreground contaminants, derived from the Besan{\c c}on Milky Way model.  The predicted Besan{\c c}on distribution along the line of sight to each dSph is normalized according to the estimated membership fraction $\hat{N}_{M}/N$.}
  \label{fig:members_carfor}
\end{figure*}

\begin{figure*}
  \plotone{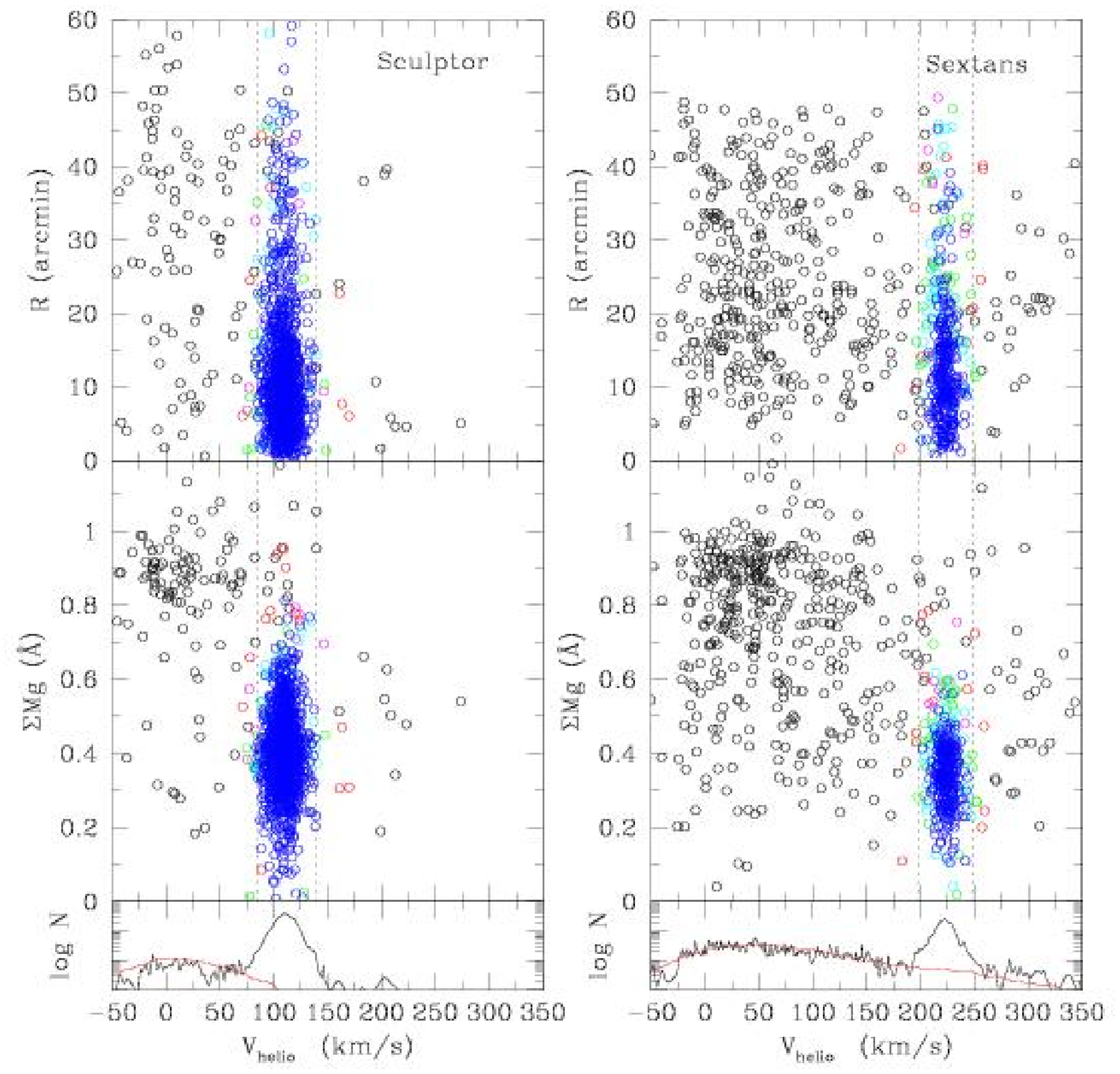}
  \caption{\scriptsize Same as Figure \ref{fig:members_carfor}, but for the Sculptor and Sextans dSphs.}
  \label{fig:members_sclsex}
\end{figure*}

Figures \ref{fig:members_carfor} and \ref{fig:members_sclsex} plot for all MMFS stars the measured velocity, $V$, magnesium index $\Sigma$Mg, and projected radius $R$ (angular distance from the dSph center).  Viewing these scatter plots, one can distinguish loci of foreground stars from those of the Carina, Sculptor and Sextans populations.  The members of these three dSphs cluster into relatively narrow velocity distributions and have systematically weaker $\Sigma$Mg than do the foreground stars.  The latter effect owes to the fact that foreground stars sufficiently faint to be included in our CMD selection regions are likely to be K-dwarfs in the MW disk, which exhibit strong magnesium absorption due to the fact that their higher surface gravities and higher metal abundances increase their atmospheric opacity \citep{cayrel91}.  Due to Fornax's relatively high surface brightness, our Fornax targets are more likely to be bona fide members, and the Fornax data do not show an obviously distinct distribution of foreground stars.  However, given Fornax's relatively broad distribution of $\Sigma$Mg and small systemic velocity of $\sim 55$ km s$^{-1}$ along the line of sight, its velocity distribution overlaps with that expected for the foreground distribution.  Therefore we must not consider the Fornax sample to be free of contamination, especially in the low-velocity tail of the distribution (see Figure \ref{fig:members_carfor}).  

EM provides a reliable and efficient means of characterizing the distributions of both member and contaminant populations.  In what follows we provide a general description of the EM approach, which has many potential applications in astronomy.  We then develop an EM algorithm for our specific purpose of measuring dSph mean velocities and velocity dispersions from the MMFS data.  Our EM algorithm incorporates stellar velocity, magnesium index and projected position to generate maximum-likelihood estimates of the desired parameters.  The algorithm also evaluates for each star the probability of dSph membership.  Adding these probabilities, we find that the MMFS sample contains more than 5000 dSph members.  Finally, we generate artificial dSph-like samples that allow us to examine the performance of the EM algorithm given a variety of sample sizes, levels of contamination, and amounts of overlap between member and contaminant velocity distributions.  We demonstrate that the EM algorithm consistently recovers the correct distribution parameters in all but the smallest and most severely contaminated samples, and dramatically outperforms conventional methods for outlier rejection.  



\section{Expectation-Maximization}
\label{sec:em}

Consider a population for which some random variable $X$ follows a probability distribution $p_{mem}(x)$ that is characterized by parameter set $\zeta_{mem}$.  For example, if the distribution is Gaussian, $p_{mem}(x)=(2\pi\sigma^2)^{-1/2}\exp[-(x-\langle X\rangle)^2/(2\sigma^2)]$, and $\zeta_{mem}=\{\langle X\rangle_{mem},\sigma^2_{mem}\}$ consists of the mean and variance.  Typically one samples the distribution with the goal of estimating $\zeta_{mem}$.  If contamination is present, then some fraction of the sample is drawn from a second, ``non-member'' population that has the probability distribution $p_{non}(x)$, with parameter set $\zeta_{non}$.  Let $p(a)$ represent the unconditional (i.e., \textit{a priori})member fraction which, unless constant over the sample, depends on some random variable $A$ that is independent of $X$.  For example, in the case of dSph data, we find that a star's \textit{a priori} probability of membership correlates with its projected distance from the dSph center, such that the member fraction, $p(a)$, is highest at the center of the galaxy and decreases toward its sparsely populated outer regions.  

Suppose the variable $M$ indicates membership and takes one of two possible values: $M=1$ indicates observation of a member, $M=0$ indicates observation of a nonmember.  Given a data set $\{X_i,M_i\}_{i=1}^N$ for which all $M_i$ are known, one can evaluate the likelihood,
\begin{eqnarray}
  L(\zeta_{mem},\zeta_{non})=\displaystyle\prod_{i=1}^N\biggl [p_{mem}(X_i)p(a_i)\biggr ]^{M_i}\hspace{0.75in}\nonumber\\
  \times \biggl [p_{non}(X_i)[1-p(a_i)]\biggr]^{1-M_i},\hspace{0.2in}
  \label{eq:likelihood}
\end{eqnarray}
or more conveniently, the log-likelihood given by
\begin{eqnarray}
  \ln[L(\zeta_{mem},\zeta_{non})]=\displaystyle \sum_{i=1}^{N}M_i \ln \biggl [p_{mem}(X_i)p(a_i)\biggr ]\hspace{0.4in}\nonumber\\
  +\displaystyle \sum_{i=1}^N(1-M_i)\ln \biggl [p_{non}(X_i)[1-p(a_i)]\biggr].\hspace{0.2in}
  \label{eq:loglikelihood}
\end{eqnarray}

However, the values $M_i$, which can be either exactly 0 or exactly 1, usually are unknown, in which case the likelihood cannot be evaluated.  Using the EM technique, one evaluates instead the \textit{expected} log-likelihood in terms of the \textit{expected} values of $M$.  Prior to observation, $M$ is a Bernoulli random variable with ``probability of sucess''  given by the \textit{a priori} membership fraction: $P(M=1)=p(a)$.  After observation, the expected value of $M$ is the probability that $M=1$, subject to the data as well as the prior constraints specified by $p(a)$.  Denoting this probability $P_{M}$, we have
\begin{eqnarray}
  P_{M_i}\equiv P(M_i=1|X_i,a_i)\hspace{1.75in}\nonumber\\
  =\frac{p_{mem}(X_i)p(a_i)}{p_{mem}(X_i)p(a_i)+p_{non}(X_i)[1-p(a_i)]}.\hspace{0.25in}
  \label{eq:estep}
\end{eqnarray}
The expected log-likelihood, given data $S\equiv\{X_i\}_{i=1}^N$, is then   
\begin{eqnarray}
  E(\ln L(\zeta_{mem},\zeta_{non})|S)=\displaystyle \sum_{i=1}^{N}P_{M_i} \ln \biggl [p_{mem}(X_i)p(a_i)\biggr ]\nonumber\\
  +\displaystyle \sum_{i=1}^N(1-P_{M_i})\ln \biggl [p_{non}(X_i)[1-p(a_i)]\biggr].\hspace{0.2in}
  \label{eq:mstep}
\end{eqnarray}
In the EM approach, maximum-likelihood estimates of the parameter sets $\zeta_{mem}$ and $\zeta_{non}$, as well as any unknown parameters in $p(a)$, take the values that maximize $E(\ln L(\zeta_{mem},\zeta_{non})|S)$.

The iterative EM algorithm derives its name from its two fundamental steps.  In the ``expectation'' step, the expected value of $M_i$ is identified with the probability $P_{M_i}$ and evaluated for all $i$ according to Equation \ref{eq:estep}.  In the ``maximization'' step, the distribution parameters are estimated by maximizing the expected log-likelihood specified by Equation \ref{eq:mstep}.  The parameter estimates are then used to update the values $P_{M_i}$ in the next expectation step, followed by the updating of parameter estimates in the subsequent maximization step, and so forth until convergence occurs.  In addition to estimates of the parameter sets $\zeta_{mem}$, $\zeta_{non}$ (and any free parameters in the function $p(a)$), the algorithm provides useful estimates of $P_{M_i}$, the probability of membership, for all $i$.

\section{EM Applied to dSph Data}
\label{sec:emdsph}

We now develop an EM algorithm for estimating parameters of the distributions sampled by the MMFS data.  In this application we evaluate the membership of stars using MMFS data that sample \textit{two} independent distributions, velocity and magnesium index.  Also, inspection of the top panels in Figures \ref{fig:members_carfor} and \ref{fig:members_sclsex} reveals that the fraction of contamination tends to increase as a function of distance from the dSph center.  Projected radius therefore serves as the independent variable in the unconditional probability function $p(a)$.  In a previous application of the EM algorithm to dSph data, \citet{sen07} assume $p(a)$ decreases exponentially, consistent with the measured surface brightness profiles of dSphs \citep{ih95}.  Here, in order to avoid bias due to selection effects, we choose not to specify a particular functional form for $p(a)$; instead, after each maximization step we estimate $p(a)$ via monotonic regression (see Appendix \ref{app:pav}).  In the end we obtain estimates of the means and variances of the velocity and $\Sigma$Mg distributions for member and nonmember populations, the membership probability of each star, and the member fraction as a function of projected radius.

We assume that for dSph members the joint distribution of velocity, $V$, and magnesium strength, henceforth denoted $W$, is a bivariate Gaussian distribution with means $\langle V \rangle_{mem}$ and $\langle W \rangle_{mem}$, variances $\sigma_{V_0,mem}^2$ and $\sigma_{W_0,mem}^2$, and covariance $\sigma_{V_0W_0,mem}=\langle (V- \langle V \rangle_{mem})(W- \langle W \rangle_{mem}) \rangle =0$ (the distributions of $V$ and $W$ are independent).  If we were to sample only the dSph population (i.e., with no contamination), the probability density for observation of a star with velocity $V_i$ and line strength $W_i$ would be 
\begin{equation}
  p_{mem}(V_i,W_i) = \frac{\exp \biggl [-\frac{1}{2} \biggl (\frac{[V_i-\langle V \rangle_{mem}]^2}{\sigma_{V_0,mem}^2+\sigma_{V_i}^2}+\frac{[W_i-\langle W \rangle_{mem}]^2}{\sigma_{W_0,mem}^2+\sigma_{W_i}^2} \biggr ) \biggr ]}{2\pi\sqrt{(\sigma_{V_0,mem}^2+\sigma_{V_i}^2)(\sigma_{W_0,mem}^2+\sigma_{W_i}^2)}},.
  \label{eq:pmem}
\end{equation}
where $\sigma_{V_i}$ and $\sigma_{W_i}$ represent measurement errors.  For the $\sim 5\%$ of stars in the MMFS sample that lack an acceptable measurement of magnesium strength we replace Equation \ref{eq:pmem} with the univariate Gaussian probability 
\begin{equation}
  p_{mem}(V_i) = \frac{\exp \biggl [-\frac{1}{2} \biggl (\frac{[V_i-\langle V \rangle_{mem}]^2}{\sigma_{V_0,mem}^2+\sigma_{V_i}^2}\biggr ) \biggr ]}{\sqrt{2\pi (\sigma_{V_0,mem}^2+\sigma_{V_i}^2)}}.
  \label{eq:pmemnomg}
\end{equation}

A virtue of the EM method lies in its ability to incorporate and to evaluate the contaminant as well as member distributions.  The efficacy of the algorithm therefore depends on the suitability of the foreground model used to characterize $p_{non}(V_i,W_i)$.  Since the distribution of the MW's disk rotational velocities is non-Gaussian, we do not model the contaminant velocity distribution as a simple Gaussian.  Instead we adopt the distribution of foreground velocities specified by the Besan{\c c}on numerical model of the MW \citep{robin03}\footnote{available at http://bison.obs-besancon.fr/modele/}.  From the Besan{\c c}on model, which includes both disk and halo components, we obtain simulated velocity and $V,I$ photometric data along the line of sight to each dSph.  After removing simulated stars that fall outside our CMD selection regions (see Figure 1 of Paper I), we use a Gaussian kernel to estimate the marginal distribution of contaminant velocities from $N_{bes}$ artificial data points:
\begin{eqnarray}
  p_{non}(v)=p_{bes}(v)=\frac{1}{N_{bes}} \sum_{i=1}^{N_{bes}}\frac{1}{\sqrt{2\pi \sigma_{V_{bes}}^2}}\hspace{0.65in} \nonumber\\
  \times \exp \biggl [ -\frac{1}{2}\frac{(V_{bes_i}-v)^2}{\sigma_{V_{bes}}^2} \biggr ].\hspace{0.2in}
  \label{eq:pbes}
  \end{eqnarray}
We adopt\footnote{We experimented with different choices for the smoothing bandwidth, including $\sigma_{V_{bes}}=6$ km s$^{-1}$ as well as a variable bandwidth with each artificial error drawn randomly from the set of MMFS velocity errors.  The EM algorithm converged on the same (to within $0.001$ km s$^{-1}$) parameter estimates and membership probabilities in all cases.} $\sigma_{V_{bes}}=2$ km s$^{-1}$, similar to the median MMFS velocity error.  Note that this smooth foreground model does \textit{not} account for known and/or potential substructures (e.g., debris from accretion events as discussed e.g., by \citealt{bell08}) in the MW halo; for some applications it may become necessary to build such features into the foreground model.  

We approximate the marginal distribution of foreground magnesium strength as Gaussian and independent of velocity, with mean $\langle W \rangle_{non}$ and variance $\sigma_{W_0,non}^2$.  Notice that this assumption ignores the fact that the MW foreground has distinct components---thin and thick disk, halo---with different Mg distributions.  In some situations, particularly when samples are small, it may become necessary to adopt a more realistic foreground model, but for our present purposes the single-Gaussian model suffices.  Under this simple model, if we were to sample only from the population of contaminating MW stars satisfying our target-selection criteria, the probability density for observation of a star with velocity $V_i$ and line strength $W_i$ would then be
\begin{equation}
  p_{non}(V_i,W_i) = \frac{{p_{bes}}(V_i) \exp \biggl [-\frac{1}{2} \frac{[W_i-\langle W \rangle_{non}]^2}{\sigma_{W,non}^2+\sigma_{W_i}^2} \biggr ]}{\sqrt{2\pi (\sigma_{W,non}^2+\sigma_{W_i}^2)}}.
  \label{eq:pnon}
\end{equation}

In reality we sample from the union of dSph and contaminant populations.  Assuming the surface brightness of nonmembers remains approximately constant over the face of a given dSph, the \textit{a priori} probability of observing a dSph member decreases in proportion to the surface density of actual members meeting our target selection criteria.  We therefore choose the function $p(a)$ to represent the fraction of selected targets at elliptical radius\footnote{A star's ``elliptical radius'' refers to the semi-major axis of the isophotal ellipse that passes through the star's position.  We make the approximation that a dSph's isophotal ellipses all have common centers and orientation, and we adopt ellipticities and position angles from \citet{ih95}.} $a$ that are actually dSph members.  One might reasonably expect $p(a)$ to decrease exponentially, since exponential profiles generally provide adequate fits to dSph surface brightness data \citep{ih95}.  However, in order to allow for sample bias introduced by our target selection procedure, we adopt the less restrictive assumption that $p(a)$ is merely a non-increasing function.  

Thus in our formulation, the EM analysis of a dSph for which we have data set $S\equiv\{V_i,W_i,a_i\}_{i=1}^N$ involves the estimation of a set of six parameters, $\zeta\equiv\zeta_{mem} \cup \zeta_{non}=\{\langle V \rangle_{mem},\sigma_{V_0,mem}^2,\langle W \rangle_{mem},\sigma_{W_0,mem}^2,\langle W \rangle_{non},\sigma_{W_0,non}^2\}$, as well as the values of $p(a_i)$ for all $i$.  Adopting the notation that $\hat{X}$ is the estimate of $X$, we let $\hat{\zeta}^{\{n\}}$ denote the set of parameter estimates obtained in the $n^{th}$ iteration of the algorithm.  In the ``expectation'' step of the next iteration, we use these estimates to obtain $\hat{p}_{mem_i}^{\{n+1\}}$, $\hat{p}_{non_i}^{\{n+1\}}$, and then $\hat{P}_{M_i}^{\{n+1\}}$ for all $i$.  From equations \ref{eq:pmem}, \ref{eq:pnon}, and \ref{eq:estep} we obtain
\begin{equation}
  \hat{p}_{mem}^{\{n+1\}}(V_i,W_i) = \frac{\exp \biggl [-\frac{1}{2} \biggl (\frac{[V_i-\langle \hat{V}\rangle^{\{n\}}_{mem}]^2}{\hat{\sigma}_{V_0,mem}^{2\{n\}}+\sigma_{V_i}^2}+\frac{[W_i-\langle \hat{W}\rangle ^{\{n\}}_{mem}]^2}{\hat{\sigma}_{W_0,mem}^{2\{n\}}+\sigma_{W_i}^2} \biggr ) \biggr ]}{2\pi\sqrt{(\hat{\sigma}_{V_0,mem}^{2\{n\}}+\sigma_{V_i}^2)(\hat{\sigma}_{W_0,mem}^{2\{n\}}+\sigma_{W_i}^2)}};
  \label{eq:pmemestimate}
\end{equation}
\begin{equation}
  \hat{p}_{non}^{\{n+1\}}(V_i,W_i) = \frac{{p_{bes}}(V_i) \exp \biggl [-\frac{1}{2} \frac{[W_i-\langle \hat{W}\rangle ^{\{n\}}_{non}]^2}{\hat{\sigma}_{W,non}^{2\{n\}}+\sigma_{W_i}^2} \biggr ]}{\sqrt{2\pi (\hat{\sigma}_{W,non}^{2\{n\}}+\sigma_{W_i}^2)}};
  \label{eq:pnonestimate}
\end{equation}
\begin{equation}
  \hat{P}_{M_i}^{\{n+1\}}=\frac{\hat{p}_{mem}^{\{n\}}(X_i)\hat{p}^{\{n\}}(a_i)}{\hat{p}^{\{n\}}_{mem}(X_i)\hat{p}^{\{n\}}(a_i)+\hat{p}^{\{n\}}_{non}(X_i)[1-\hat{p}^{\{n\}}(a_i)]}.
  \label{eq:estepestimate}
\end{equation}

In the ``maximization'' step of iteration $n+1$ we maximize the expected log-likelihood (Equation \ref{eq:mstep}).  Because the form of $p(a)$ is unknown, our maximization step has two parts.  In the first part we estimate the parameter set $\zeta$.  By setting equal to zero the partial derivatives of the expected log-likelihood with respect to each parameter, we obtain six equations: 
\begin{eqnarray}
  \label{eq:emderivatives}
  \displaystyle \sum_{i=1}^{N}P_{M_i}\biggl (\frac{1}{\sigma_{V_0,mem}^2+\sigma_{V_i}^2}[V_i-\langle V \rangle_{mem}] \biggr )=0;\hspace{0.3in}\\
  \displaystyle \sum_{i=1}^{N}P_{M_i}\biggl (\frac{[V_i-\langle V \rangle_{mem}]^2}{[\sigma_{V_0,mem}^2+\sigma_{V_i}^2]^2}-\frac{1}{[\sigma_{V_0,mem}^2+\sigma_{V_i}^2]}   \biggr )=0;\nonumber\\
  \displaystyle \sum_{i=1}^{N}P_{M_i}\biggl (\frac{1}{\sigma_{W_0,mem}^2+\sigma_{W_i}^2}[W_i-\langle W \rangle_{mem}] \biggr )=0;\nonumber\\
  \displaystyle \sum_{i=1}^{N}P_{M_i}\biggl (\frac{[W_i-\langle W \rangle_{mem}]^2}{[\sigma_{W_0,mem}^2+\sigma_{W_i}^2]^2}-\frac{1}{[\sigma_{W_0,mem}^2+\sigma_{W_i}^2]}   \biggr )=0;\nonumber\\
  \displaystyle \sum_{i=1}^{N}(1-P_{M_i})\biggl (\frac{1}{\sigma_{W_0,non}^2+\sigma_{W_i}^2}[W_i-\langle W \rangle_{non}] \biggr )=0;\nonumber\\
  \displaystyle \sum_{i=1}^{N}(1-P_{M_i})\biggl (\frac{[W_i-\langle W \rangle_{non}]^2}{[\sigma_{W_0,non}^2+\sigma_{W_i}^2]^2}-\frac{1}{[\sigma_{W_0,non}^2+\sigma_{W_i}^2]}   \biggr )=0.\nonumber
\end{eqnarray}
These equations do not have analytic solutions, but we obtain consistent estimates of the means and variances using the estimates from the previous iteration.  For example, 
\begin{equation}
  \langle \hat{V} \rangle^{\{n+1\}}=\frac{\displaystyle\sum_{i=1}^{N}\frac{\hat{P}_{M_i}^{\{n\}}V_i}{1+\sigma_{V_i}^2/\hat{\sigma}_{V_0}^{2\{n\}}}}{\displaystyle\sum_{i=1}^N\frac{\hat{P}_{M_i}^{\{n\}}}{1+\sigma_{V_i}^2/\hat{\sigma}_{V_0}^{2\{n\}}}}
  \label{eq:estimatemean}
\end{equation}
and
\begin{equation}
  \hat{\sigma}_{V_0}^{2\{n+1\}}=\frac{\displaystyle\sum_{i=1}^{N}\frac{\hat{P}_{M_i}^{\{n\}}[V_i-\langle \hat{V}\rangle^{\{n+1\}}]^2}{(1+\sigma_{V_i}^2/\hat{\sigma}_{V_0}^{2\{n\}})^2}}{\displaystyle\sum_{i=1}^N\frac{\hat{P}_{M_i}^{\{n\}}}{1+\sigma_{V_i}^2/\hat{\sigma}_{V_0}^{2\{n\}}}}.
  \label{eq:estimatedisp}
\end{equation}
Iterative solutions for the means and variances of the magnesium distributions take the same form (with $(1-\hat{P}_{M_i}^{\{n\}})$ replacing $\hat{P}_{M_i}^{\{n\}}$ for estimates pertaining to the contaminant distribution).  We calculate the error associated with each parameter estimate by propagating the measurement error as well as the parameter errors from the previous iteration.  Formulae for computing the sizes of $1\sigma$ errorbars are derived in Appendix \ref{app:errors}.  

Recall that we assume about $p(a)$ only that it is non-increasing.  Therefore in the second part of the maximization step, we maximize the expected log-likelihood (Equation \ref{eq:mstep}) with respect to all non-increasing functions $p(a)$.  \citet{robertson88} show that the solution is the isotonic estimator of the form 
\begin{equation}
\hat{p}^{\{n+1\}}(a_i)=\displaystyle \min_{1\leq u \leq i}\biggl [\displaystyle\max_{i\leq v \leq N}\frac{\Sigma_{j=u}^v\hat{P}_{M_j}^{\{n\}}}{v-u+1}\biggr ],
  \label{eq:monotonic}
\end{equation}
where the notation $\max_{a\leq z \leq b}f(z)$ specifies the maximum value of $f(z)$ in the interval $a\leq z \leq b$, and $\min_{c\leq x \leq d}g(x)$ specifies the minimum value of $g(x)$ in the interval $c\leq x \leq d$.  It is computationally expensive to perform an exhaustive search for the minimum among the maxima at each data point; however, the monotonic regression is performed efficiently using the ``Pool-Adjacent-Violators'' algorithm \citep{grotzinger84}, which we describe in Appendix \ref{app:pav}.  

\section{Procedure}
\label{sec:procedure}

\begin{figure}
  \epsscale{1.2}
  \plotone{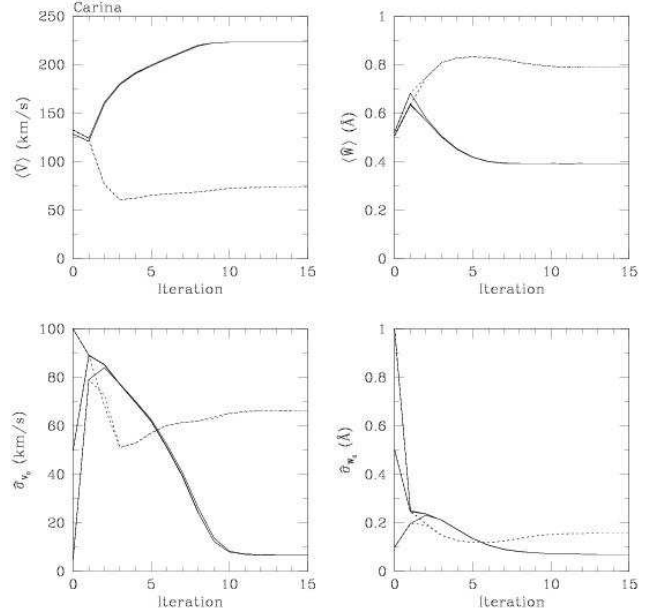}
  \caption{\scriptsize Parameter estimates $\hat{\zeta}\equiv$$\{\langle \hat{V} \rangle_{mem}$,$\hat{\sigma}_{V_0,mem}$,$\langle \hat{W} \rangle_{mem}$,$\hat{\sigma}_{W_0,mem}$,$\langle \hat{W} \rangle_{non}$,$\hat{\sigma}_{W_0,non}\}$ obtained in the first 15 iterations of the EM algorithm, applied here to the MMFS sample for the Carina dSph.  In each panel, solid curves represent the member population and dotted curves represent the nonmember population.  We have run the algorithm three times, using initial values $\{\sigma_{V_0,mem},\sigma_{W_0,mem},\sigma_{W_0,non}\}=\{5,0.1,0.1\}$, $\{50,0.5,0.5\}$, and $\{100,1.0,1.0\}$ in units of $\{\mathrm{km s}^{-1},\mathrm{\AA},\mathrm{\AA}\}$.  The final estimates are insensitive to the initialization, even for the large range of values considered.  For illustrative purposes only, the velocity mean and dispersion shown for nonmembers are calculated as in Equations \ref{eq:estimatemean}; the nonmember velocity distribution is non-Gaussian, and during the EM algorithm $p_{non}(v)$ is evaluated according to Equation \ref{eq:pbes} (see Section \ref{sec:emdsph}).} 
  \label{fig:car_members_iterations}
\end{figure}

\begin{figure}
  \epsscale{1.2}
  \plotone{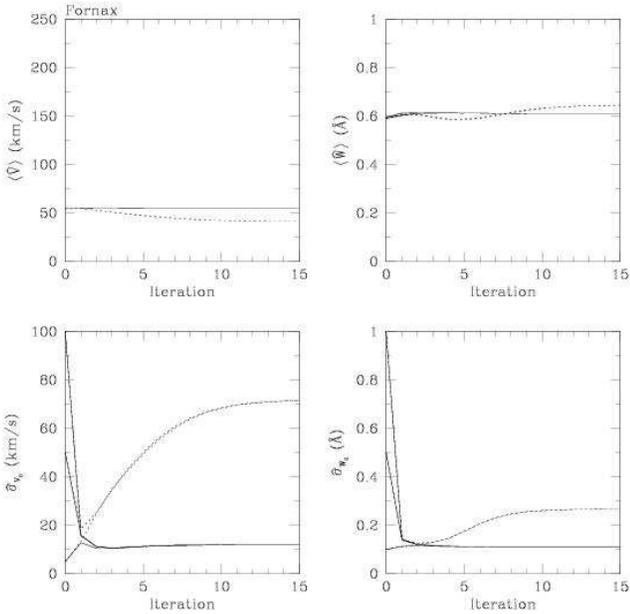}
  \caption{\scriptsize Same as Figure \ref{fig:car_members_iterations} but for Fornax data.}
  \label{fig:for_members_iterations}
\end{figure}

\begin{figure}
  \epsscale{1.2}
  \plotone{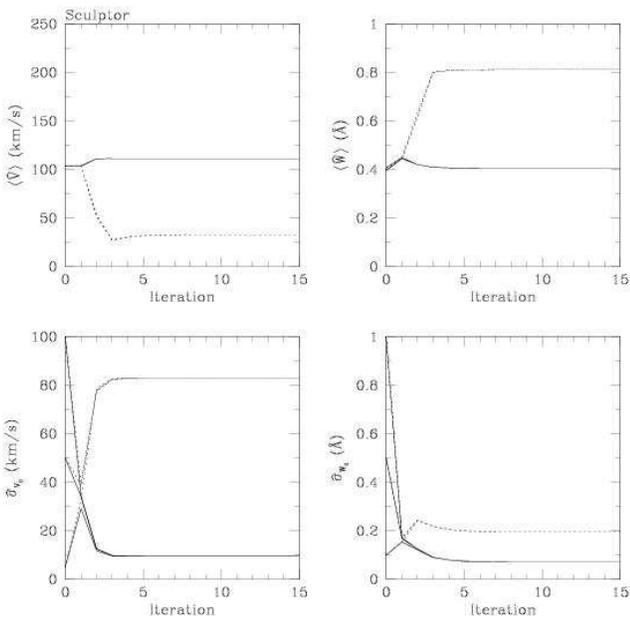}
  \caption{\scriptsize Same as Figure \ref{fig:car_members_iterations} but for Sculptor data.} 
  \label{fig:scl_members_iterations}
\end{figure}

\begin{figure}
  \epsscale{1.2}
  \plotone{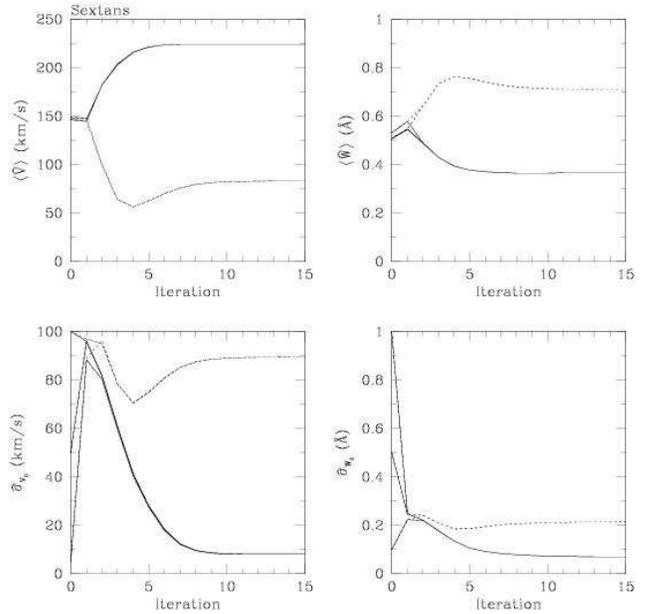}
  \caption{\scriptsize Same as Figure \ref{fig:car_members_iterations} but for Sextans data.} 
  \label{fig:sex_members_iterations}
\end{figure}

In our implementation of the EM algorithm, we initialize $\hat{p}^{\{0\}}(a_i)=0.5$ and $\hat{P}^{\{0\}}_{M_i}=0.5$ for all $i$.  We then initialize the variances $\sigma_{V_0,mem}^2$, $\sigma_{W_0,mem}^2$ and $\sigma_{W_0,non}^2$ and evaluate Equations \ref{eq:estimatemean} and \ref{eq:estimatedisp} to obtain intial estimates of all parameters $\hat{\zeta}^{\{0\}}$.  As Figures \ref{fig:car_members_iterations} - \ref{fig:sex_members_iterations} show, the final estimates are insensitive to the values used to initialize the variances.  

After initialization, the algorithm begins with the expectation step.  We use Equations \ref{eq:pmemestimate} - \ref{eq:estepestimate} and the initial estimates in $\hat{\zeta}^{\{0\}}$ to calculate $\hat{p}_{mem}^{\{1\}}(V_i,W_i)$, $\hat{p}_{non}^{\{1\}}(V_i,W_i)$, and $\hat{P}_{M_i}^{\{1\}}$ for all $i$.  Next, in the maximization step, we use Equations \ref{eq:estimatemean} and \ref{eq:estimatedisp} to update the parameter estimates in $\hat{\zeta}^{\{1\}}$.  After maximization, we use monotonic regression (see Appendix \ref{app:pav}) to evaluate Equation \ref{eq:monotonic}; the regression provides estimates $\hat{p}^{\{1\}}(a_i)$ for all $i$.  We then proceed to the next iteration, which begins again with the expectation step.  Parameter estimates typically stabilize after $\sim 15-20$ iterations.  Because the algorithm runs quickly, we iterate 50 times.  

Figures \ref{fig:car_members_iterations} - \ref{fig:sex_members_iterations} depict the evolution of the parameters in $\hat{\zeta}^{\{n\}}$ for each dSph and for several choices of variance initialization.  The estimates in $\hat{\zeta}^{\{n\}}$ obtained after 15-20 iterations do not depend on the initialization.  


\section{Results}
\label{sec:results}

 Table \ref{tab:membership} summarizes results from our application of the EM algorithm to the MMFS data for each dSph.  Column 2 lists the number of observed stars, $N$, and column 3 lists the number of members, defined by $\hat{N}_{M}\equiv \sum_{i=1}^{N} \hat{P}_{M_i}$.  Columns 4 - 9 then list the parameter estimates in $\hat{\zeta}$.  

The EM algorithm effectively separates likely members from nonmembers in the MMFS data.  Applying the algorithm to our entire MMFS sample, we find that $4946$ stars have $\hat{P}_{M}\geq 0.9$, $1918$ stars have $\hat{P}_{M}\leq 0.1$, and $239$ stars have $\hat{P}_{M}$ between these limits.  Just 97 stars have what we might call ``ambiguous'' membership, $0.25 \leq \hat{P}_{M} \leq 0.75$. Due to the bimodal distribution of $\hat{P}_{M}$ values, $\hat{N}_{M}$ typically exceeds by less than $1\%$ the number of stars having $\hat{P}_{M}\geq 0.9$.  Adding all probabilities, the entire MMFS data set contains $\hat{N}_{M}=5063$ dSph member stars.  

Marker colors in Figures \ref{fig:members_carfor} and \ref{fig:members_sclsex} indicate the values of the membership probabilities, $\hat{P}_M$, returned by the EM algorithm.  Black points signify the most likely nonmembers, with $\hat{P}_{M} \leq 0.01$.  For the remaining stars, bluer colors signify more likely membership: red, magenta, green, cyan, and blue points correspond to stars having $\hat{P}_{M}$ exceeding 0.01, 0.50, 0.68, 0.95, and 0.99, respectively.  Pairs of dotted, vertical lines enclose stars that would be accepted as members using a conventional, $3\sigma$ velocity threshold (Section \ref{subsec:sigmaclip}).  The nonmember velocity distributions predicted by the Besan{\c c}on models are normalized for each dSph according to the final membership fraction and plotted in red in the bottom subpanels in Figures \ref{fig:members_carfor} and \ref{fig:members_sclsex}.  We find that the velocity distributions of the observed nonmembers generally agree with the Besan{\c c}on predictions.  

\renewcommand{\arraystretch}{0.6}
\begin{deluxetable*}{lrrrrrrrrrrr}
  \tabletypesize{\scriptsize}
  \tablewidth{0pc}
  \tablecaption{ Results from EM algorithm for MMFS data}
  \tablehead{\colhead{Galaxy}&\colhead{$N$}&\colhead{$\hat{N}_{M}$}&\colhead{$\langle \hat{V} \rangle_{mem}$}&\colhead{$\hat{\sigma}_{V_0,mem}$}&\colhead{$\langle \hat{W} \rangle_{mem}$}&\colhead{$\hat{\sigma}_{W_0,mem}$}&\colhead{$\langle \hat{W} \rangle_{non}$}&\colhead{$\hat{\sigma}_{W_0,non}$}\\
    \colhead{}&\colhead{}&\colhead{}&\colhead{(km s$^{-1}$)}&\colhead{(km s$^{-1}$)}&\colhead{(\AA)}&\colhead{(\AA)}&\colhead{(\AA)}&\colhead{(\AA)}
  }
\startdata
Carina&1982&$774$&$222.9\pm 0.1$&$6.6\pm 1.2$&$0.40\pm 0.01$&$0.07\pm 0.02$&$0.83\pm 0.01$&$0.13\pm 0.02$\\
Fornax&2633&$2483$&$55.2\pm 0.1$&$11.7\pm 0.9$&$0.59\pm 0.01$&$0.12\pm 0.02$&$0.51\pm 0.01$&$0.26\pm 0.06$\\
Sculptor&1541&$1365$&$111.4\pm 0.1$&$9.2\pm 1.1$&$0.39\pm 0.01$&$0.07\pm 0.01$&$0.77\pm 0.01$&$0.22\pm 0.05$\\
Sextans&947&$441$&$224.3\pm 0.1$&$7.9\pm 1.3$&$0.36\pm 0.01$&$0.05\pm 0.02$&$0.72\pm 0.01$&$0.21\pm 0.03$\\
\enddata
\label{tab:membership}
\end{deluxetable*}

\begin{figure*}
  \plotone{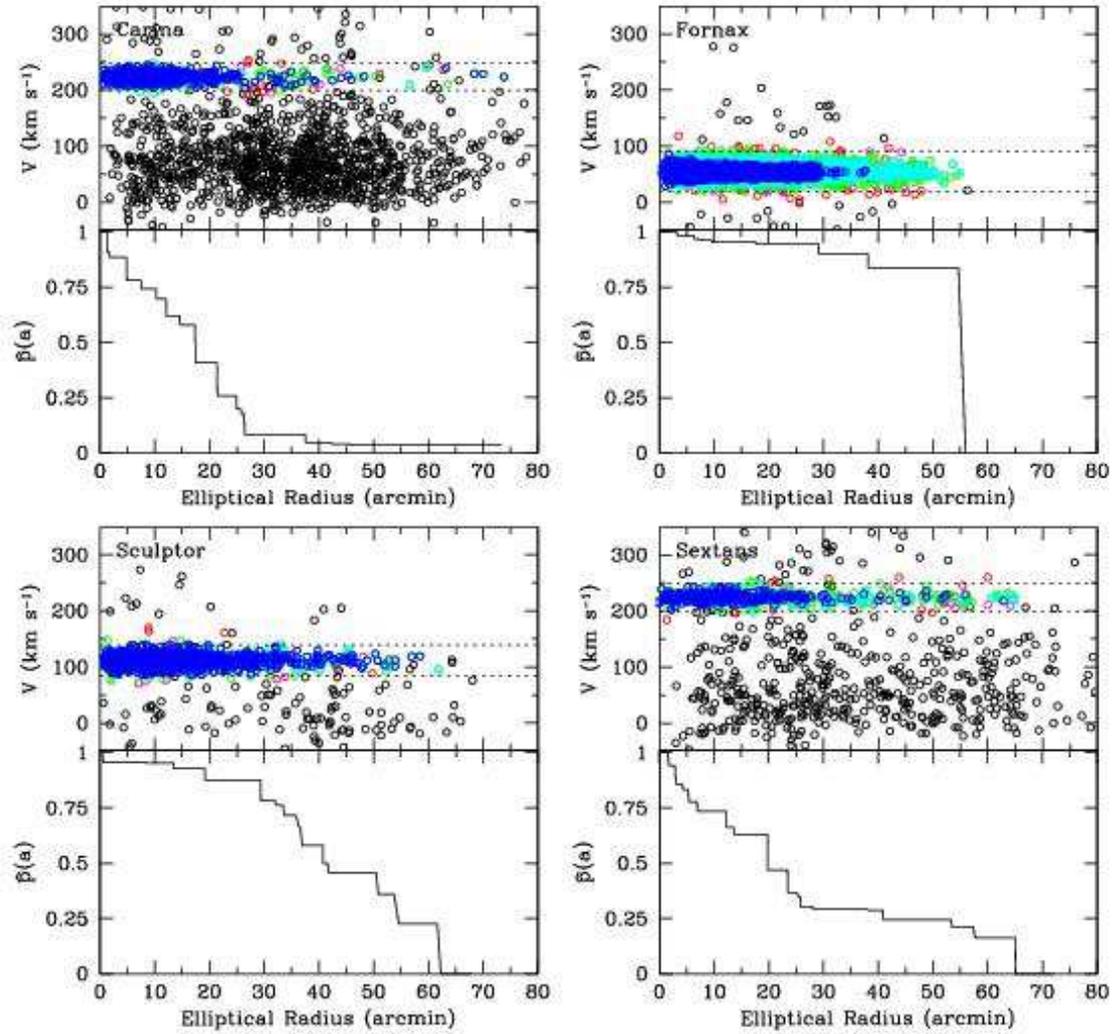}
  \caption{\scriptsize Velocity distribution and membership fraction versus elliptical radius.  The top sub-panels for each galaxy plot stellar velocity versus elliptical radius.  Marker colors indicate $\hat{P}_{M}$ as in Figure \ref{fig:members_carfor}.  The bottom sub-panels plot the membership fraction versus elliptical radius, estimated via monotonic regression. Dotted horizontal lines enclose the conventional 3$\sigma$ velocity range.}  
  \label{fig:members_vr}
\end{figure*}
\begin{figure*}
  \plotone{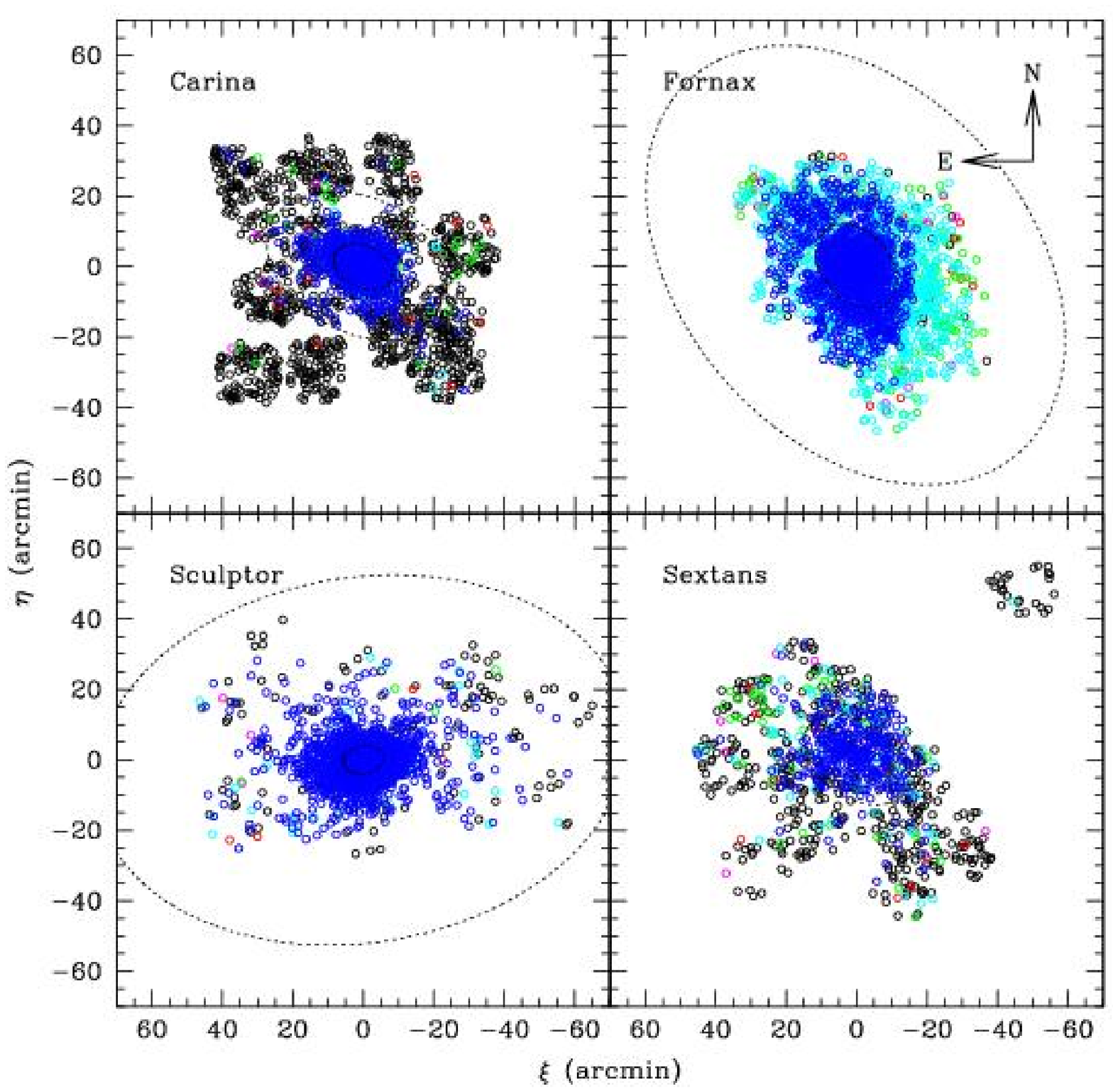}
  \caption{\scriptsize Maps indicating probability of membership.  As in Figure \ref{fig:members_carfor}, marker color indicates probability, $P_{M}$, that the star is a dSph member.  Black (red; magenta; green; cyan; blue) markers signify $\hat{P}_{M} \leq 0.01$ ($\hat{P}_{M} > 0.01; > 0.50; > 0.68; > 0.95; > 0.99$).  Ellipses mark King core and ``tidal'' radii (the King outer radius of Sextans is beyond the window dimensions), adopted from IH95.}  
  \label{fig:members_map}
\end{figure*}

Upper sub-panels in Figure \ref{fig:members_vr} plot the measured velocities as a function of elliptical radius.  Bottom sub-panels plot the monotonic regression estimate $\hat{p}(a)$, the member fraction as a function of elliptical radius.  We note that $\hat{p}(a)$ need not have a steep slope.  Our simple assumption that $p(a)$ is non-increasing allows the data to determine the dependence of membership on position.  Figure \ref{fig:members_map} maps sky positions of the observed stars, again using marker color to indicate membership.


\section{Performance of the EM Algorithm Using Simulated Data}
\label{sec:performance}

In order to evaluate the performance of the EM algorithm under a variety of conditions, we test it on simulated data sets that differ in sample size, level of contamination, and amount of velocity overlap between member and contaminant distributions.  In all simulations we draw a fraction $f_{mem}$ of $N$ artificial data points from a dSph member population that we assume to have a Gaussian velocity distribution with dispersion $10$ km s$^{-1}$, characteristic of the dSphs observed with MMFS.  We further assume that the dSph has a Plummer surface brightness profile, such that members have projected radii drawn with probability $p(R)\propto (1+R^2/r_p^2)^{-2}$ and sampled to a maximum radius of $R\leq 5r_p$.  We draw magnesium indices for the dSph members directly from the real MMFS measurements for the most likely ($\hat{P}_M \geq 0.9$) Carina, Sculptor and Sextans members (we exclude Fornax, the most metal-rich of the observed dSphs, because of its systematically larger Mg values).  

We draw a fraction $1-f_{mem}$ of the simulated data points from a contaminant population that we assume to follow a uniform spatial distribution over the sampled region.  We draw contaminant velocities from the Besan{\c c}on model of the Carina foreground (Equation \ref{eq:pbes}, see also the bottom-left panel of Figure \ref{fig:members_carfor}).  We draw contaminant $\Sigma$Mg values from the real MMFS measurements of the least likely ($\hat{P}_M \leq 0.1$) Carina, Sculptor and Sextans members.  To each artificial data point we add a small scatter (in velocity and $\Sigma$Mg) that corresponds to measurement errors drawn randomly from the real MMFS data.  

We consider the performance of the EM algorithm while varying the following three quantities: sample size $N$, member fraction $f_{mem}$, and member mean velocity $\langle V\rangle_{mem}$.  The mean velocity controls the amount of overlap between member and contaminant distributions.  We consider three sample sizes to represent realistic dSph data sets.  The smallest have 30 stars, typical of the samples now available for the faintest nearby dSphs (e.g., \citealt{simon07,martin07}).  The intermediate-size samples have 300 stars and are thus similar to many of the samples published for brighter dSphs within the past five years (e.g., \citealt{kleyna02,walker06a,munoz06a,koch07a,mateo08}).  The largest simulated samples have 3000 stars, of the same magnitude as the MMFS samples.  For each sample size, we consider three possible member fractions --- $f_{mem}=0.2,0.5,0.8$ --- that range from contaminant-dominated to member-dominated.  For each case of sample size and member fraction, we then consider three possible values for the dSph mean velocity: $\langle V\rangle_{mem}=50$, $100$, and $200$ km s$^{-1}$.  Since the artificial foreground peaks at a velocity of $\sim 50$ km s$^{-1}$, these values represent a range in degree of overlap between dSph and contaminant velocities.  

We apply the EM algorithm to each simulated data set and record the resulting estimates of the member mean velocity and velocity dispersion, as well as the membership probabilities determined for each star.  Figures \ref{fig:sim_members_30_50}$-$\ref{fig:sim_members_3000_200} in Appendix \ref{app:sim} display the artificial data and demonstrate the effectiveness with which the EM algorithm distinguishes members from contaminants in each of the 27 simulations.  Left-hand panels plot projected radius and $\Sigma$Mg against velocity, using blue circles to represent simulated dSph members and black squares to represent contaminants.  Middle panels show the same scatter plots, but with marker color now indicating, as in Figures \ref{fig:members_carfor} and \ref{fig:members_sclsex}, the membership probabilities returned by the EM algorithm.  Right-hand panels identify any stars that are mis-classified by the EM algorithm---i.e., member stars for which the EM algorithm gives $\hat{P}_M < 0.5$ and/or contaminants for which $\hat{P}_M > 0.5$. 

In 26 of the 27 simulations, the number of mis-classified stars amounts to less than $10\%$ of the total sample, and the typical number of mis-classified stars is between $1\%-3\%$ of the total sample.  In 23 of the 27 simulations, the EM algorithm returns accurate (i.e., the error bar includes the true value) estimates of the dSph mean velocity and velocity dispersion.  Among the intermediate-sized and large samples ($N\geq 300$), the algorithm's lone failure occurs for the heavily contaminated ($f_{mem}=0.2$) $N=300$ sample for which the dSph mean velocity of $\langle V\rangle_{mem}=50$ km s$^{-1}$ coincides with the peak of the contaminant distribution.  Under those circumstances the algorithm mis-classifies nearly two-thirds of all stars, and the measured velocity dispersion is effectively that of the foreground (see Figure \ref{fig:sim_members_300_50}, bottom-left).   

The three other cases of inaccurate estimates all occur for simulations of small ($N=30$) samples.  As one might expect, the case with with the heaviest contamination ($f_{mem}=0.2$) and most velocity overlap ($\langle V\rangle_{mem}=50$ km s$^{-1}$) yields an inflated estimate of the velocity dispersion (see Figure \ref{fig:sim_members_30_50}, bottom left).  The two remaining cases of failure at $N=30$ each merit some discussion.  The simulation with heavy contamination ($f_{mem}=0.2$) and moderate overlap ($\langle V\rangle_{mem}=100$ km s$^{-1}$) yields an \textit{under}-estimate of the velocity dispersion.  This particular failure results from a mis-classification of just two stars (among just six members).  Inspection of Figure \ref{fig:sim_members_30_100} (bottom left) reveals that the member that is mistakenly classified as a contaminant occupies the extreme tails in the member distributions of velocity, position and Mg, while the contaminant mistakenly identified as a member lies near the center of the member distributions.  A reasonable person examining the data shown in Figure \ref{fig:sim_members_30_100} (bottom left) would be likely to mis-classify these two stars in the same way as the algorithm.  In the simulation with moderate contamination ($f_{mem}=0.5$) and minimal overlap ($\langle V\rangle_{mem}=200$ km s$^{-1}$) the algorithm correctly classifies all stars but, as happens with non-negligible probability for data sets of $N\leq 15$, the sample simply exhibits a smaller velocity dispersion than the population from which it is drawn (see Figure \ref{fig:sim_members_30_200}, top right).  Thus these final two cases of failure do little to incriminate the EM algorithm, but rather remind us of the perils of working with small samples.  We conclude that, barring a conspiracy among heavy contamination, close velocity overlap and small-number statistics, the EM algorithm provides an excellent tool for quantifying the properties of dSph kinematic samples.

\subsection{On the Faintest dSphs}

Mining of data from the Sloan Digital Sky Survey (SDSS) has recently uncovered a new population of extremely faint dSphs in the Local Group, with absolute magnitudes as small as $M_V\sim -2$ (e.g., \citealt{belokurov07}).  Initial kinematic studies of these systems suggest that they have velocity dispersions as small as $\sim 4$ km s$^{-1}$, systematically colder than those of brighter dSphs \citep{simon07,martin07}.  In order to test the efficacy of the EM algorithm in the regime relevant to these systems, we consider nine additional simulations that use samples of $N=30$ stars drawn from member distributions with velocity dispersions of $4$ km s$^{-1}$.  We again consider member fractions of $f_{mem}=0.2,0.5,0.8$ and mean velocities $\langle V\rangle_{mem}=50,100,200$ km s$^{-1}$).  

The results of these nine simulations (see Figures \ref{fig:sim_smalldisp_members_30_50}$-$\ref{fig:sim_smalldisp_members_30_200} of Appendix \ref{app:sim}) are similar to those for the small samples previously discussed.  The EM algorithm returns accurate estimates in six of the nine simulations, the same rate of success as in the nine previous simulations with $N=30$ and $\sigma_{V_0,mem}=10$ km s$^{-1}$.  Two of the three failures occur for the samples that have the heaviest contamination ($f_{mem}=0.2$) and most velocity overlap ($\langle V\rangle=50,100$ km s$^{-1}$).  The third ($f_{mem}=0.5$, $\langle V\rangle_{mem}=100$ km s$^{-1}$) barely fails, returning a velocity dispersion of $\sigma_{V_0,mem}=8.5\pm 4.0$ km s$^{-1}$ after mis-classifying just one star.  With these simulations we find that the efficacy of the EM algorithm is insensitive to the member velocity dispersion, so long as that dispersion is sufficiently smaller---as is the case for all dSphs---than that of the contaminant population.  Of course the coldest systems require extreme caution if the velocity dispersion is similar to the velocity errors, as is the case for some dSphs in the samples of \citet{simon07} and \citet{martin07}.  In such situations the algorithm's performance depends on external factors affecting the validity of the velocity dispersion estimate itself, such as the degree to which the measurement errors are known (see Walker et al. in prep).  

\subsection{Improvement via a velocity filter}
\label{subsec:filter}

Because kinematic samples for the faintest dSphs are likely to suffer heavy contamination, one hopes to improve the reliability of the EM algorithm in that regime.  Often it is possible to produce a reasonable guess of the mean velocity upon visual inspection of the $V$, $R$, and (if available) $\Sigma$Mg scatter plots (e.g., Figures \ref{fig:sim_smalldisp_members_30_50}$-$\ref{fig:sim_smalldisp_members_30_200}), even for small and contaminated samples.  In such cases, application of an initial velocity ``filter'' can restore the effectiveness of the algorithm.  We have repeated all 36 simulations after initializing all stars within $40$ km s$^{-1}$ of the dSph mean velocity to have $\hat{P}_{M}^{\{0\}}=1$, while all stars outside this range receive $\hat{P}_{M}^{\{0\}}=0$ (recall that in the ``unfiltered'' algorithm, all stars receive initialization $\hat{P}_{M}^{\{0\}}=0.5$).  We find that the filter corrects all noted cases of overestimated velocity dispersions.  However, the filter does not correct the noted cases of underestimated dispersions for the $N=30$ samples, as these failures continue to reflect the difficulties that arise from small-number statistics.  Tables \ref{tab:sim} and \ref{tab:sim_smalldisp} list parenthetically the parameter estimates obtained after applying the velocity filter.

\section{Comparison with Other Methods}

We now use the simulated data sets to compare parameter and membership estimates obtained from the EM algorithm to those obtained using more conventional methods for contaminant removal.  We consider two such methods.  The first is the simple and widely used sigma-clipping algorithm.  The second, advocated recently by \citet{klimentowski07} and \citet{wojtak07}, identifies and rejects stars that are not gravitationally bound to the dSph, according to a mass estimator based on the virial theorem.  Like the EM algorithm, both methods require estimation of the mean velocity, and the sigma-clipping algorithm also requires estimation of the velocity dispersion.  For consistency we continue to use Equations \ref{eq:estimatemean} and \ref{eq:estimatedisp} to estimate these parameters.  Here it is important to remember, however, one of the key differences between the EM and the alternative algorithms.  While the EM algorithm deals with membership probabilities, the non-EM algorithms consider a given star as either a member or a nonmember.  Therefore, in executing either of the non-EM algorithms, we initialize $P_{M_i}^{\{0\}}=1$ for all $i$, and stars that are rejected during the course of the algorithm receive $P_{M}=0$ during all subsequent iterations.

\subsection{Sigma Clipping}
\label{subsec:sigmaclip}

Sigma-clipping algorithms iteratively estimate the variance of some variable among member stars and then remove stars with outlying values.  Historically, most dSph kinematic samples provide only velocity information (no line strength information).  Hence a sigma-clipping routine typically involves estimating the member mean velocity and velocity dispersion, then rejecting all stars with velocities that differ from the mean by more than some multiple of the velocity dispersion.  The choice of threshold is arbitrary, with typical values between $2.5\sigma_{V_0}$ and $4\sigma_{V_0}$.  Here we consider a $3\sigma$ clipping algorithm.  In executing the (unfiltered) algorithm we initialize $P_{M_i}^{\{0\}}=1$ for all $i$.  Then each iteration of the algorithm consists of the following two steps: 1) estimate the mean velocity and velocity dispersion via 50 iterations of equations \ref{eq:estimatemean} and \ref{eq:estimatedisp}, then 2) assign $P_{M}=0$ to all stars with velocities more than $3\hat{\sigma}_{V_0}$ from the mean.  We terminate the algorithm after the iteration in which no stars are newly rejected.  We record the estimates of the mean velocity and velocity dispersion obtained in the last iteration, and we estimate the number of members as $\sum_{i=1}^N\hat{P}_{M_i}$, which in this case equals the number of stars for which $\hat{P}_{M}=1$.  

\subsection{Virial Theorem}
\label{subsec:vt}

\citet{klimentowski07} and \citet{wojtak07} show that a second technique, which they apply after an initial sigma clip, can remove contaminants more reliably than sigma-clipping alone.  This method, originally proposed in an anlaysis by \citet{denhartog96} of galaxy cluster kinematics, seeks explicitly to distinguish stars that are gravitationally bound to the dSph from those that are unbound.  It incorporates a dynamical analysis that employs a mass estimator derived from the virial theorem.  After sorting stars in sequence from smallest to largest projected radius and removing nonmembers (identified here as stars for which $\hat{P}_{M}=0$), the mass estimator is given by \citep{heisler85}
\begin{equation}
  M_{VT}(r)=\frac{3\pi N \sum_{i=1}^{N}(V_i-\langle V\rangle_{mem})^2}{2G\sum_{j=2}^{N}\sum_{i=1}^{j-1}1/D_{i,j}},  
  \label{eq:vt}
\end{equation}
where $N$ is the number of stars with projected radius $R<r$  and $D_{i,j}$ is the projected distance between the $i^{th}$ and $j^{th}$ stars.  Each rejection iteration is preceded by estimation of the mean velocity (and velocity dispersion, although this quantity is not used by the algorithm) using Equations \ref{eq:estimatemean} and \ref{eq:estimatedisp}.  In this ``virial theorem'' (VT) technique, a star at radius $R$ is rejected (and assigned $\hat{P}_{M}=0$) if its velocity differs from $\langle V\rangle_{mem}$ by more than $\sqrt{2GM_{VT}(r)/r}$ evaluated at $r=R$.  Thus the VT algorithm goes beyond the sigma-clipping algorithm in that it considers the projected positions of the stars as well as their velocities.  As in the EM algorithm, marginal outliers are less likely to be considered members if they are projected large distances from the dSph center, but here the explicit reason is that such stars are less likely to be gravitationally bound to the dSph.  

The VT algorithm requires initialization of $\langle \hat{V}\rangle_{mem}$ and the values $\hat{P}_{M_i}$ for all $i$ (effectively a velocity cut); we consider two possible routes.  First, in what we denote as the ``VT$_{3\sigma}$'' algorithm, we follow \citet{klimentowski07} in using the $3\sigma$ clipping algorithm to provide the initialization.  Second, in what we call the ``VT$_{EM}$'' algorithm, we intialize $\langle \hat{V}\rangle_{mem}$ with the value obtained from the EM algorithm, and then assign $\hat{P}_{M}=0$ to all stars for which the EM algorithm returned $\hat{P}_{M}<0.5$, and $\hat{P}_{M}=1$ to all stars for which the EM algorithm returned $\hat{P}_{M}\geq 0.5$.

Each iteration of the VT algorithm then consists of the following three steps: 1) estimate the mean velocity and velocity dispersion via 50 iterations of equations \ref{eq:estimatemean} and \ref{eq:estimatedisp}, then 2) calculate $M_{VT}(r)$ from Equation \ref{eq:vt}, and finally 3) assign $P_{M}=0$ to all stars with velocities more than $\sqrt{2GM_{VT}(r)/r}$ from the mean.  As with the sigma clipping algorithm, we terminate the VT algorithm after the iteration in which no stars are newly rejected, recording final estimates of the mean velocity and velocity dispersion as well as the number of members, $\sum_{i=1}^N\hat{P}_{M_i}$.

\subsection{Alternative Versions of the EM algorithm}

Finally, we consider alternative versions of the EM algorithm.  The full version that we advocate and have described  above uses velocity, magnesium and position information.  But not all kinematic data sets have associated line strength information, and one may reasonably wonder about the effect of considering the stellar positions.  In order to examine the performance of the EM algorithm in the absence of line-strength data, and in order to examine the relevance of the positional information, we consider versions of the EM algorithm that do not use these bits of information.  

\subsubsection{EM$_V$: Velocity Only}

We denote as ``EM$_V$'' a version of the EM algorithm that considers only the velocity data.  Execution of the algorithm differs from that of the full EM algorithm in just two respects.  In the expectation step we evaluate $p_{mem}(V_i)$ using the univariate Gaussian probability given by Equation \ref{eq:pmemnomg}, rather than the bivariate probability $p_{mem}(V_i,W_i)$ given by Equation \ref{eq:pmem}.  Second, we do not perform the monotonic regression estimate of $p(a)$; instead, after the maximization step in the $n^{th}$ iteration we simply set $\hat{p}^{\{n\}}(a_i)$ equal to the global member fraction, $N^{-1}\Sigma_{i=1}^N\hat{P}_{M_i}^{\{n\}}$, for all $i$.  This effectively removes any influence of position on the membership probabilities and parameter estimates.  Thus the EM$_V$ algorithm operates only on velocity data, similar in that respect to a conventional sigma clipping algorithm.  The EM$_V$ algorithm differs from the sigma clip in that it assigns membership probabilities rather than enforcing an arbitrary threshold, and it uses the expected distribution of foreground velocities to help evaluate these probabilities.  

\subsubsection{EM$_{VR}$: Velocity and Position Only}

We denote as ``EM$_{VR}$'' a version of the EM algorithm that considers the velocity and position data but ignores the magnesium index.  This method is therefore applicable to most kinematic data sets.  Its execution differs from that of the full EM algorithm only in the expectation step.  There, we evaluate $p_{mem}(V_i)$ using the univariate Gaussian probability given by Equation \ref{eq:pmemnomg}, rather than the bivariate probability $p_{mem}(V_i,W_i)$ given by Equation \ref{eq:pmem}.  

\subsubsection{EM$_{VW}$: Velocity and Magnesium Only}

Finally, we denote as ``EM$_{VW}$'' a version of the EM algorithm that considers the velocity and magnesium data but ignores the stellar positions.  By comparing its effectiveness to that of the full EM algorithm we can judge the the importance of including the position data.  The execution of the EM$_{VW}$ algorithm differs from that of the full EM algorithm only in that we do not perform the monotonic regression estimate of $p(a)$.  Instead, as in the EM$_{V}$ algorithm, after the maximization step in the $n$th iteration we set $\hat{p}^{\{n\}}(a_i)$ equal to the global member fraction, $N^{-1}\Sigma_{i=1}^N\hat{P}_{M_i}^{\{n\}}$, for all $i$, removing the influence of position on the membership probabilities and parameter estimates.  

\subsection{Comparison of Results Using Simulated Data}

Tables \ref{tab:sim} (for the 27 simulations in which $\sigma_{V_0,mem}=10$ km s$^{-1}$) and \ref{tab:sim_smalldisp} (for the nine simulations in which $\sigma_{V_0,mem}=4$ km s$^{-1}$) list the results of applying each of the algorithms described above to the artificial data sets.  Underneath the row that identifies the data set, columns list the estimated number of members, $N_{mem}=\sum_{i=1}^N\hat{P}_{M_i}$, followed by the number of mis-classified stars---i.e., the number of members for which $\hat{P}_{M}<0.5$ plus the number of nonmembers for which $\hat{P}_{M}>0.5$.  Columns 5 and 6 list estimates of the dSph mean velocity and velocity dispersion, obtained from iterating Equations \ref{eq:estimatemean} and \ref{eq:estimatedisp} using the membership probabilities obtained from the algorithm.  Parenthetical values indicate estimates obtained after implementing the initial velocity filter described in Section \ref{subsec:filter}.

We judge the performance of each algorithm according to whether or not it recovers (within the error bars) the mean velocity and velocity dispersion of the simulated member population.  The final columns in Tables \ref{tab:sim} and \ref{tab:sim_smalldisp} indicate with either a ``Y'' or ``N'' whether the algorithm satisfies this criterion.  As previously discussed (Section \ref{sec:performance}), the unfiltered EM algorithm succeeds in 23 of 27 simulations with $\sigma_{V_0,mem}=10$ km s$^{-1}$ (and in 6 of 9 with $\sigma_{V_0,mem}=4$ km s$^{-1}$), and performs well in all simulations in which the sample has either $N > 30$ or $f_{mem} > 0.2$.  Use of an initial velocity filter helps the algorithm's performance in cases of heavy contamination---the filtered EM algorithm succeeds in 32 of 36 simulations.

\subsubsection{EM versus sigma clipping and virial theorem}

In contrast, the unfiltered $3\sigma$ algorithm succeeds in only four of 27 simulations with $\sigma_{V_0,mem}=10$ km s$^{-1}$, and in only two of nine with $\sigma_{V_0,mem}=4$ km s$^{-1}$.  The unfiltered $3\sigma$ algorithm clearly has difficulty in the presence of moderate to heavy contamination, failing in every simulation in which $f_{mem} < 0.8$.  Because the $3\sigma$ algorithm uses only velocity information, its reliability depends critically on the contrast between member and contaminant velocity distributions.  In all cases of its failure, the $3\sigma$ algorithm errs by failing to classify nonmembers as such, resulting in overestimates of the velocity dispersion.  

The $3\sigma$ algorithm performs more reliably if we supply an initial velocity filter (see Section \ref{subsec:filter}).  The filtered $3\sigma$ algorithm succeeds in nine simulations for which it otherwise fails.  However, we note that even without a filter, the EM algorithm generally performs well and distinguishes members from contaminants more reliably than even the filtered $3\sigma$ algorithm.  Clearly, one benefits by considering the contaminant distribution and using the additional information from positions and/or spectral indices when it is available.

The performance of the VT$_{3\sigma}$ algorithm generally depends on the performance of the $3\sigma$ algorithm that provides its initialization.  Counting all 72 filtered and unfiltered runs, the VT$_{3\sigma}$ algorithm succeeds in all 21 cases in which the $3\sigma$ algorithm succeeds, and fails in 43 of the 51 cases in which the $3\sigma$ algorithm fails.  In most cases the VT$_{3\sigma}$ algorithm manages to reject a handful of nonmembers that are unrecognized as such by the $3\sigma$ algorithm, and so it yields slightly less inaccurate parameter estimates.  In eight cases the VT$_{3\sigma}$ algorithm is able to compensate sufficiently to recover accurate paramter estimates where the $3\sigma$ algorithm fails.  Similarly, the performance of the VT$_{EM}$ algorithm traces that of the EM algorithm, succeeding in all but one case in which the EM algorithm succeeds, and failing in all but one case in which the EM algorithm fails.

It is important to note that the VT method has a different specific purpose than either of the methods chosen for its initialization.  It seeks explicitly to isolate a clean sample of \textit{gravitationally bound} dSph members \citep{klimentowski07,wojtak07}.  If such a sample is desired, as it is when applying equilibrium models, then the VT is appropriate to use after identifying likely foreground stars using the EM algorithm.  The VT method is then complementary to the EM algorithm in this application.

\subsubsection{Which EM algorithm?}
\label{subsubsec:whichem}

The EM algorithm clearly outperforms the sigma clipping algorithm, and provides the appropriate initialization for the VT algorithm when desired.  But, assuming Mg data are available, which EM algorithm is best?  Interestingly, the EM$_{VR}$ algorithm, which excludes the magnesium information, enjoys the same rate of success as the full EM algorithm.  The EM$_{VR}$ algorithm returns accurate parameter estimates in 23 of the 27 simulations with $\sigma_{V_0,mem}=10$ km s$^{-1}$ (and in six of the nine simulations with $\sigma_{V_0,mem}=4$ km s$^{-1}$).  The EM$_{V}$ and EM$_{VW}$ algorithms both succeed in 15 of 27 simulations with $\sigma_{V_0,mem}=10$ km s$^{-1}$ (and six of the nine simulations with $\sigma_{V_0,mem}=4$ km s$^{-1}$). 

These results prompt two conclusions.  First, the superior performance of the full EM and EM$_{VR}$ algorithms imply that it is important to consider the positions of stars when identifying contaminants.  Given that the projected density of members declines as the density of contaminants remains approximately uniform, the \textit{a priori} probability of membership decreases with projected distance from the dSph center.  Algorithms that incorporate this trend outperform those that do not.  Second, the ability of the EM$_{VR}$ algorithm to recover the correct parameters in some cases where the full EM algorithm fails suggests that there is room to improve our model of the Mg index distribution.  Where the full EM algorithm fails, it fails because it mis-classifies contaminants with small Mg-index values as members.  These contaminants correspond to metal-poor giants in the Milky Way halo, whereas our Gaussian model of the contaminant Mg distribution allows for only a single component that is dominated by metal-rich dwarfs in the Milky Way disk.  The full EM algorithm may therefore benefit from a contaminant model in which the Mg distribution is a double-Gaussian, allowing for a distinct halo component.  We have not adopted such a model here because the single-component model performs well in all simulated data sets of comparable size to our MMFS samples.  A two-component foreground model may become necessary, however, to optimize performance of the EM algorithm with small, heavily contaminated samples typical of those obtained for ultra-faint dSphs \citep{simon07,martin07}.  


\section{Discussion and Summary}

We have introduced the EM method as an effective tool with which to attack the problem of sample contamination.  In our implementation, the EM algorithm operates on all the available velocity, magnesium-index and positional information.  This marks a significant improvement over contaminant-removal methods that rely on velocity alone, particularly the conventional sigma-clipping method.  Further, with appropriate modification of the likelihood function (Equation \ref{eq:likelihood}), the EM algorithm can easily be tailored to incorporate other sorts of data as diagnostics of membership (e.g., photometric data, as in \citealt{gilbert06}).  Applied to our MMFS data, the EM algorithm evaluates the probability that a given star is a dSph member, while simultaneously estimating the mean velocity, velocity dispersion, and fraction of members as a function of position.  


Within the context of applying equilibrium dynamical models, \citet{klimentowski07} and \citet{wojtak07} discuss the importance of removing a second type of contaminant---former dSph members that have been tidally stripped and are thus no longer gravitationally bound to the dSph.  These authors prescribe an iterative procedure that uses the virial theorem to provide a nonparameteric estimate of the dSph mass profile, and then removes stars with velocities greater than the local circular velocity (Section \ref{subsec:vt}).  For the following reasons, this method should not be viewed as an ``alternative'' to the EM algorithm described above.  First, while it uses velocity and position in applying the virial theorem, it does not use $\Sigma$Mg, which we have shown to be a powerful, independent indicator of dSph membership.  Second, and more fundamentally, it has a different and more specific purpose---the identification of a clean sample of gravitationally bound dSph members.  If such a sample is desired, as it is when applying equilibrium models, then the virial theorem method of \citet{klimentowski07} and \citet{wojtak07} can still be applied appropriately after identifying likely foreground stars using the EM algorithm.  

MGW and MM thank the Horace H. Rackham Graduate School at the University of Michigan for generous support, including funds for travel to Las Campanas Observatory.  MM acknowledges support from NSF grants AST-0206081  0507453, and 0808043.  EO acknowledges support from NSF Grants AST-0205790, 0505711, and 0807498.


\bibliography{ref}

\input{table2}

\input{table3}
\input{astroph_appendix}

\end{document}

%% file: table2.tex
\clearpage
\renewcommand{\arraystretch}{0.6}
\begin{longtable*}{lccccccccccl}
  \tabletypesize{\scriptsize}
  \tablewidth{0pc}
  \tablecaption{\scriptsize Results of Algorithms\tablenotemark{a} with Simulated Data and $\sigma_{V_0}=10$ km s$^{-1}$}
  \tablehead{
\colhead{}&\colhead{$N$}&\colhead{$\hat{N}_{mem}$}&\colhead{$N_{wrong}$}&\colhead{$\langle \hat{V}\rangle_{mem}$}&\colhead{$\hat{\sigma}_{V_0}$}&\colhead{Success?}\\
    \colhead{}&\colhead{}&\colhead{}&\colhead{}&\colhead{(km s$^{-1}$)}&\colhead{(km s$^{-1}$)}
  }
\textbf{Simulation}&$\mathbf{N=   30}$&$\mathbf{N_{mem}=   24}$&&$\mathbf{\langle V\rangle_{mem}=   50\mathrm{km s}^{-1}}$&$\mathbf{\sigma_{V_{0,mem}}=   10\mathrm{km s}^{-1}}$\\
EM             &\nodata&$   23(   23)$&$    0(    0)$&$    48.8\pm      0.6(    48.8\pm      0.6)$&$    12.2\pm      3.2(    12.2\pm      3.2)$&Y(Y)\\
EM$_{V}$       &\nodata&$   23(   23)$&$    2(    2)$&$    47.9\pm      0.5(    47.8\pm      0.5)$&$    14.0\pm      3.0(    13.7\pm      3.0)$&N(N)\\
EM$_{VR}$      &\nodata&$   23(   23)$&$    1(    1)$&$    48.9\pm      0.6(    48.9\pm      0.6)$&$    12.8\pm      3.3(    12.8\pm      3.3)$&Y(Y)\\
EM$_{VW}$      &\nodata&$   23(   23)$&$    0(    0)$&$    48.7\pm      0.6(    48.7\pm      0.6)$&$    12.2\pm      3.2(    12.2\pm      3.2)$&Y(Y)\\
$3\sigma$&\nodata&$   29(   27)$&$    5(    3)$&$    51.0\pm      0.5(    47.0\pm      0.5)$&$    20.6\pm      3.4(    15.3\pm      3.2)$&N(N)\\
VT$_{3\sigma}$&\nodata&$   26(   25)$&$    2(    3)$&$    48.5\pm      0.5(    46.1\pm      0.5)$&$    13.7\pm      3.1(    15.1\pm      3.3)$&N(N)\\
VT$_{EM}$&\nodata&$   22(   22)$&$    2(    2)$&$    48.6\pm      0.6(    48.6\pm      0.6)$&$    12.6\pm      3.3(    12.6\pm      3.3)$&Y(Y)\\
 \\

\textbf{Simulation}&$\mathbf{N=   30}$&$\mathbf{N_{mem}=   15}$&&$\mathbf{\langle V\rangle_{mem}=   50\mathrm{km s}^{-1}}$&$\mathbf{\sigma_{V_{0,mem}}=   10\mathrm{km s}^{-1}}$\\
EM             &\nodata&$   16(   16)$&$    2(    2)$&$    46.1\pm      0.5(    46.1\pm      0.5)$&$     7.9\pm      2.7(     7.9\pm      2.7)$&Y(Y)\\
EM$_{V}$       &\nodata&$   12(   12)$&$    5(    5)$&$    48.2\pm      0.5(    48.2\pm      0.5)$&$     5.3\pm      1.9(     5.3\pm      1.9)$&N(N)\\
EM$_{VR}$      &\nodata&$   15(   15)$&$    5(    5)$&$    46.3\pm      0.4(    46.3\pm      0.4)$&$     7.4\pm      2.6(     7.4\pm      2.6)$&Y(Y)\\
EM$_{VW}$      &\nodata&$   16(   16)$&$    2(    2)$&$    46.2\pm      0.5(    46.2\pm      0.5)$&$     7.9\pm      2.7(     7.9\pm      2.7)$&Y(Y)\\
$3\sigma$&\nodata&$   30(   24)$&$   15(    9)$&$    48.8\pm      0.5(    47.6\pm      0.4)$&$    42.6\pm      6.5(    15.9\pm      3.4)$&N(N)\\
VT$_{3\sigma}$&\nodata&$   28(   21)$&$   13(    8)$&$    48.8\pm      0.5(    46.8\pm      0.4)$&$    28.5\pm      6.4(    13.2\pm      3.1)$&N(N)\\
VT$_{EM}$&\nodata&$   13(   13)$&$    2(    2)$&$    45.1\pm      0.4(    45.1\pm      0.4)$&$     8.0\pm      2.6(     8.0\pm      2.6)$&Y(Y)\\
 \\

\textbf{Simulation}&$\mathbf{N=   30}$&$\mathbf{N_{mem}=    6}$&&$\mathbf{\langle V\rangle_{mem}=   50\mathrm{km s}^{-1}}$&$\mathbf{\sigma_{V_{0,mem}}=   10\mathrm{km s}^{-1}}$\\
EM             &\nodata&$    8(    5)$&$    2(    0)$&$    34.5\pm      1.3(    49.2\pm      1.0)$&$    24.2\pm      8.1(     5.1\pm      3.1)$&N(N)\\
EM$_{V}$       &\nodata&$   13(    6)$&$   20(    5)$&$    20.2\pm      0.9(    43.5\pm      0.9)$&$    29.7\pm      8.1(     7.9\pm      3.2)$&N(Y)\\
EM$_{VR}$      &\nodata&$   14(    5)$&$   10(    0)$&$    28.5\pm      0.9(    49.4\pm      1.1)$&$    22.3\pm      6.5(     5.2\pm      3.1)$&N(N)\\
EM$_{VW}$      &\nodata&$   12(    4)$&$    5(    1)$&$    32.0\pm      1.0(    48.5\pm      0.8)$&$    27.0\pm      7.4(     4.7\pm      2.5)$&N(N)\\
$3\sigma$&\nodata&$   29(   16)$&$   23(   10)$&$    45.0\pm      0.7(    40.1\pm      0.8)$&$    48.7\pm      8.4(    13.5\pm      4.3)$&N(Y)\\
VT$_{3\sigma}$&\nodata&$   28(    9)$&$   22(    3)$&$    42.2\pm      0.8(    43.8\pm      1.0)$&$    47.3\pm      8.2(     8.9\pm      3.9)$&N(Y)\\
VT$_{EM}$&\nodata&$    6(    4)$&$    0(    2)$&$    49.2\pm      1.0(    51.2\pm      1.4)$&$     5.1\pm      3.0(     5.8\pm      3.7)$&N(N)\\
 \\

\textbf{Simulation}&$\mathbf{N=   30}$&$\mathbf{N_{mem}=   24}$&&$\mathbf{\langle V\rangle_{mem}=  100\mathrm{km s}^{-1}}$&$\mathbf{\sigma_{V_{0,mem}}=   10\mathrm{km s}^{-1}}$\\
EM             &\nodata&$   23(   23)$&$    0(    0)$&$   102.6\pm      0.5(   102.6\pm      0.5)$&$     7.8\pm      2.7(     7.8\pm      2.7)$&Y(Y)\\
EM$_{V}$       &\nodata&$   23(   23)$&$    1(    1)$&$   101.8\pm      0.5(   101.8\pm      0.5)$&$     6.7\pm      2.3(     6.7\pm      2.3)$&N(N)\\
EM$_{VR}$      &\nodata&$   24(   24)$&$    1(    1)$&$   102.3\pm      0.5(   102.3\pm      0.5)$&$     7.5\pm      2.5(     7.5\pm      2.5)$&Y(Y)\\
EM$_{VW}$      &\nodata&$   23(   23)$&$    0(    0)$&$   102.6\pm      0.5(   102.6\pm      0.5)$&$     7.8\pm      2.7(     7.8\pm      2.7)$&Y(Y)\\
$3\sigma$&\nodata&$   30(   27)$&$    6(    3)$&$    93.4\pm      0.5(    99.8\pm      0.5)$&$    22.3\pm      4.0(    11.7\pm      2.9)$&N(Y)\\
VT$_{3\sigma}$&\nodata&$   27(   25)$&$    3(    1)$&$    99.8\pm      0.5(   101.2\pm      0.5)$&$    11.7\pm      2.9(    10.0\pm      2.9)$&N(Y)\\
VT$_{EM}$&\nodata&$   22(   22)$&$    2(    2)$&$   102.8\pm      0.5(   102.8\pm      0.5)$&$     8.0\pm      2.7(     8.0\pm      2.7)$&Y(Y)\\
 \\

\textbf{Simulation}&$\mathbf{N=   30}$&$\mathbf{N_{mem}=   15}$&&$\mathbf{\langle V\rangle_{mem}=  100\mathrm{km s}^{-1}}$&$\mathbf{\sigma_{V_{0,mem}}=   10\mathrm{km s}^{-1}}$\\
EM             &\nodata&$   15(   15)$&$    1(    1)$&$   102.9\pm      0.6(   102.9\pm      0.6)$&$    10.5\pm      3.3(    10.5\pm      3.3)$&Y(Y)\\
EM$_{V}$       &\nodata&$   15(   15)$&$    5(    5)$&$    99.1\pm      0.6(    99.1\pm      0.6)$&$    10.5\pm      3.2(    10.5\pm      3.2)$&Y(Y)\\
EM$_{VR}$      &\nodata&$   15(   15)$&$    1(    1)$&$   102.7\pm      0.6(   102.7\pm      0.6)$&$    10.8\pm      3.3(    10.8\pm      3.3)$&Y(Y)\\
EM$_{VW}$      &\nodata&$   14(   14)$&$    2(    2)$&$   101.5\pm      0.6(   101.5\pm      0.6)$&$     9.6\pm      2.9(     9.6\pm      2.9)$&Y(Y)\\
$3\sigma$&\nodata&$   30(   21)$&$   15(    6)$&$    75.3\pm      0.5(    97.7\pm      0.5)$&$    38.6\pm      6.2(    14.0\pm      3.9)$&N(N)\\
VT$_{3\sigma}$&\nodata&$   30(   15)$&$   15(    2)$&$    75.3\pm      0.5(   103.0\pm      0.6)$&$    38.6\pm      6.2(    10.8\pm      3.4)$&N(Y)\\
VT$_{EM}$&\nodata&$   14(   14)$&$    3(    3)$&$   102.7\pm      0.6(   102.7\pm      0.6)$&$    11.1\pm      3.5(    11.1\pm      3.5)$&Y(Y)\\
 \\

\textbf{Simulation}&$\mathbf{N=   30}$&$\mathbf{N_{mem}=    6}$&&$\mathbf{\langle V\rangle_{mem}=  100\mathrm{km s}^{-1}}$&$\mathbf{\sigma_{V_{0,mem}}=   10\mathrm{km s}^{-1}}$\\
EM             &\nodata&$    5(    5)$&$    2(    2)$&$    89.6\pm      0.9(    89.6\pm      0.9)$&$     5.3\pm      2.7(     5.3\pm      2.7)$&N(N)\\
EM$_{V}$       &\nodata&$   23(    8)$&$   24(   10)$&$    85.2\pm      0.6(   105.8\pm      0.8)$&$    43.1\pm      7.2(    18.3\pm      4.6)$&N(N)\\
EM$_{VR}$      &\nodata&$   11(   11)$&$    8(    8)$&$   104.2\pm      0.6(   104.3\pm      0.7)$&$    18.3\pm      4.6(    18.0\pm      4.6)$&N(N)\\
EM$_{VW}$      &\nodata&$    5(    5)$&$    1(    1)$&$    90.6\pm      0.9(    90.6\pm      0.9)$&$     4.7\pm      2.6(     4.7\pm      2.6)$&N(N)\\
$3\sigma$&\nodata&$   30(   17)$&$   24(   11)$&$    81.4\pm      0.6(   102.1\pm      0.7)$&$    44.9\pm      7.8(    20.6\pm      5.2)$&N(N)\\
VT$_{3\sigma}$&\nodata&$   30(   11)$&$   24(    5)$&$    81.4\pm      0.6(   104.6\pm      0.8)$&$    44.9\pm      7.8(    18.3\pm      5.0)$&N(N)\\
VT$_{EM}$&\nodata&$    5(    5)$&$    3(    3)$&$    87.8\pm      0.9(    87.8\pm      0.9)$&$     4.2\pm      2.6(     4.2\pm      2.6)$&N(N)\\
 \\

\textbf{Simulation}&$\mathbf{N=   30}$&$\mathbf{N_{mem}=   24}$&&$\mathbf{\langle V\rangle_{mem}=  200\mathrm{km s}^{-1}}$&$\mathbf{\sigma_{V_{0,mem}}=   10\mathrm{km s}^{-1}}$\\
EM             &\nodata&$   23(   23)$&$    0(    0)$&$   199.7\pm      0.6(   199.7\pm      0.6)$&$     8.6\pm      3.2(     8.6\pm      3.2)$&Y(Y)\\
EM$_{V}$       &\nodata&$   24(   24)$&$    1(    1)$&$   199.6\pm      0.6(   199.6\pm      0.6)$&$     8.4\pm      3.1(     8.4\pm      3.1)$&Y(Y)\\
EM$_{VR}$      &\nodata&$   24(   24)$&$    1(    1)$&$   199.6\pm      0.6(   199.6\pm      0.6)$&$     8.5\pm      3.1(     8.5\pm      3.1)$&Y(Y)\\
EM$_{VW}$      &\nodata&$   23(   23)$&$    0(    0)$&$   199.7\pm      0.6(   199.7\pm      0.6)$&$     8.6\pm      3.2(     8.6\pm      3.2)$&Y(Y)\\
$3\sigma$&\nodata&$   30(   24)$&$    6(    2)$&$   174.8\pm      0.6(   200.2\pm      0.6)$&$    57.3\pm      6.4(     8.1\pm      2.9)$&N(Y)\\
VT$_{3\sigma}$&\nodata&$   28(   22)$&$    4(    2)$&$   184.9\pm      0.7(   199.9\pm      0.6)$&$    44.2\pm      5.1(     8.1\pm      2.8)$&N(Y)\\
VT$_{EM}$&\nodata&$   23(   23)$&$    1(    1)$&$   199.3\pm      0.6(   199.3\pm      0.6)$&$     8.5\pm      3.1(     8.5\pm      3.1)$&Y(Y)\\
 \\

\textbf{Simulation}&$\mathbf{N=   30}$&$\mathbf{N_{mem}=   15}$&&$\mathbf{\langle V\rangle_{mem}=  200\mathrm{km s}^{-1}}$&$\mathbf{\sigma_{V_{0,mem}}=   10\mathrm{km s}^{-1}}$\\
EM             &\nodata&$   14(   14)$&$    0(    0)$&$   202.7\pm      0.4(   202.7\pm      0.4)$&$     4.5\pm      1.8(     4.5\pm      1.8)$&N(N)\\
EM$_{V}$       &\nodata&$   14(   14)$&$    0(    0)$&$   202.7\pm      0.4(   202.7\pm      0.4)$&$     4.4\pm      1.8(     4.4\pm      1.8)$&N(N)\\
EM$_{VR}$      &\nodata&$   14(   14)$&$    0(    0)$&$   202.5\pm      0.4(   202.5\pm      0.4)$&$     4.4\pm      1.7(     4.4\pm      1.7)$&N(N)\\
EM$_{VW}$      &\nodata&$   14(   14)$&$    0(    0)$&$   202.8\pm      0.4(   202.8\pm      0.4)$&$     4.5\pm      1.8(     4.5\pm      1.8)$&N(N)\\
$3\sigma$&\nodata&$   30(   15)$&$   15(    0)$&$   123.9\pm      0.5(   202.7\pm      0.4)$&$    86.8\pm      9.6(     4.5\pm      1.8)$&N(N)\\
VT$_{3\sigma}$&\nodata&$   30(   13)$&$   15(    2)$&$   123.9\pm      0.5(   201.9\pm      0.4)$&$    86.8\pm      9.6(     4.0\pm      1.4)$&N(N)\\
VT$_{EM}$&\nodata&$   13(   13)$&$    2(    2)$&$   201.9\pm      0.4(   201.9\pm      0.4)$&$     4.0\pm      1.4(     4.0\pm      1.4)$&N(N)\\
 \\

\textbf{Simulation}&$\mathbf{N=   30}$&$\mathbf{N_{mem}=    6}$&&$\mathbf{\langle V\rangle_{mem}=  200\mathrm{km s}^{-1}}$&$\mathbf{\sigma_{V_{0,mem}}=   10\mathrm{km s}^{-1}}$\\
EM             &\nodata&$    5(    5)$&$    0(    0)$&$   198.2\pm      0.7(   198.2\pm      0.7)$&$     9.6\pm      4.2(     9.6\pm      4.2)$&Y(Y)\\
EM$_{V}$       &\nodata&$   13(    5)$&$    4(    0)$&$   125.8\pm      0.5(   197.9\pm      0.7)$&$    88.2\pm      9.4(     9.1\pm      4.0)$&N(Y)\\
EM$_{VR}$      &\nodata&$   14(    5)$&$    7(    0)$&$   128.3\pm      0.6(   198.5\pm      0.7)$&$    90.8\pm     11.1(     9.6\pm      4.2)$&N(Y)\\
EM$_{VW}$      &\nodata&$    9(    5)$&$    4(    0)$&$   154.3\pm      0.7(   197.6\pm      0.7)$&$    60.0\pm      8.3(     9.1\pm      3.9)$&N(Y)\\
$3\sigma$&\nodata&$   30(    6)$&$   24(    0)$&$    89.7\pm      0.5(   198.3\pm      0.7)$&$    81.7\pm      8.7(     9.7\pm      4.2)$&N(Y)\\
VT$_{3\sigma}$&\nodata&$   30(    3)$&$   24(    3)$&$    89.7\pm      0.5(   201.7\pm      1.1)$&$    81.7\pm      8.7(    12.0\pm      5.5)$&N(Y)\\
VT$_{EM}$&\nodata&$    3(    3)$&$    3(    3)$&$   201.7\pm      1.1(   201.7\pm      1.1)$&$    12.0\pm      5.5(    12.0\pm      5.5)$&Y(Y)\\
 \\

\textbf{Simulation}&$\mathbf{N=  300}$&$\mathbf{N_{mem}=  240}$&&$\mathbf{\langle V\rangle_{mem}=   50\mathrm{km s}^{-1}}$&$\mathbf{\sigma_{V_{0,mem}}=   10\mathrm{km s}^{-1}}$\\
EM             &\nodata&$  242(  241)$&$    6(    6)$&$    48.7\pm      0.2(    48.7\pm      0.2)$&$    10.8\pm      1.9(    10.7\pm      1.9)$&Y(Y)\\
EM$_{V}$       &\nodata&$  235(  234)$&$   25(   25)$&$    48.7\pm      0.2(    48.7\pm      0.2)$&$    10.7\pm      1.8(    10.6\pm      1.7)$&Y(Y)\\
EM$_{VR}$      &\nodata&$  238(  238)$&$   15(   15)$&$    48.6\pm      0.2(    48.6\pm      0.2)$&$    10.7\pm      1.8(    10.7\pm      1.8)$&Y(Y)\\
EM$_{VW}$      &\nodata&$  243(  241)$&$    9(    8)$&$    48.9\pm      0.2(    48.7\pm      0.2)$&$    11.2\pm      1.9(    10.8\pm      1.9)$&Y(Y)\\
$3\sigma$&\nodata&$  271(  271)$&$   31(   31)$&$    48.4\pm      0.2(    48.4\pm      0.2)$&$    12.6\pm      2.1(    12.6\pm      2.1)$&N(Y)\\
VT$_{3\sigma}$&\nodata&$  255(  255)$&$   23(   23)$&$    48.5\pm      0.2(    48.5\pm      0.2)$&$    11.9\pm      2.0(    11.9\pm      2.0)$&Y(Y)\\
VT$_{EM}$&\nodata&$  223(  223)$&$   21(   21)$&$    48.7\pm      0.2(    48.7\pm      0.2)$&$    10.8\pm      1.9(    10.8\pm      1.9)$&Y(Y)\\
 \\

\textbf{Simulation}&$\mathbf{N=  300}$&$\mathbf{N_{mem}=  150}$&&$\mathbf{\langle V\rangle_{mem}=   50\mathrm{km s}^{-1}}$&$\mathbf{\sigma_{V_{0,mem}}=   10\mathrm{km s}^{-1}}$\\
EM             &\nodata&$  153(  153)$&$   10(   10)$&$    50.3\pm      0.2(    50.3\pm      0.2)$&$    10.9\pm      2.0(    10.9\pm      2.0)$&Y(Y)\\
EM$_{V}$       &\nodata&$  149(  148)$&$   51(   51)$&$    50.0\pm      0.2(    50.0\pm      0.2)$&$     9.7\pm      1.7(     9.7\pm      1.7)$&Y(Y)\\
EM$_{VR}$      &\nodata&$  147(  147)$&$   37(   37)$&$    50.1\pm      0.2(    50.1\pm      0.2)$&$     9.2\pm      1.8(     9.2\pm      1.8)$&Y(Y)\\
EM$_{VW}$      &\nodata&$  156(  152)$&$   17(   15)$&$    50.1\pm      0.2(    50.4\pm      0.2)$&$    12.2\pm      2.1(    11.4\pm      2.0)$&N(Y)\\
$3\sigma$&\nodata&$  273(  238)$&$  123(   88)$&$    52.9\pm      0.2(    50.8\pm      0.2)$&$    24.7\pm      2.9(    16.1\pm      2.2)$&N(N)\\
VT$_{3\sigma}$&\nodata&$  223(  195)$&$   79(   53)$&$    51.4\pm      0.2(    50.4\pm      0.2)$&$    16.9\pm      2.3(    13.5\pm      2.1)$&N(N)\\
VT$_{EM}$&\nodata&$  130(  130)$&$   26(   26)$&$    50.1\pm      0.3(    50.1\pm      0.3)$&$    10.1\pm      2.0(    10.1\pm      2.0)$&Y(Y)\\
 \\

\textbf{Simulation}&$\mathbf{N=  300}$&$\mathbf{N_{mem}=   60}$&&$\mathbf{\langle V\rangle_{mem}=   50\mathrm{km s}^{-1}}$&$\mathbf{\sigma_{V_{0,mem}}=   10\mathrm{km s}^{-1}}$\\
EM             &\nodata&$  128(   69)$&$  197(   11)$&$    70.9\pm      0.2(    50.5\pm      0.3)$&$    60.6\pm      4.9(     8.8\pm      2.3)$&N(Y)\\
EM$_{V}$       &\nodata&$   72(   72)$&$   66(   66)$&$    48.4\pm      0.2(    48.4\pm      0.2)$&$     8.2\pm      1.6(     8.2\pm      1.6)$&N(N)\\
EM$_{VR}$      &\nodata&$   74(   74)$&$   21(   21)$&$    50.4\pm      0.3(    50.4\pm      0.3)$&$     8.6\pm      2.0(     8.6\pm      2.0)$&Y(Y)\\
EM$_{VW}$      &\nodata&$  129(   67)$&$  199(   17)$&$    70.9\pm      0.2(    49.7\pm      0.3)$&$    60.5\pm      4.9(     8.8\pm      2.2)$&N(Y)\\
$3\sigma$&\nodata&$  293(  188)$&$  233(  128)$&$    63.3\pm      0.2(    49.2\pm      0.2)$&$    48.2\pm      4.0(    17.4\pm      2.6)$&N(N)\\
VT$_{3\sigma}$&\nodata&$  280(  128)$&$  220(   68)$&$    63.4\pm      0.2(    50.7\pm      0.2)$&$    45.0\pm      3.9(    16.7\pm      2.7)$&N(N)\\
VT$_{EM}$&\nodata&$   46(   51)$&$  106(   13)$&$    67.2\pm      0.3(    52.1\pm      0.4)$&$    48.9\pm      6.1(     8.9\pm      2.6)$&N(Y)\\
 \\

\textbf{Simulation}&$\mathbf{N=  300}$&$\mathbf{N_{mem}=  240}$&&$\mathbf{\langle V\rangle_{mem}=  100\mathrm{km s}^{-1}}$&$\mathbf{\sigma_{V_{0,mem}}=   10\mathrm{km s}^{-1}}$\\
EM             &\nodata&$  241(  241)$&$    8(    8)$&$    99.6\pm      0.2(    99.6\pm      0.2)$&$    10.2\pm      1.8(    10.2\pm      1.8)$&Y(Y)\\
EM$_{V}$       &\nodata&$  244(  244)$&$   23(   22)$&$    99.7\pm      0.2(    99.7\pm      0.2)$&$    10.1\pm      1.7(    10.1\pm      1.7)$&Y(Y)\\
EM$_{VR}$      &\nodata&$  241(  241)$&$   17(   17)$&$    99.7\pm      0.2(    99.7\pm      0.2)$&$     9.9\pm      1.7(     9.9\pm      1.7)$&Y(Y)\\
EM$_{VW}$      &\nodata&$  240(  240)$&$   11(   11)$&$    99.5\pm      0.2(    99.5\pm      0.2)$&$    10.2\pm      1.8(    10.2\pm      1.8)$&Y(Y)\\
$3\sigma$&\nodata&$  261(  261)$&$   23(   23)$&$    99.3\pm      0.2(    99.3\pm      0.2)$&$    10.7\pm      1.8(    10.7\pm      1.8)$&Y(Y)\\
VT$_{3\sigma}$&\nodata&$  237(  237)$&$   21(   21)$&$    99.4\pm      0.2(    99.4\pm      0.2)$&$     9.9\pm      1.7(     9.9\pm      1.7)$&Y(Y)\\
VT$_{EM}$&\nodata&$  219(  219)$&$   21(   21)$&$    99.2\pm      0.2(    99.2\pm      0.2)$&$     9.6\pm      1.7(     9.6\pm      1.7)$&Y(Y)\\
 \\

\textbf{Simulation}&$\mathbf{N=  300}$&$\mathbf{N_{mem}=  150}$&&$\mathbf{\langle V\rangle_{mem}=  100\mathrm{km s}^{-1}}$&$\mathbf{\sigma_{V_{0,mem}}=   10\mathrm{km s}^{-1}}$\\
EM             &\nodata&$  151(  151)$&$    9(    9)$&$    99.0\pm      0.2(    99.0\pm      0.2)$&$    10.4\pm      1.9(    10.4\pm      1.9)$&Y(Y)\\
EM$_{V}$       &\nodata&$  156(  155)$&$   37(   36)$&$    98.4\pm      0.2(    98.5\pm      0.2)$&$    11.9\pm      1.9(    11.8\pm      1.9)$&Y(Y)\\
EM$_{VR}$      &\nodata&$  147(  147)$&$   22(   22)$&$    99.0\pm      0.2(    99.0\pm      0.2)$&$    10.4\pm      1.8(    10.4\pm      1.8)$&Y(Y)\\
EM$_{VW}$      &\nodata&$  155(  154)$&$   16(   16)$&$    98.6\pm      0.2(    98.7\pm      0.2)$&$    11.3\pm      2.0(    11.3\pm      2.0)$&Y(Y)\\
$3\sigma$&\nodata&$  296(  216)$&$  146(   66)$&$    82.9\pm      0.2(    96.6\pm      0.2)$&$    33.3\pm      3.4(    15.5\pm      2.4)$&N(N)\\
VT$_{3\sigma}$&\nodata&$  282(  173)$&$  132(   41)$&$    84.1\pm      0.2(    97.6\pm      0.2)$&$    29.8\pm      3.2(    13.8\pm      2.3)$&N(N)\\
VT$_{EM}$&\nodata&$  123(  123)$&$   33(   33)$&$    99.6\pm      0.2(    99.6\pm      0.2)$&$    10.1\pm      2.1(    10.1\pm      2.1)$&Y(Y)\\
 \\

\textbf{Simulation}&$\mathbf{N=  300}$&$\mathbf{N_{mem}=   60}$&&$\mathbf{\langle V\rangle_{mem}=  100\mathrm{km s}^{-1}}$&$\mathbf{\sigma_{V_{0,mem}}=   10\mathrm{km s}^{-1}}$\\
EM             &\nodata&$   67(   67)$&$    9(    9)$&$    98.0\pm      0.3(    98.0\pm      0.3)$&$    10.7\pm      2.5(    10.7\pm      2.5)$&Y(Y)\\
EM$_{V}$       &\nodata&$   79(   70)$&$   56(   58)$&$    94.7\pm      0.3(    95.8\pm      0.3)$&$    12.9\pm      2.2(     9.9\pm      1.9)$&N(Y)\\
EM$_{VR}$      &\nodata&$   78(   78)$&$   18(   19)$&$    97.2\pm      0.3(    97.2\pm      0.3)$&$    11.8\pm      2.4(    11.7\pm      2.4)$&Y(Y)\\
EM$_{VW}$      &\nodata&$  120(   65)$&$   60(   18)$&$    87.7\pm      0.3(    97.9\pm      0.3)$&$    39.6\pm      4.2(    11.5\pm      2.5)$&N(Y)\\
$3\sigma$&\nodata&$  294(  169)$&$  234(  109)$&$    76.7\pm      0.2(    93.2\pm      0.2)$&$    48.9\pm      4.2(    18.6\pm      2.8)$&N(N)\\
VT$_{3\sigma}$&\nodata&$  280(  105)$&$  220(   45)$&$    76.2\pm      0.2(    94.0\pm      0.3)$&$    44.7\pm      3.9(    16.0\pm      2.9)$&N(N)\\
VT$_{EM}$&\nodata&$   54(   54)$&$   12(   12)$&$    99.1\pm      0.4(    99.1\pm      0.4)$&$    10.6\pm      2.8(    10.6\pm      2.8)$&Y(Y)\\
 \\

\textbf{Simulation}&$\mathbf{N=  300}$&$\mathbf{N_{mem}=  240}$&&$\mathbf{\langle V\rangle_{mem}=  200\mathrm{km s}^{-1}}$&$\mathbf{\sigma_{V_{0,mem}}=   10\mathrm{km s}^{-1}}$\\
EM             &\nodata&$  242(  242)$&$    3(    3)$&$   198.7\pm      0.2(   198.7\pm      0.2)$&$    10.2\pm      1.7(    10.2\pm      1.7)$&Y(Y)\\
EM$_{V}$       &\nodata&$  242(  242)$&$    5(    5)$&$   198.5\pm      0.2(   198.5\pm      0.2)$&$    10.4\pm      1.7(    10.4\pm      1.7)$&Y(Y)\\
EM$_{VR}$      &\nodata&$  241(  241)$&$    4(    4)$&$   198.7\pm      0.2(   198.7\pm      0.2)$&$    10.2\pm      1.7(    10.2\pm      1.7)$&Y(Y)\\
EM$_{VW}$      &\nodata&$  242(  242)$&$    3(    3)$&$   198.7\pm      0.2(   198.7\pm      0.2)$&$    10.2\pm      1.7(    10.2\pm      1.7)$&Y(Y)\\
$3\sigma$&\nodata&$  243(  243)$&$    5(    5)$&$   198.5\pm      0.2(   198.5\pm      0.2)$&$    10.2\pm      1.7(    10.2\pm      1.7)$&Y(Y)\\
VT$_{3\sigma}$&\nodata&$  225(  225)$&$   23(   23)$&$   198.7\pm      0.2(   198.7\pm      0.2)$&$    10.1\pm      1.7(    10.1\pm      1.7)$&Y(Y)\\
VT$_{EM}$&\nodata&$  225(  225)$&$   21(   21)$&$   198.9\pm      0.2(   198.9\pm      0.2)$&$    10.2\pm      1.7(    10.2\pm      1.7)$&Y(Y)\\
 \\

\textbf{Simulation}&$\mathbf{N=  300}$&$\mathbf{N_{mem}=  150}$&&$\mathbf{\langle V\rangle_{mem}=  200\mathrm{km s}^{-1}}$&$\mathbf{\sigma_{V_{0,mem}}=   10\mathrm{km s}^{-1}}$\\
EM             &\nodata&$  150(  150)$&$    3(    3)$&$   198.7\pm      0.2(   198.7\pm      0.2)$&$     9.5\pm      1.8(     9.5\pm      1.8)$&Y(Y)\\
EM$_{V}$       &\nodata&$  149(  149)$&$    5(    5)$&$   198.6\pm      0.2(   198.6\pm      0.2)$&$     9.3\pm      1.8(     9.3\pm      1.8)$&Y(Y)\\
EM$_{VR}$      &\nodata&$  151(  151)$&$    3(    3)$&$   198.4\pm      0.2(   198.4\pm      0.2)$&$     9.5\pm      1.8(     9.5\pm      1.8)$&Y(Y)\\
EM$_{VW}$      &\nodata&$  150(  150)$&$    2(    2)$&$   198.8\pm      0.2(   198.8\pm      0.2)$&$     9.4\pm      1.8(     9.4\pm      1.8)$&Y(Y)\\
$3\sigma$&\nodata&$  300(  155)$&$  150(    5)$&$   134.6\pm      0.2(   198.5\pm      0.2)$&$    82.3\pm      5.3(     9.6\pm      1.9)$&N(Y)\\
VT$_{3\sigma}$&\nodata&$  299(  126)$&$  149(   24)$&$   133.8\pm      0.2(   198.8\pm      0.2)$&$    81.2\pm      5.4(     9.7\pm      2.0)$&N(Y)\\
VT$_{EM}$&\nodata&$  124(  124)$&$   26(   26)$&$   198.8\pm      0.2(   198.8\pm      0.2)$&$     9.8\pm      2.0(     9.8\pm      2.0)$&Y(Y)\\
 \\

\textbf{Simulation}&$\mathbf{N=  300}$&$\mathbf{N_{mem}=   60}$&&$\mathbf{\langle V\rangle_{mem}=  200\mathrm{km s}^{-1}}$&$\mathbf{\sigma_{V_{0,mem}}=   10\mathrm{km s}^{-1}}$\\
EM             &\nodata&$   60(   60)$&$    0(    0)$&$   199.5\pm      0.3(   199.5\pm      0.3)$&$     9.2\pm      2.1(     9.2\pm      2.1)$&Y(Y)\\
EM$_{V}$       &\nodata&$   62(   62)$&$    9(    9)$&$   198.1\pm      0.3(   198.1\pm      0.3)$&$     9.7\pm      2.3(     9.7\pm      2.3)$&Y(Y)\\
EM$_{VR}$      &\nodata&$   60(   60)$&$    1(    1)$&$   199.7\pm      0.3(   199.7\pm      0.3)$&$     9.3\pm      2.1(     9.3\pm      2.1)$&Y(Y)\\
EM$_{VW}$      &\nodata&$   85(   61)$&$   23(    2)$&$   168.4\pm      0.3(   199.0\pm      0.3)$&$    63.2\pm      7.0(     9.5\pm      2.2)$&N(Y)\\
$3\sigma$&\nodata&$  298(   73)$&$  238(   13)$&$    95.9\pm      0.2(   196.4\pm      0.3)$&$    70.9\pm      5.1(    12.1\pm      2.5)$&N(Y)\\
VT$_{3\sigma}$&\nodata&$  296(   54)$&$  236(    6)$&$    96.4\pm      0.2(   199.3\pm      0.3)$&$    70.8\pm      5.1(     9.4\pm      2.2)$&N(Y)\\
VT$_{EM}$&\nodata&$   51(   51)$&$    9(    9)$&$   199.3\pm      0.3(   199.3\pm      0.3)$&$     9.7\pm      2.3(     9.7\pm      2.3)$&Y(Y)\\
 \\

\textbf{Simulation}&$\mathbf{N= 3000}$&$\mathbf{N_{mem}= 2400}$&&$\mathbf{\langle V\rangle_{mem}=   50\mathrm{km s}^{-1}}$&$\mathbf{\sigma_{V_{0,mem}}=   10\mathrm{km s}^{-1}}$\\
EM             &\nodata&$ 2417( 2416)$&$   71(   71)$&$    50.2\pm      0.1(    50.2\pm      0.1)$&$    10.1\pm      1.0(    10.1\pm      1.0)$&Y(Y)\\
EM$_{V}$       &\nodata&$ 2432( 2430)$&$  277(  277)$&$    50.1\pm      0.1(    50.1\pm      0.1)$&$    10.2\pm      0.9(    10.2\pm      0.9)$&Y(Y)\\
EM$_{VR}$      &\nodata&$ 2421( 2421)$&$  198(  199)$&$    50.1\pm      0.1(    50.1\pm      0.1)$&$    10.1\pm      1.0(    10.1\pm      1.0)$&Y(Y)\\
EM$_{VW}$      &\nodata&$ 2416( 2413)$&$  109(  107)$&$    50.2\pm      0.1(    50.2\pm      0.1)$&$    10.2\pm      1.0(    10.2\pm      1.0)$&Y(Y)\\
$3\sigma$&\nodata&$ 2696( 2696)$&$  300(  300)$&$    50.1\pm      0.1(    50.1\pm      0.1)$&$    11.2\pm      1.0(    11.2\pm      1.0)$&N(Y)\\
VT$_{3\sigma}$&\nodata&$ 2492( 2492)$&$  236(  236)$&$    50.1\pm      0.1(    50.1\pm      0.1)$&$    10.4\pm      1.0(    10.4\pm      1.0)$&Y(Y)\\
VT$_{EM}$&\nodata&$ 2260( 2260)$&$  196(  196)$&$    50.2\pm      0.1(    50.2\pm      0.1)$&$    10.0\pm      1.0(    10.0\pm      1.0)$&Y(Y)\\
 \\

\textbf{Simulation}&$\mathbf{N= 3000}$&$\mathbf{N_{mem}= 1500}$&&$\mathbf{\langle V\rangle_{mem}=   50\mathrm{km s}^{-1}}$&$\mathbf{\sigma_{V_{0,mem}}=   10\mathrm{km s}^{-1}}$\\
EM             &\nodata&$ 1540( 1538)$&$  121(  121)$&$    49.6\pm      0.1(    49.6\pm      0.1)$&$    10.6\pm      1.1(    10.6\pm      1.1)$&Y(Y)\\
EM$_{V}$       &\nodata&$ 1569( 1565)$&$  583(  584)$&$    49.9\pm      0.1(    49.9\pm      0.1)$&$    10.5\pm      1.0(    10.4\pm      1.0)$&Y(Y)\\
EM$_{VR}$      &\nodata&$ 1545( 1544)$&$  295(  296)$&$    49.6\pm      0.1(    49.6\pm      0.1)$&$    10.2\pm      1.0(    10.2\pm      1.0)$&Y(Y)\\
EM$_{VW}$      &\nodata&$ 1566( 1555)$&$  192(  191)$&$    49.8\pm      0.1(    49.7\pm      0.1)$&$    11.1\pm      1.1(    10.9\pm      1.1)$&Y(Y)\\
$3\sigma$&\nodata&$ 2783( 2340)$&$ 1283(  840)$&$    54.3\pm      0.1(    50.1\pm      0.1)$&$    27.3\pm      1.7(    15.0\pm      1.3)$&N(N)\\
VT$_{3\sigma}$&\nodata&$ 2341( 1927)$&$  849(  499)$&$    51.7\pm      0.1(    50.0\pm      0.1)$&$    18.7\pm      1.4(    13.1\pm      1.2)$&N(N)\\
VT$_{EM}$&\nodata&$ 1305( 1305)$&$  253(  253)$&$    49.7\pm      0.1(    49.7\pm      0.1)$&$    10.3\pm      1.1(    10.3\pm      1.1)$&Y(Y)\\
 \\

\textbf{Simulation}&$\mathbf{N= 3000}$&$\mathbf{N_{mem}=  600}$&&$\mathbf{\langle V\rangle_{mem}=   50\mathrm{km s}^{-1}}$&$\mathbf{\sigma_{V_{0,mem}}=   10\mathrm{km s}^{-1}}$\\
EM             &\nodata&$  676(  674)$&$  134(  134)$&$    50.1\pm      0.1(    50.1\pm      0.1)$&$    10.3\pm      1.3(    10.2\pm      1.3)$&Y(Y)\\
EM$_{V}$       &\nodata&$  662(  650)$&$  581(  587)$&$    51.0\pm      0.1(    51.0\pm      0.1)$&$    10.1\pm      1.0(     9.8\pm      1.0)$&Y(Y)\\
EM$_{VR}$      &\nodata&$  653(  650)$&$  241(  241)$&$    50.6\pm      0.1(    50.6\pm      0.1)$&$     9.9\pm      1.1(     9.8\pm      1.1)$&Y(Y)\\
EM$_{VW}$      &\nodata&$  718(  710)$&$  210(  209)$&$    49.9\pm      0.1(    50.0\pm      0.1)$&$    11.5\pm      1.3(    11.1\pm      1.3)$&N(Y)\\
$3\sigma$&\nodata&$ 2894( 1936)$&$ 2294( 1336)$&$    62.5\pm      0.1(    50.6\pm      0.1)$&$    43.7\pm      2.2(    18.8\pm      1.5)$&N(N)\\
VT$_{3\sigma}$&\nodata&$ 2719( 1337)$&$ 2123(  777)$&$    60.8\pm      0.1(    50.5\pm      0.1)$&$    40.5\pm      2.1(    17.6\pm      1.6)$&N(N)\\
VT$_{EM}$&\nodata&$  446(  446)$&$  204(  204)$&$    50.5\pm      0.1(    50.5\pm      0.1)$&$     9.7\pm      1.4(     9.7\pm      1.4)$&Y(Y)\\
 \\

\textbf{Simulation}&$\mathbf{N= 3000}$&$\mathbf{N_{mem}= 2400}$&&$\mathbf{\langle V\rangle_{mem}=  100\mathrm{km s}^{-1}}$&$\mathbf{\sigma_{V_{0,mem}}=   10\mathrm{km s}^{-1}}$\\
EM             &\nodata&$ 2403( 2402)$&$   47(   47)$&$   100.2\pm      0.1(   100.2\pm      0.1)$&$    10.1\pm      1.0(    10.1\pm      1.0)$&Y(Y)\\
EM$_{V}$       &\nodata&$ 2410( 2408)$&$  190(  190)$&$   100.2\pm      0.1(   100.2\pm      0.1)$&$    10.2\pm      1.0(    10.2\pm      1.0)$&Y(Y)\\
EM$_{VR}$      &\nodata&$ 2406( 2405)$&$  143(  144)$&$   100.2\pm      0.1(   100.2\pm      0.1)$&$    10.2\pm      1.0(    10.1\pm      1.0)$&Y(Y)\\
EM$_{VW}$      &\nodata&$ 2399( 2398)$&$   71(   72)$&$   100.2\pm      0.1(   100.2\pm      0.1)$&$    10.1\pm      1.0(    10.1\pm      1.0)$&Y(Y)\\
$3\sigma$&\nodata&$ 2599( 2599)$&$  207(  207)$&$    99.8\pm      0.1(    99.8\pm      0.1)$&$    11.0\pm      1.0(    11.0\pm      1.0)$&Y(Y)\\
VT$_{3\sigma}$&\nodata&$ 2399( 2399)$&$  187(  187)$&$    99.9\pm      0.1(    99.9\pm      0.1)$&$    10.5\pm      1.0(    10.5\pm      1.0)$&Y(Y)\\
VT$_{EM}$&\nodata&$ 2226( 2226)$&$  196(  196)$&$   100.2\pm      0.1(   100.2\pm      0.1)$&$    10.0\pm      1.0(    10.0\pm      1.0)$&Y(Y)\\
 \\

\textbf{Simulation}&$\mathbf{N= 3000}$&$\mathbf{N_{mem}= 1500}$&&$\mathbf{\langle V\rangle_{mem}=  100\mathrm{km s}^{-1}}$&$\mathbf{\sigma_{V_{0,mem}}=   10\mathrm{km s}^{-1}}$\\
EM             &\nodata&$ 1516( 1516)$&$   86(   86)$&$    99.7\pm      0.1(    99.7\pm      0.1)$&$    10.1\pm      1.1(    10.0\pm      1.1)$&Y(Y)\\
EM$_{V}$       &\nodata&$ 1522( 1521)$&$  394(  394)$&$    99.7\pm      0.1(    99.7\pm      0.1)$&$     9.9\pm      1.0(     9.9\pm      1.0)$&Y(Y)\\
EM$_{VR}$      &\nodata&$ 1528( 1527)$&$  230(  230)$&$    99.6\pm      0.1(    99.6\pm      0.1)$&$     9.9\pm      1.1(     9.9\pm      1.1)$&Y(Y)\\
EM$_{VW}$      &\nodata&$ 1520( 1518)$&$  125(  127)$&$    99.6\pm      0.1(    99.7\pm      0.1)$&$    10.2\pm      1.1(    10.2\pm      1.1)$&Y(Y)\\
$3\sigma$&\nodata&$ 2917( 2104)$&$ 1417(  604)$&$    83.8\pm      0.1(    98.0\pm      0.1)$&$    37.6\pm      2.0(    14.7\pm      1.3)$&N(N)\\
VT$_{3\sigma}$&\nodata&$ 2630( 1689)$&$ 1140(  303)$&$    87.0\pm      0.1(    98.6\pm      0.1)$&$    30.1\pm      1.8(    12.4\pm      1.2)$&N(N)\\
VT$_{EM}$&\nodata&$ 1283( 1283)$&$  241(  241)$&$    99.8\pm      0.1(    99.8\pm      0.1)$&$     9.9\pm      1.2(     9.9\pm      1.2)$&Y(Y)\\
 \\

\textbf{Simulation}&$\mathbf{N= 3000}$&$\mathbf{N_{mem}=  600}$&&$\mathbf{\langle V\rangle_{mem}=  100\mathrm{km s}^{-1}}$&$\mathbf{\sigma_{V_{0,mem}}=   10\mathrm{km s}^{-1}}$\\
EM             &\nodata&$  641(  640)$&$   99(   98)$&$    99.5\pm      0.1(    99.5\pm      0.1)$&$    10.6\pm      1.4(    10.6\pm      1.4)$&Y(Y)\\
EM$_{V}$       &\nodata&$  649(  620)$&$  447(  439)$&$    98.7\pm      0.1(    99.0\pm      0.1)$&$    11.3\pm      1.2(    10.3\pm      1.1)$&N(Y)\\
EM$_{VR}$      &\nodata&$  601(  599)$&$  198(  198)$&$   100.1\pm      0.1(   100.2\pm      0.1)$&$     9.7\pm      1.2(     9.7\pm      1.2)$&Y(Y)\\
EM$_{VW}$      &\nodata&$  933(  655)$&$  362(  154)$&$    90.7\pm      0.1(    99.0\pm      0.1)$&$    32.8\pm      2.4(    11.5\pm      1.4)$&N(N)\\
$3\sigma$&\nodata&$ 2910( 1570)$&$ 2310(  970)$&$    73.4\pm      0.1(    95.3\pm      0.1)$&$    47.4\pm      2.3(    18.5\pm      1.6)$&N(N)\\
VT$_{3\sigma}$&\nodata&$ 2804(  955)$&$ 2206(  421)$&$    73.5\pm      0.1(    97.1\pm      0.1)$&$    46.3\pm      2.3(    16.4\pm      1.7)$&N(N)\\
VT$_{EM}$&\nodata&$  427(  427)$&$  205(  205)$&$   100.1\pm      0.1(   100.1\pm      0.1)$&$    10.3\pm      1.5(    10.3\pm      1.5)$&Y(Y)\\
 \\

\textbf{Simulation}&$\mathbf{N= 3000}$&$\mathbf{N_{mem}= 2400}$&&$\mathbf{\langle V\rangle_{mem}=  200\mathrm{km s}^{-1}}$&$\mathbf{\sigma_{V_{0,mem}}=   10\mathrm{km s}^{-1}}$\\
EM             &\nodata&$ 2402( 2402)$&$   12(   12)$&$   200.0\pm      0.1(   200.0\pm      0.1)$&$    10.0\pm      1.0(    10.0\pm      1.0)$&Y(Y)\\
EM$_{V}$       &\nodata&$ 2393( 2392)$&$   24(   24)$&$   200.1\pm      0.1(   200.1\pm      0.1)$&$     9.9\pm      1.0(     9.9\pm      1.0)$&Y(Y)\\
EM$_{VR}$      &\nodata&$ 2393( 2393)$&$   19(   19)$&$   200.0\pm      0.1(   200.0\pm      0.1)$&$     9.9\pm      1.0(     9.9\pm      1.0)$&Y(Y)\\
EM$_{VW}$      &\nodata&$ 2402( 2401)$&$   10(   10)$&$   200.0\pm      0.1(   200.0\pm      0.1)$&$    10.0\pm      1.0(    10.0\pm      1.0)$&Y(Y)\\
$3\sigma$&\nodata&$ 2401( 2401)$&$   31(   31)$&$   200.0\pm      0.1(   200.0\pm      0.1)$&$     9.8\pm      1.0(     9.8\pm      1.0)$&Y(Y)\\
VT$_{3\sigma}$&\nodata&$ 2231( 2231)$&$  175(  175)$&$   200.0\pm      0.1(   200.0\pm      0.1)$&$     9.7\pm      1.0(     9.7\pm      1.0)$&Y(Y)\\
VT$_{EM}$&\nodata&$ 2237( 2237)$&$  169(  169)$&$   200.0\pm      0.1(   200.0\pm      0.1)$&$     9.8\pm      1.0(     9.8\pm      1.0)$&Y(Y)\\
 \\

\textbf{Simulation}&$\mathbf{N= 3000}$&$\mathbf{N_{mem}= 1500}$&&$\mathbf{\langle V\rangle_{mem}=  200\mathrm{km s}^{-1}}$&$\mathbf{\sigma_{V_{0,mem}}=   10\mathrm{km s}^{-1}}$\\
EM             &\nodata&$ 1501( 1501)$&$   15(   15)$&$   199.9\pm      0.1(   199.9\pm      0.1)$&$    10.0\pm      1.1(    10.0\pm      1.1)$&Y(Y)\\
EM$_{V}$       &\nodata&$ 1502( 1502)$&$   57(   57)$&$   200.1\pm      0.1(   200.1\pm      0.1)$&$    10.0\pm      1.1(     9.9\pm      1.1)$&Y(Y)\\
EM$_{VR}$      &\nodata&$ 1500( 1500)$&$   45(   45)$&$   200.0\pm      0.1(   200.0\pm      0.1)$&$     9.9\pm      1.1(     9.9\pm      1.1)$&Y(Y)\\
EM$_{VW}$      &\nodata&$ 1503( 1503)$&$   22(   22)$&$   199.9\pm      0.1(   199.9\pm      0.1)$&$    10.1\pm      1.1(    10.1\pm      1.1)$&Y(Y)\\
$3\sigma$&\nodata&$ 2994( 1545)$&$ 1494(   55)$&$   136.6\pm      0.1(   199.8\pm      0.1)$&$    76.8\pm      2.9(    10.0\pm      1.1)$&N(Y)\\
VT$_{3\sigma}$&\nodata&$ 2978( 1289)$&$ 1478(  221)$&$   136.8\pm      0.1(   200.0\pm      0.1)$&$    75.9\pm      2.9(     9.8\pm      1.1)$&N(Y)\\
VT$_{EM}$&\nodata&$ 1280( 1280)$&$  224(  224)$&$   200.0\pm      0.1(   200.0\pm      0.1)$&$     9.9\pm      1.2(     9.9\pm      1.2)$&Y(Y)\\
 \\
\pagebreak
\textbf{Simulation}&$\mathbf{N= 3000}$&$\mathbf{N_{mem}=  600}$&&$\mathbf{\langle V\rangle_{mem}=  200\mathrm{km s}^{-1}}$&$\mathbf{\sigma_{V_{0,mem}}=   10\mathrm{km s}^{-1}}$\\
EM             &\nodata&$  601(  601)$&$   17(   17)$&$   200.9\pm      0.1(   200.9\pm      0.1)$&$    10.4\pm      1.4(    10.4\pm      1.4)$&Y(Y)\\
EM$_{V}$       &\nodata&$  597(  596)$&$   76(   76)$&$   200.6\pm      0.1(   200.6\pm      0.1)$&$    10.2\pm      1.3(    10.2\pm      1.3)$&Y(Y)\\
EM$_{VR}$      &\nodata&$  597(  597)$&$   59(   59)$&$   200.8\pm      0.1(   200.8\pm      0.1)$&$    10.2\pm      1.4(    10.2\pm      1.4)$&Y(Y)\\
EM$_{VW}$      &\nodata&$  926(  596)$&$  340(   29)$&$   157.1\pm      0.1(   200.8\pm      0.1)$&$    70.8\pm      3.6(    10.4\pm      1.4)$&N(Y)\\
$3\sigma$&\nodata&$ 2982(  722)$&$ 2382(  124)$&$    96.4\pm      0.1(   198.4\pm      0.1)$&$    73.8\pm      2.9(    13.7\pm      1.6)$&N(N)\\
VT$_{3\sigma}$&\nodata&$ 2964(  469)$&$ 2364(  167)$&$    96.6\pm      0.1(   200.1\pm      0.1)$&$    73.9\pm      2.9(    11.2\pm      1.6)$&N(Y)\\
VT$_{EM}$&\nodata&$  405(  405)$&$  197(  197)$&$   201.0\pm      0.1(   201.0\pm      0.1)$&$    10.3\pm      1.6(    10.3\pm      1.6)$&Y(Y)\\
 \tablenotetext{a}{Table values in parentheses are obtained after using an initial velocity filter.}
 \label{tab:sim}
\end{longtable*}

%% file: table3.tex
\renewcommand{\arraystretch}{0.6}
\begin{longtable*}{lcccccccccc}
  \tabletypesize{\scriptsize}
  \tablewidth{0pc}
  \tablecaption{\scriptsize Results of Algorithms \tablenotemark{a} with Simulated Data and $\sigma_{V_0}=4$ km s$^{-1}$}
  \tablehead{
\colhead{}&\colhead{$N$}&\colhead{$\hat{N}_{mem}$}&\colhead{$N_{wrong}$}&\colhead{$\langle \hat{V}\rangle_{mem}$}&\colhead{$\hat{\sigma}_{V_0}$}&\colhead{Success?}\\
    \colhead{}&\colhead{}&\colhead{}&\colhead{}&\colhead{(km s$^{-1}$)}&\colhead{(km s$^{-1}$)}
  }
\textbf{Simulation}&$\mathbf{N=   30}$&$\mathbf{N_{mem}=   24}$&&$\mathbf{\langle V\rangle_{mem}=   50\mathrm{km s}^{-1}}$&$\mathbf{\sigma_{V_{0,mem}}=    4\mathrm{km s}^{-1}}$\\
EM             &\nodata&$   23(   23)$&$    0(    0)$&$    50.2\pm      0.4(    50.2\pm      0.4)$&$     3.2\pm      1.3(     3.2\pm      1.3)$&Y(Y)\\
EM$_{V}$       &\nodata&$   25(   25)$&$    2(    2)$&$    49.8\pm      0.3(    49.8\pm      0.3)$&$     3.4\pm      1.2(     3.4\pm      1.2)$&Y(Y)\\
EM$_{VR}$      &\nodata&$   24(   24)$&$    1(    1)$&$    50.1\pm      0.4(    50.1\pm      0.4)$&$     3.1\pm      1.2(     3.1\pm      1.2)$&Y(Y)\\
EM$_{VW}$      &\nodata&$   23(   23)$&$    0(    0)$&$    50.2\pm      0.4(    50.2\pm      0.4)$&$     3.2\pm      1.3(     3.2\pm      1.3)$&Y(Y)\\
$3\sigma$&\nodata&$   25(   25)$&$    3(    3)$&$    50.0\pm      0.3(    50.0\pm      0.3)$&$     3.3\pm      1.1(     3.3\pm      1.1)$&Y(Y)\\
VT$_{3\sigma}$&\nodata&$   23(   23)$&$    3(    3)$&$    50.5\pm      0.4(    50.5\pm      0.4)$&$     2.9\pm      1.1(     2.9\pm      1.1)$&Y(Y)\\
VT$_{EM}$&\nodata&$   23(   23)$&$    1(    1)$&$    50.3\pm      0.4(    50.3\pm      0.4)$&$     3.3\pm      1.3(     3.3\pm      1.3)$&Y(Y)\\
 \\

\textbf{Simulation}&$\mathbf{N=   30}$&$\mathbf{N_{mem}=   15}$&&$\mathbf{\langle V\rangle_{mem}=   50\mathrm{km s}^{-1}}$&$\mathbf{\sigma_{V_{0,mem}}=    4\mathrm{km s}^{-1}}$\\
EM             &\nodata&$   14(   14)$&$    0(    0)$&$    49.2\pm      0.4(    49.2\pm      0.4)$&$     4.0\pm      1.5(     4.0\pm      1.5)$&Y(Y)\\
EM$_{V}$       &\nodata&$   20(   14)$&$   12(    2)$&$    38.8\pm      0.5(    48.3\pm      0.4)$&$    40.5\pm      4.7(     4.1\pm      1.5)$&N(Y)\\
EM$_{VR}$      &\nodata&$   15(   14)$&$    2(    2)$&$    45.3\pm      0.6(    48.9\pm      0.4)$&$    13.8\pm      2.1(     4.2\pm      1.6)$&N(Y)\\
EM$_{VW}$      &\nodata&$   14(   14)$&$    0(    0)$&$    49.1\pm      0.4(    49.1\pm      0.4)$&$     3.9\pm      1.5(     3.9\pm      1.5)$&Y(Y)\\
$3\sigma$&\nodata&$   30(   21)$&$   15(    6)$&$    48.5\pm      0.5(    50.6\pm      0.5)$&$    47.6\pm      5.5(    11.8\pm      2.9)$&N(N)\\
VT$_{3\sigma}$&\nodata&$   26(   18)$&$   11(    5)$&$    57.1\pm      0.5(    51.4\pm      0.6)$&$    39.0\pm      5.8(    11.2\pm      3.0)$&N(N)\\
VT$_{EM}$&\nodata&$   13(   13)$&$    2(    2)$&$    49.3\pm      0.5(    49.3\pm      0.5)$&$     4.4\pm      1.7(     4.4\pm      1.7)$&Y(Y)\\
 \\

\textbf{Simulation}&$\mathbf{N=   30}$&$\mathbf{N_{mem}=    6}$&&$\mathbf{\langle V\rangle_{mem}=   50\mathrm{km s}^{-1}}$&$\mathbf{\sigma_{V_{0,mem}}=    4\mathrm{km s}^{-1}}$\\
EM             &\nodata&$   21(    5)$&$   28(    0)$&$    65.6\pm      0.6(    48.9\pm      0.8)$&$    53.0\pm      8.0(     4.3\pm      2.5)$&N(Y)\\
EM$_{V}$       &\nodata&$    6(    3)$&$    6(    6)$&$    36.9\pm      0.5(    41.5\pm      0.7)$&$    33.7\pm      5.1(    11.0\pm      3.5)$&N(N)\\
EM$_{VR}$      &\nodata&$   10(    8)$&$    7(    3)$&$    50.7\pm      0.6(    48.4\pm      0.7)$&$    17.0\pm      4.4(    13.1\pm      3.6)$&N(N)\\
EM$_{VW}$      &\nodata&$   20(    5)$&$   28(    1)$&$    65.8\pm      0.6(    49.0\pm      0.8)$&$    53.3\pm      8.1(     4.3\pm      2.5)$&N(Y)\\
$3\sigma$&\nodata&$   29(   18)$&$   23(   12)$&$    60.2\pm      0.5(    50.0\pm      0.6)$&$    48.1\pm      7.3(    20.4\pm      4.8)$&N(N)\\
VT$_{3\sigma}$&\nodata&$   28(   14)$&$   22(    8)$&$    63.4\pm      0.5(    52.3\pm      0.7)$&$    45.7\pm      7.3(    18.5\pm      5.2)$&N(N)\\
VT$_{EM}$&\nodata&$   15(    3)$&$   21(    3)$&$    70.3\pm      0.7(    47.8\pm      0.7)$&$    36.9\pm      6.9(     4.5\pm      2.6)$&N(Y)\\
 \\

\textbf{Simulation}&$\mathbf{N=   30}$&$\mathbf{N_{mem}=   24}$&&$\mathbf{\langle V\rangle_{mem}=  100\mathrm{km s}^{-1}}$&$\mathbf{\sigma_{V_{0,mem}}=    4\mathrm{km s}^{-1}}$\\
EM             &\nodata&$   23(   23)$&$    0(    0)$&$    99.4\pm      0.5(    99.4\pm      0.5)$&$     5.3\pm      2.0(     5.3\pm      2.0)$&Y(Y)\\
EM$_{V}$       &\nodata&$   25(   25)$&$    2(    2)$&$    99.8\pm      0.4(    99.8\pm      0.4)$&$     5.4\pm      1.9(     5.4\pm      1.9)$&Y(Y)\\
EM$_{VR}$      &\nodata&$   25(   25)$&$    1(    1)$&$    99.5\pm      0.4(    99.5\pm      0.4)$&$     5.4\pm      2.0(     5.4\pm      2.0)$&Y(Y)\\
EM$_{VW}$      &\nodata&$   24(   24)$&$    1(    1)$&$    99.7\pm      0.5(    99.7\pm      0.5)$&$     5.5\pm      2.0(     5.5\pm      2.0)$&Y(Y)\\
$3\sigma$&\nodata&$   26(   26)$&$    2(    2)$&$    99.8\pm      0.4(    99.8\pm      0.4)$&$     5.5\pm      2.0(     5.5\pm      2.0)$&Y(Y)\\
VT$_{3\sigma}$&\nodata&$   24(   24)$&$    2(    2)$&$    99.3\pm      0.4(    99.3\pm      0.4)$&$     5.2\pm      2.0(     5.2\pm      2.0)$&Y(Y)\\
VT$_{EM}$&\nodata&$   22(   22)$&$    2(    2)$&$    99.3\pm      0.5(    99.3\pm      0.5)$&$     5.5\pm      2.1(     5.5\pm      2.1)$&Y(Y)\\
 \\

\textbf{Simulation}&$\mathbf{N=   30}$&$\mathbf{N_{mem}=   15}$&&$\mathbf{\langle V\rangle_{mem}=  100\mathrm{km s}^{-1}}$&$\mathbf{\sigma_{V_{0,mem}}=    4\mathrm{km s}^{-1}}$\\
EM             &\nodata&$   15(   14)$&$    1(    0)$&$    95.7\pm      0.5(    97.8\pm      0.4)$&$     8.5\pm      4.0(     3.2\pm      1.3)$&N(Y)\\
EM$_{V}$       &\nodata&$   14(   14)$&$    1(    1)$&$    97.6\pm      0.4(    97.6\pm      0.4)$&$     3.4\pm      1.4(     3.4\pm      1.4)$&Y(Y)\\
EM$_{VR}$      &\nodata&$   13(   13)$&$    1(    1)$&$    98.1\pm      0.4(    98.1\pm      0.4)$&$     2.9\pm      1.3(     2.9\pm      1.3)$&Y(Y)\\
EM$_{VW}$      &\nodata&$   15(   14)$&$    1(    0)$&$    95.7\pm      0.5(    97.8\pm      0.4)$&$     8.6\pm      4.0(     3.2\pm      1.3)$&N(Y)\\
$3\sigma$&\nodata&$   29(   20)$&$   14(    5)$&$    80.4\pm      0.5(    97.0\pm      0.5)$&$    39.4\pm      6.8(    10.5\pm      3.5)$&N(N)\\
VT$_{3\sigma}$&\nodata&$   28(   15)$&$   13(    2)$&$    82.6\pm      0.5(    96.1\pm      0.4)$&$    38.3\pm      6.9(     6.7\pm      2.1)$&N(N)\\
VT$_{EM}$&\nodata&$   13(   12)$&$    2(    3)$&$    98.0\pm      0.4(    97.8\pm      0.4)$&$     3.2\pm      1.4(     3.3\pm      1.4)$&Y(Y)\\
 \\

\textbf{Simulation}&$\mathbf{N=   30}$&$\mathbf{N_{mem}=    6}$&&$\mathbf{\langle V\rangle_{mem}=  100\mathrm{km s}^{-1}}$&$\mathbf{\sigma_{V_{0,mem}}=    4\mathrm{km s}^{-1}}$\\
EM             &\nodata&$    8(    6)$&$    2(    1)$&$    73.6\pm      1.4(    99.3\pm      0.6)$&$    45.6\pm     12.1(     1.9\pm      1.2)$&N(N)\\
EM$_{V}$       &\nodata&$   13(    6)$&$   13(    2)$&$    43.0\pm      0.9(    98.4\pm      0.6)$&$    54.3\pm     10.5(     2.2\pm      1.3)$&N(N)\\
EM$_{VR}$      &\nodata&$   11(    6)$&$   13(    1)$&$    36.6\pm      0.9(    99.2\pm      0.6)$&$    52.0\pm     10.5(     2.1\pm      1.3)$&N(N)\\
EM$_{VW}$      &\nodata&$   13(    6)$&$   12(    1)$&$    43.4\pm      0.9(    99.2\pm      0.6)$&$    54.4\pm     10.6(     1.8\pm      1.2)$&N(N)\\
$3\sigma$&\nodata&$   29(   11)$&$   23(    5)$&$    55.5\pm      0.7(   107.2\pm      0.9)$&$    53.7\pm      9.2(    13.7\pm      4.1)$&N(N)\\
VT$_{3\sigma}$&\nodata&$   28(    7)$&$   22(    1)$&$    53.1\pm      0.8(   104.4\pm      1.1)$&$    53.1\pm      9.2(    10.5\pm      3.5)$&N(N)\\
VT$_{EM}$&\nodata&$    6(    5)$&$    0(    1)$&$    99.5\pm      0.6(    99.8\pm      0.6)$&$     1.9\pm      1.3(     2.0\pm      1.4)$&N(N)\\
 \\

\textbf{Simulation}&$\mathbf{N=   30}$&$\mathbf{N_{mem}=   24}$&&$\mathbf{\langle V\rangle_{mem}=  200\mathrm{km s}^{-1}}$&$\mathbf{\sigma_{V_{0,mem}}=    4\mathrm{km s}^{-1}}$\\
EM             &\nodata&$   23(   23)$&$    0(    0)$&$   200.8\pm      0.4(   200.8\pm      0.4)$&$     3.2\pm      1.5(     3.2\pm      1.5)$&Y(Y)\\
EM$_{V}$       &\nodata&$   23(   23)$&$    0(    0)$&$   200.8\pm      0.4(   200.8\pm      0.4)$&$     3.2\pm      1.5(     3.2\pm      1.5)$&Y(Y)\\
EM$_{VR}$      &\nodata&$   23(   23)$&$    0(    0)$&$   200.8\pm      0.4(   200.8\pm      0.4)$&$     3.2\pm      1.5(     3.2\pm      1.5)$&Y(Y)\\
EM$_{VW}$      &\nodata&$   23(   23)$&$    0(    0)$&$   200.8\pm      0.4(   200.8\pm      0.4)$&$     3.2\pm      1.5(     3.2\pm      1.5)$&Y(Y)\\
$3\sigma$&\nodata&$   30(   23)$&$    6(    1)$&$   171.8\pm      0.5(   201.0\pm      0.4)$&$    58.4\pm      6.5(     3.1\pm      1.5)$&N(Y)\\
VT$_{3\sigma}$&\nodata&$   27(   22)$&$    3(    2)$&$   186.0\pm      0.5(   200.9\pm      0.4)$&$    42.0\pm      6.4(     3.1\pm      1.5)$&N(Y)\\
VT$_{EM}$&\nodata&$   22(   22)$&$    2(    2)$&$   200.9\pm      0.4(   200.9\pm      0.4)$&$     3.1\pm      1.5(     3.1\pm      1.5)$&Y(Y)\\
 \\

\textbf{Simulation}&$\mathbf{N=   30}$&$\mathbf{N_{mem}=   15}$&&$\mathbf{\langle V\rangle_{mem}=  200\mathrm{km s}^{-1}}$&$\mathbf{\sigma_{V_{0,mem}}=    4\mathrm{km s}^{-1}}$\\
EM             &\nodata&$   14(   14)$&$    0(    0)$&$   201.1\pm      0.5(   201.1\pm      0.5)$&$     4.3\pm      1.8(     4.3\pm      1.8)$&Y(Y)\\
EM$_{V}$       &\nodata&$   14(   14)$&$    0(    0)$&$   201.1\pm      0.5(   201.1\pm      0.5)$&$     4.3\pm      1.8(     4.3\pm      1.8)$&Y(Y)\\
EM$_{VR}$      &\nodata&$   14(   14)$&$    0(    0)$&$   201.1\pm      0.5(   201.1\pm      0.5)$&$     4.3\pm      1.8(     4.3\pm      1.8)$&Y(Y)\\
EM$_{VW}$      &\nodata&$   13(   13)$&$    1(    1)$&$   200.6\pm      0.4(   200.6\pm      0.4)$&$     4.0\pm      1.5(     4.0\pm      1.5)$&Y(Y)\\
$3\sigma$&\nodata&$   30(   15)$&$   15(    0)$&$   124.9\pm      0.5(   201.1\pm      0.5)$&$    81.4\pm      8.6(     4.3\pm      1.8)$&N(Y)\\
VT$_{3\sigma}$&\nodata&$   30(   12)$&$   15(    3)$&$   124.9\pm      0.5(   201.2\pm      0.5)$&$    81.4\pm      8.6(     4.7\pm      2.0)$&N(Y)\\
VT$_{EM}$&\nodata&$   12(   12)$&$    3(    3)$&$   201.2\pm      0.5(   201.2\pm      0.5)$&$     4.7\pm      2.0(     4.7\pm      2.0)$&Y(Y)\\
 \\
\textbf{Simulation}&$\mathbf{N=   30}$&$\mathbf{N_{mem}=    6}$&&$\mathbf{\langle V\rangle_{mem}=  200\mathrm{km s}^{-1}}$&$\mathbf{\sigma_{V_{0,mem}}=    4\mathrm{km s}^{-1}}$\\
EM             &\nodata&$    5(    5)$&$    0(    0)$&$   198.0\pm      0.9(   198.0\pm      0.9)$&$     3.6\pm      2.3(     3.6\pm      2.3)$&Y(Y)\\
EM$_{V}$       &\nodata&$    5(    5)$&$    0(    0)$&$   198.0\pm      0.9(   198.0\pm      0.9)$&$     3.6\pm      2.3(     3.6\pm      2.3)$&Y(Y)\\
EM$_{VR}$      &\nodata&$    4(    4)$&$    1(    1)$&$   195.4\pm      0.6(   195.3\pm      0.6)$&$     0.4\pm      0.6(     0.2\pm      0.4)$&N(N)\\
EM$_{VW}$      &\nodata&$    5(    5)$&$    0(    0)$&$   198.2\pm      0.9(   198.2\pm      0.9)$&$     3.8\pm      2.4(     3.8\pm      2.4)$&Y(Y)\\
$3\sigma$&\nodata&$   30(    8)$&$   24(    2)$&$   101.9\pm      0.6(   192.0\pm      0.9)$&$    65.4\pm      9.1(    11.8\pm      4.8)$&N(N)\\
VT$_{3\sigma}$&\nodata&$   30(    5)$&$   24(    1)$&$   101.9\pm      0.6(   195.2\pm      0.6)$&$    65.4\pm      9.1(     0.1\pm      0.2)$&N(N)\\
VT$_{EM}$&\nodata&$    4(    4)$&$    2(    2)$&$   194.9\pm      0.6(   194.9\pm      0.6)$&$     0.0\pm      0.0(     0.0\pm      0.0)$&N(N)\\
 \tablenotetext{a}{Table values in parentheses are obtained after using an initial velocity filter.}
 \label{tab:sim_smalldisp}
\end{longtable*}

%% file: astroph_appendix.tex
\begin{appendix}

\section{Errors for the Estimates of Mean and Variance}
\label{app:errors}
Suppose $X$ is a random variable that follows a Gaussian distribution with mean $\langle X\rangle$ and variance $\sigma_{X_0}^2$.  We sample $X$ such that our data set of $N$ observations is given by $\{(X_i,\sigma_{X_i})\}_{i=1}^N$, where $\sigma_{X_i}$ is the measurement error.  We calculate estimates $\langle \hat{X}\rangle$ and $\hat{\sigma}_{X_0}$ iteratively from Equations \ref{eq:estimatemean}-\ref{eq:estimatedisp}.  To obtain the $1\sigma$ errorbar associated with each estimate we propagate the measurement errors as well as the parameter errors from the previous iteration.  The variance of any non-linear function of $z$ variables, $f(x_1,x_2,...,x_z)$, can be approximated by 
\begin{equation}
  \sigma_f^2=\biggl (\frac{\partial f}{\partial x_1}\biggr )^2\sigma_{x_1}^2 + \biggl (\frac{\partial f}{\partial x_2}\biggr )^2\sigma_{x_2}^2 + \dots + \biggl (\frac{\partial f}{\partial x_z}\biggr )^2\sigma_{x_z}^2
  \label{eq:propagate}
\end{equation}
so long as $x_1...x_z$ are uncorrelated.  From Equations \ref{eq:estimatemean}-\ref{eq:estimatedisp}, the parameter estimates in iteration $n+1$ are functions of the data as well as the parameter estimates obtained the $n^{th}$ iteration.  We must therefore propagate the variances $\sigma_{X_i}^2$, $\sigma_{\langle \hat{X}^{\{n\}}\rangle}^2$ and $\sigma^2_{\hat{\sigma}_{X_0}^{\{n\}}}$.  From Equation \ref{eq:propagate}, the variances associated with the parameter estimates obtained in iteration $n+1$ are
\begin{equation}
\sigma_{\langle \hat{X}\rangle^{\{n+1\}}}^2=\frac{A}{F^2}+\biggl [\frac{B}{F}-\frac{CD}{F^2}\biggr ]^2 \sigma_{\hat{\sigma}_{X_0}^{2\{n\}}}^2
  \label{eq:properrormean2}
\end{equation}
and
\begin{equation}
  \sigma_{\hat{\sigma}_{X_0}^{2\{n+1\}}}^2=\frac{4G}{F^2}+\frac{4H^2}{F^2}\sigma_{\langle \hat{X}\rangle^{\{n\}}}^2+\biggl [\frac{2J}{F}-\frac{KL}{F^2}\biggr ]^2\sigma_{\hat{\sigma}_{X_0}^{2\{n\}}}^2,
  \label{eq:properrordisp2}
\end{equation}
where
\begin{eqnarray}
  A\equiv \displaystyle\sum_{i=1}^N\biggl [\frac{\hat{P}_{M_i}^{\{n\}}\sigma_{X_i}}{1+\sigma_{X_i}^2/\hat{\sigma}_{X_0}^{2\{n\}}}\biggr ]^2;\\
  B\equiv \displaystyle\sum_{i=1}^N\frac{\hat{P}_{M_i}^{\{n\}}X_i\sigma_{X_i}^2}{\hat{\sigma}_{X_0}^{4\{n\}}(1+\sigma_{X_i}^2/\hat{\sigma}_{X_0}^{2\{n\}})^2};\nonumber\\
  C\equiv \displaystyle\sum_{i=1}^N\frac{\hat{P}_{M_i}^{\{n\}}X_i}{1+\sigma_{X_i}^2/\hat{\sigma}_{X_0}^{2\{n\}}};\nonumber\\
  D\equiv \displaystyle\sum_{i=1}^N\frac{\hat{P}_{M_i}^{\{n\}}\sigma_{X_i}^2}{\hat{\sigma}_{X_0}^{4\{n\}}(1+\sigma_{X_i}^2/\hat{\sigma}_{X_0}^{2\{n\}})^2};\nonumber\\
  F\equiv \displaystyle\sum_{i=1}^N\frac{\hat{P}_{M_i}^{\{n\}}}{1+\sigma_{X_i}^2/\hat{\sigma}_{X_0}^{2\{n\}}};\nonumber\\
  G\equiv \displaystyle\sum_{i=1}^N\biggl [ \frac{\hat{P}_{M_i}^{\{n\}}(X_i-\langle \hat{X}\rangle^{\{n\}} )\sigma_{X_i}}{(1+\sigma_{X_i}^2/\hat{\sigma}_{X_0}^{2\{n\}})^2}  \biggr ]^2;\nonumber\\
  H\equiv \displaystyle\sum_{i=1}^N \frac{\hat{P}_{M_i}^{\{n\}}(X_i-\langle \hat{X}\rangle^{\{n\}} )}{(1+\sigma_{X_i}^2/\hat{\sigma}_{X_0}^{2\{n\}})^2};\nonumber\\
  J\equiv \displaystyle\sum_{i=1}^N \frac{\hat{P}_{M_i}^{\{n\}}(X_i-\langle \hat{X}\rangle^{\{n\}} )^2\sigma_{X_i}^2}{\hat{\sigma}_{X_0}^{4\{n\}}(1+\sigma_{X_i}^2/\hat{\sigma}_{X_0}^{2\{n\}})^3};\nonumber\\
  K\equiv \displaystyle\sum_{i=1}^N \frac{\hat{P}_{M_i}^{\{n\}}(X_i- \langle \hat{X}\rangle^{\{n\}} )^2}{(1+\sigma_{X_i}^2/\hat{\sigma}_{X_0}^{2\{n\}})^2};\nonumber\\
   L\equiv \displaystyle\sum_{i=1}^N \frac{\hat{P}_{M_i}^{\{n\}}\sigma_{X_i}^2}{\hat{\sigma}_{X_0}^{4\{n\}}(1+\sigma_{X_i}^2/\hat{\sigma}_{X_0}^{2\{n\}})^2}.\nonumber
  \label{eq:abcdefgh}
\end{eqnarray}
After 15-20 iterations, parameter estimates and the associated errors are insensitive to their (arbitrary) initial values (see Section \ref{sec:procedure}).

\section{Pool-Adjacent-Violators Algorithm for Monotonic Regression}
\label{app:pav}

Monotonic regression provides a nonparametric, least-squares fit to an ordered set of data points, subject to the constraint that the fit must be either non-increasing or non-decreasing.  In our application, we have estimates $\hat{P}_{M_i}$ of the probability that the $i^{th}$ star is a dSph member, and we wish to estimate a function $p(a)$, the unconditional probability of membership as a function of elliptical radius $a$.  We assume $p(a)$ is a non-increasing function.  If we sort data points by order of increasing $a_i$, the expression for the monotonic regression estimate of $p(a)$ is 
\begin{equation}
  \hat{p}(a_i)=\displaystyle \min_{1\leq u \leq i}\displaystyle\max_{i\leq v \leq N}\frac{\Sigma_{j=u}^v\hat{P}_{M_i}}{v-u+1},
  \label{eq:monotonic2}
\end{equation}
In the case that $\hat{P}_{M_{i}}\leq \hat{P}_{M_{i-1}}$ for all $i$, then we obtain the simple result $\hat{p}(a_i)=\hat{P}_{M_i}$.  However, when the data ``violate'' the shape restriction, i.e., when $\hat{P}_{M_{i}}> \hat{P}_{M_{i-1}}$ for some $i$, monotonic regression determines the non-increasing function that provides a least-squares fit to the data.  

The ``Pool-Adjacent-Violators'' (PAV) algorithm calculates monotonic regression estimates simply and efficiently.  For example, suppose we have a data set of $N=6$ stars and our ordered set of $\hat{P}_{M_i}$ is given by $\{1.0,0.9,0.8,0.5,0.6,0.2\}$.  The adjacent probabilities $\{0.5,0.6\}$ ``violate'' our assumption that $p(a)$ is non-increasing.  The PAV algorithm identifies such ``blocks'' of violators and replaces them with the average value in the block.  In this case, PAV would give $\hat{p}=\{1.0,0.9,0.8,0.55,0.55,0.2\}$.

To implement monotonic regression we use the standard PAV algorithm of \citet{grotzinger84}.  We search the ordered data points from left (smaller $a_i$) to right (larger $a_i$) for violations between successive ``blocks.''  When a violation is discovered, the offending points are ``coalesced'' into a single block having the average value of the contributing points.  In the case that the newly formed block causes a violation with the block on its immediate left, these two blocks are coalesced into a still larger block, and so on until there is no violation with previously searched blocks.  When no violation occurs, we advance the search again to the right.  The resulting estimate, $\hat{p}(a)$, is a decreasing step function.  Our PAV estimates for each galaxy are displayed in Figure \ref{fig:members_vr}.

\section{Results from EM Algorithm Using Simulated Data}
\label{app:sim}

\begin{figure*}
  \plotone{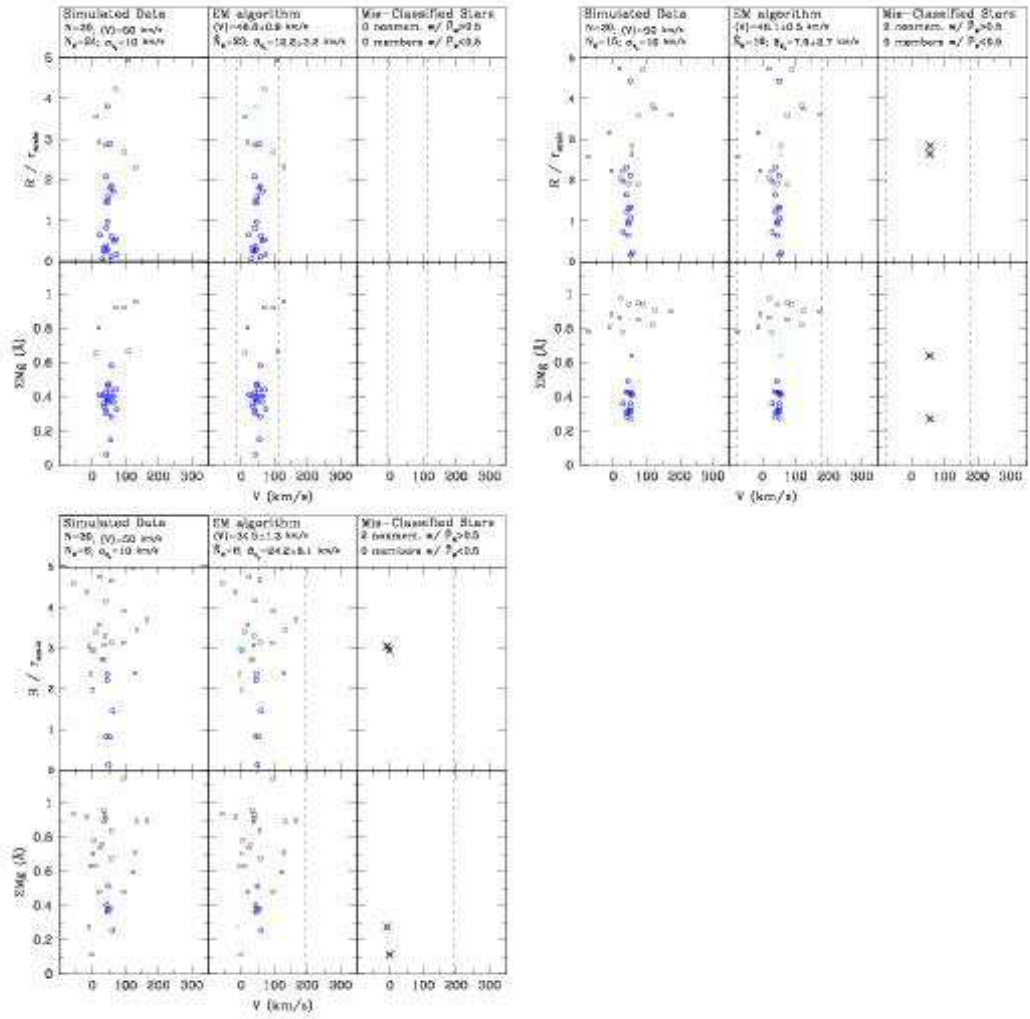}
  \caption{\scriptsize Performance of EM algorithm with simulated dSph data.  Shown here are the artificial data and EM results for simulated samples of $N=30$ stars and assuming a dSph velocity distribution with mean $50 $ km s$^{-1}$ and dispersion $10$ km s$^{-1}$.  The three main panels depict simulations with different degrees of contamination: upper left, upper right, and lower left panels correspond to simulations with member fractions $f_{mem}=0.2,0.5,0.8$, respectively.  \textit{Left sub-panels:} angular distance (top) and magnesium index (bottom) versus velocity for the simulated data set.  Text indicates the number of simulated data points, $N$, (including contamination), and the number, $N_M$, of those that are drawn from a dSph-like member population with mean velocity $\langle V\rangle_{mem}$ and velocity dispersion $\sigma_{V_0}=10$ km s$^{-1}$.  Blue circles identify the simulated members; black squares identify the simulated contaminants.  See Section \ref{sec:performance} for further details of the dSph and contaminant distributions.  \textit{Middle sub-panels:} squares/circles represent the same simulated data, but color indicates the value of $P_M$ resulting from the EM algorithm.  Black (red; magenta; green; cyan; blue) markers signify $\hat{P}_{M} \leq 0.01$ ($\hat{P}_{M} > 0.01; > 0.50; > 0.68; > 0.95; > 0.99$).  Dotted vertical lines enclose velocities that satisfy a conventional $3\sigma$ clipping algorithm.  In some of Figures \ref{fig:sim_members_30_50} - \ref{fig:sim_smalldisp_members_30_200}, one or both of these limits lies outside the plotting window.  Text indicates estimates of the number of members, mean velocity and velocity dispersion returned by the EM algorithm.  \textit{Right sub-panels:} contaminants for which the EM algorithm gives $P_{M} > 0.5$ (black squares with X) or members for which it gives $P_{M} < 0.5$ (blue circles with X).}
  \label{fig:sim_members_30_50}
\end{figure*}
\clearpage
\begin{figure*}
  \plotone{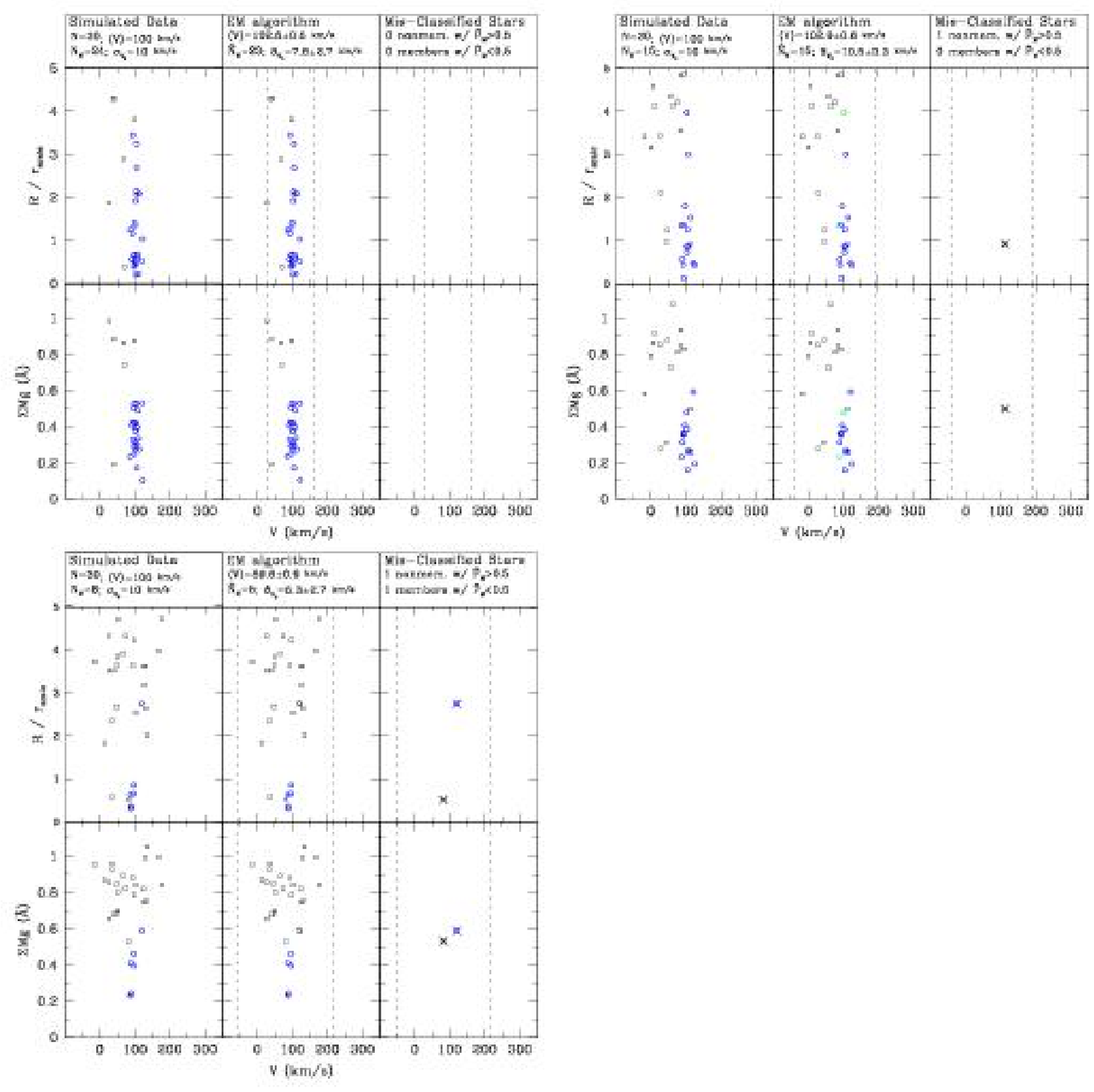}
  \caption{\scriptsize Same as Figure \ref{fig:sim_members_30_50}, but for simulated data sets with $N=30$, $\langle V\rangle_{mem}=100$ km s$^{-1}$ and $\sigma_{V_0,mem}=10$ km s$^{-1}$.}
  \label{fig:sim_members_30_100}
\end{figure*}
\clearpage
\begin{figure*}
  \plotone{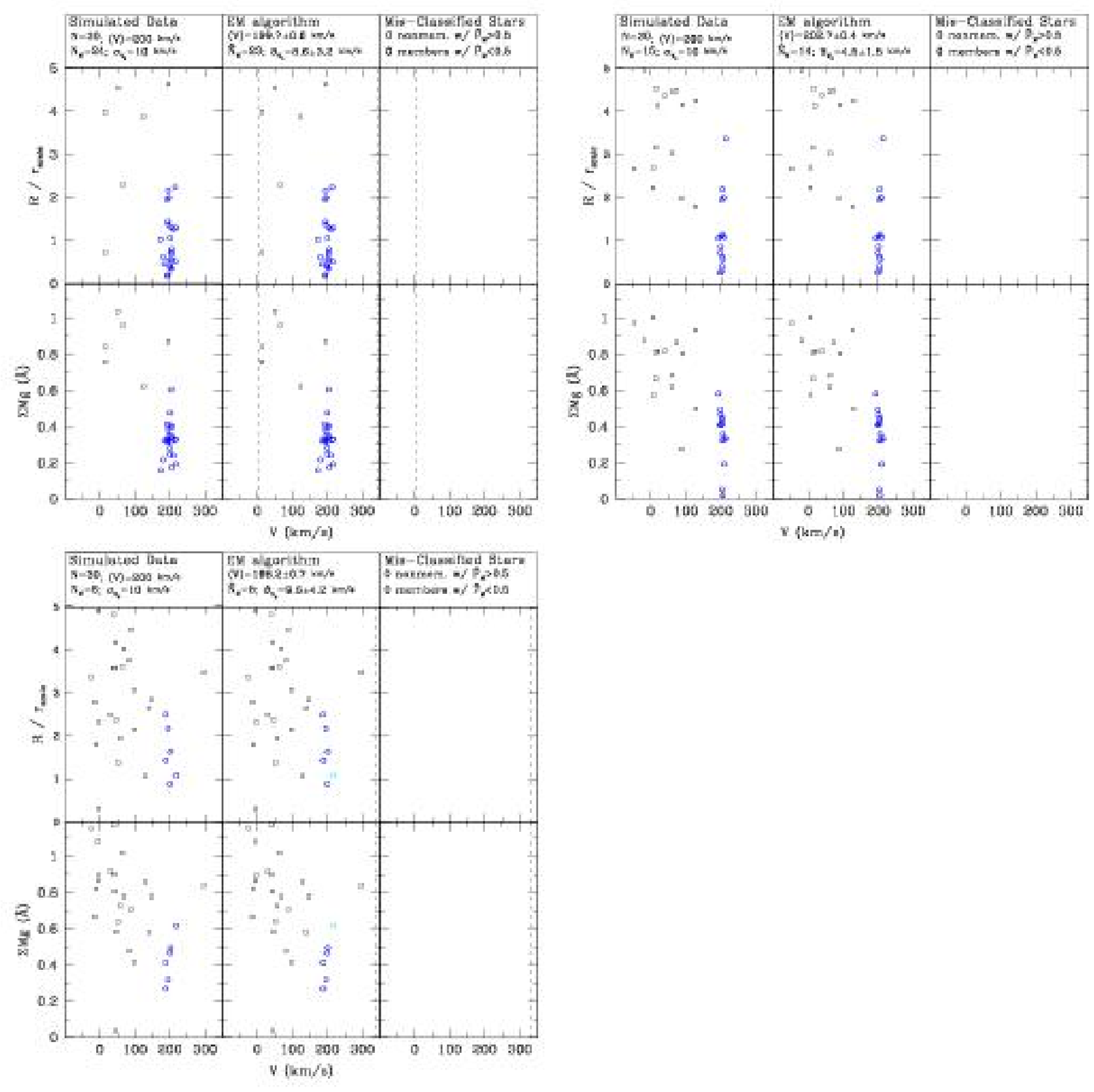}
  \caption{\scriptsize Same as Figure \ref{fig:sim_members_30_50}, but for simulated data sets with $N=30$, $\langle V\rangle_{mem}=200$ km s$^{-1}$ and $\sigma_{V_0,mem}=10$ km s$^{-1}$.}
  \label{fig:sim_members_30_200}
\end{figure*}
\clearpage
\begin{figure*}
  \plotone{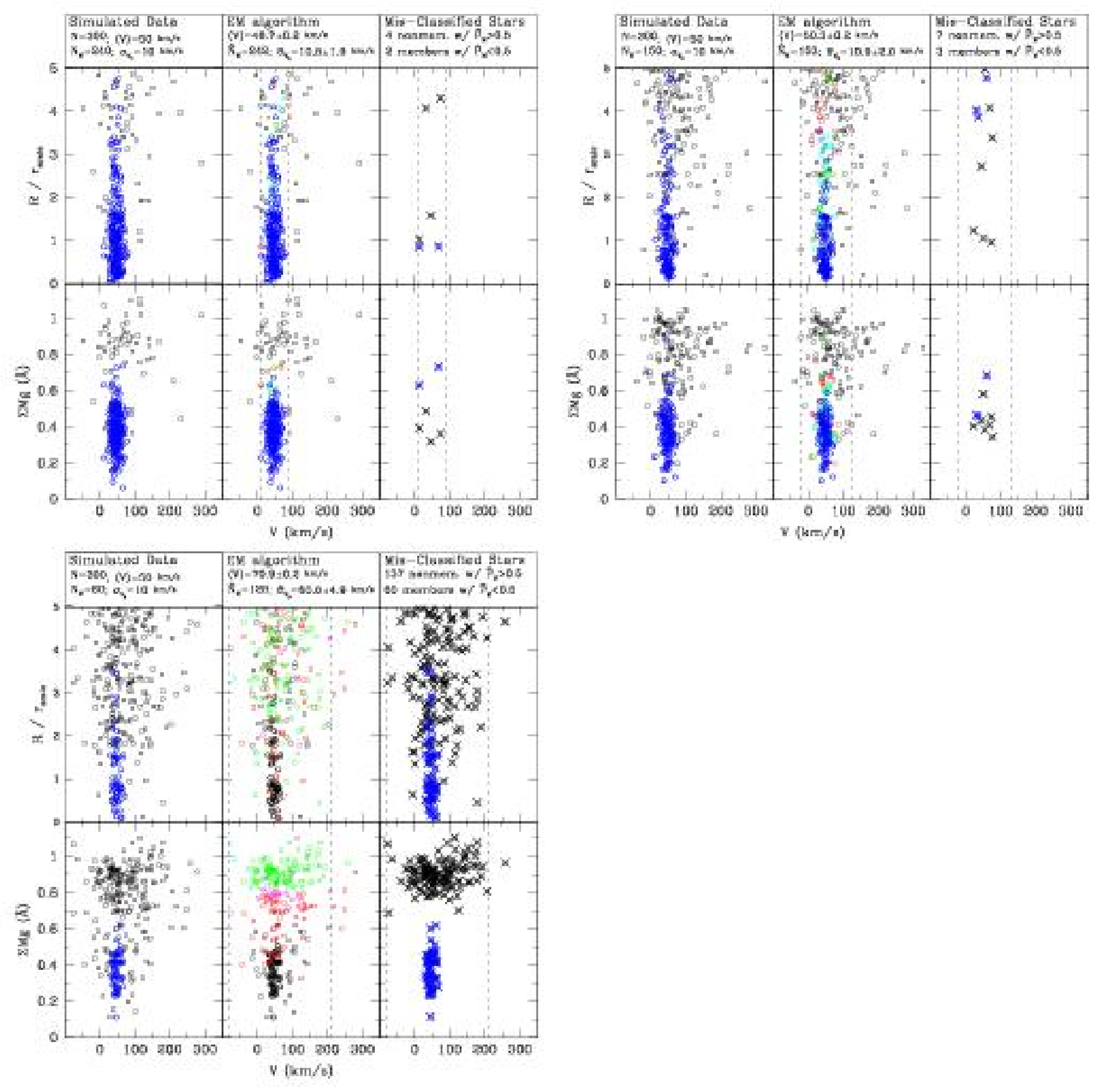}
  \caption{\scriptsize Same as Figure \ref{fig:sim_members_30_50}, but for simulated data sets with $N=300$, $\langle V\rangle_{mem}=50$ km s$^{-1}$ and $\sigma_{V_0,mem}=10$ km s$^{-1}$.}
  \label{fig:sim_members_300_50}
\end{figure*}
\clearpage
\begin{figure*}
  \plotone{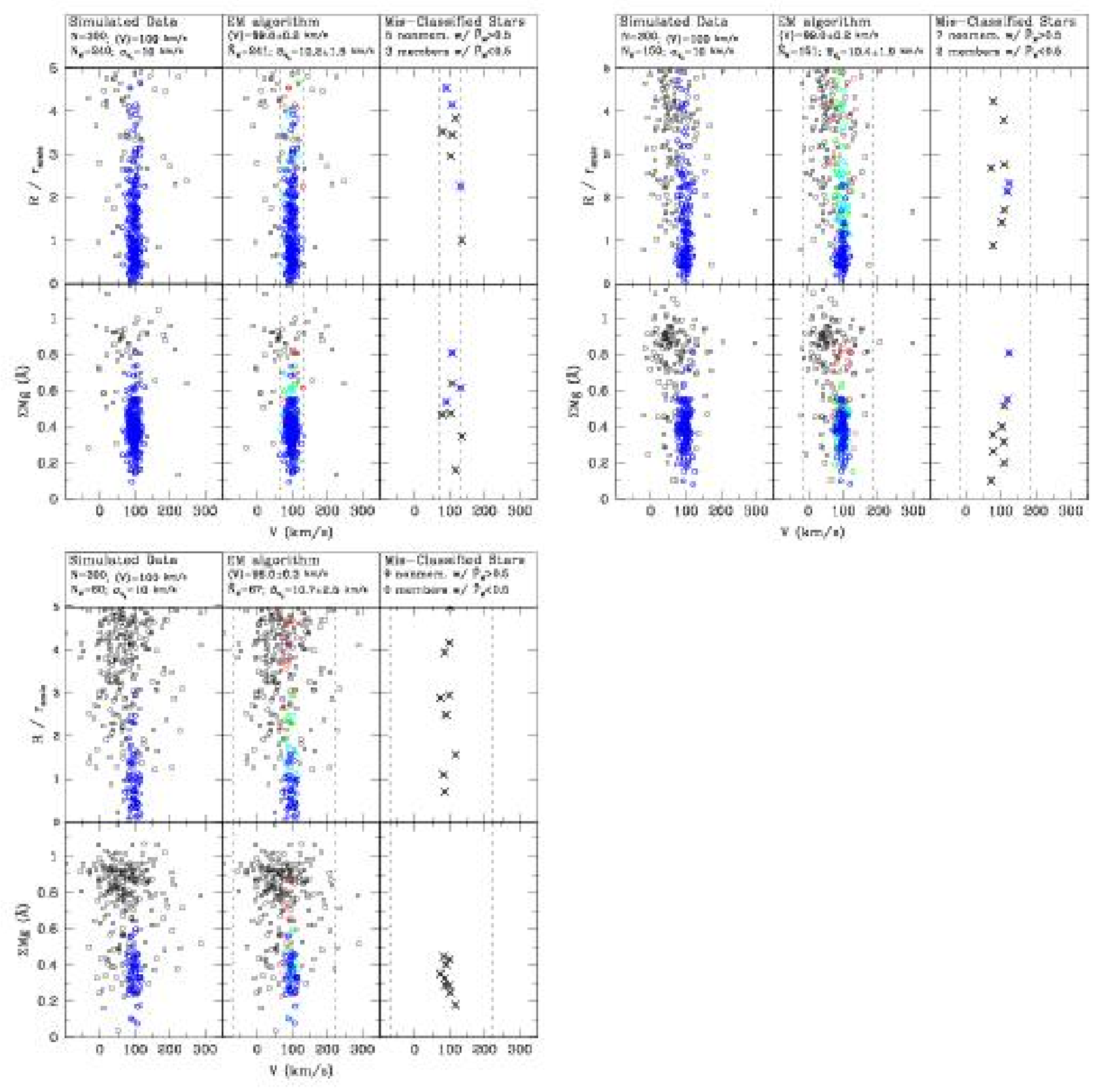}
  \caption{\scriptsize Same as Figure \ref{fig:sim_members_30_50}, but for simulated data sets with $N=300$, $\langle V\rangle_{mem}=100$ km s$^{-1}$ and $\sigma_{V_0,mem}=10$ km s$^{-1}$.}
  \label{fig:sim_members_300_100}
\end{figure*}
\clearpage
\begin{figure*}
  \plotone{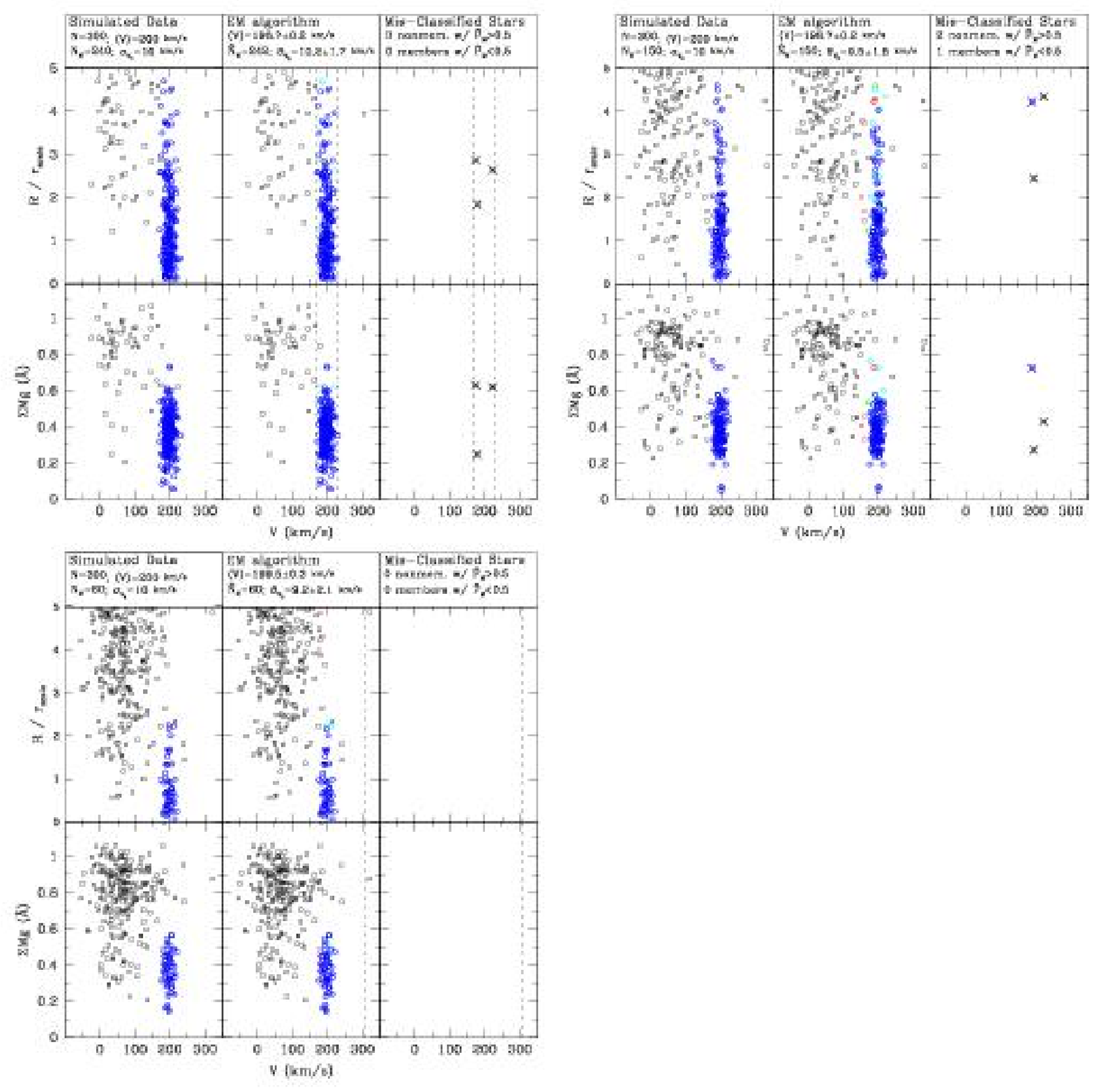}
  \caption{\scriptsize Same as Figure \ref{fig:sim_members_30_50}, but for simulated data sets with $N=300$, $\langle V\rangle_{mem}=200$ km s$^{-1}$ and $\sigma_{V_0,mem}=10$ km s$^{-1}$.}
  \label{fig:sim_members_300_200}
\end{figure*}
\clearpage
\begin{figure*}
  \plotone{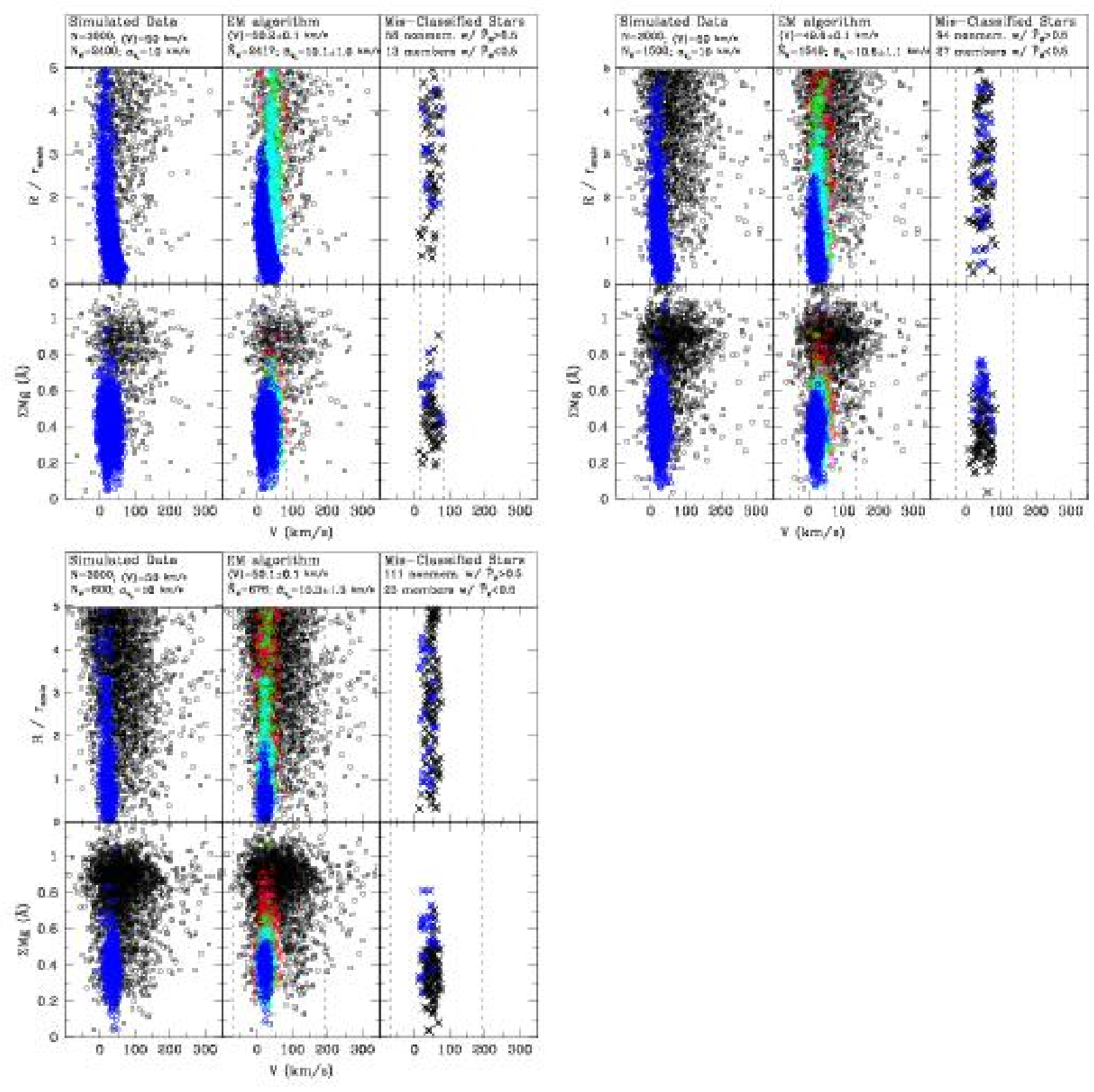}
  \caption{\scriptsize Same as Figure \ref{fig:sim_members_30_50}, but for simulated data sets with $N=3000$, $\langle V\rangle_{mem}=50$ km s$^{-1}$ and $\sigma_{V_0,mem}=10$ km s$^{-1}$.}
  \label{fig:sim_members_3000_50}
\end{figure*}
\clearpage
\begin{figure*}
  \plotone{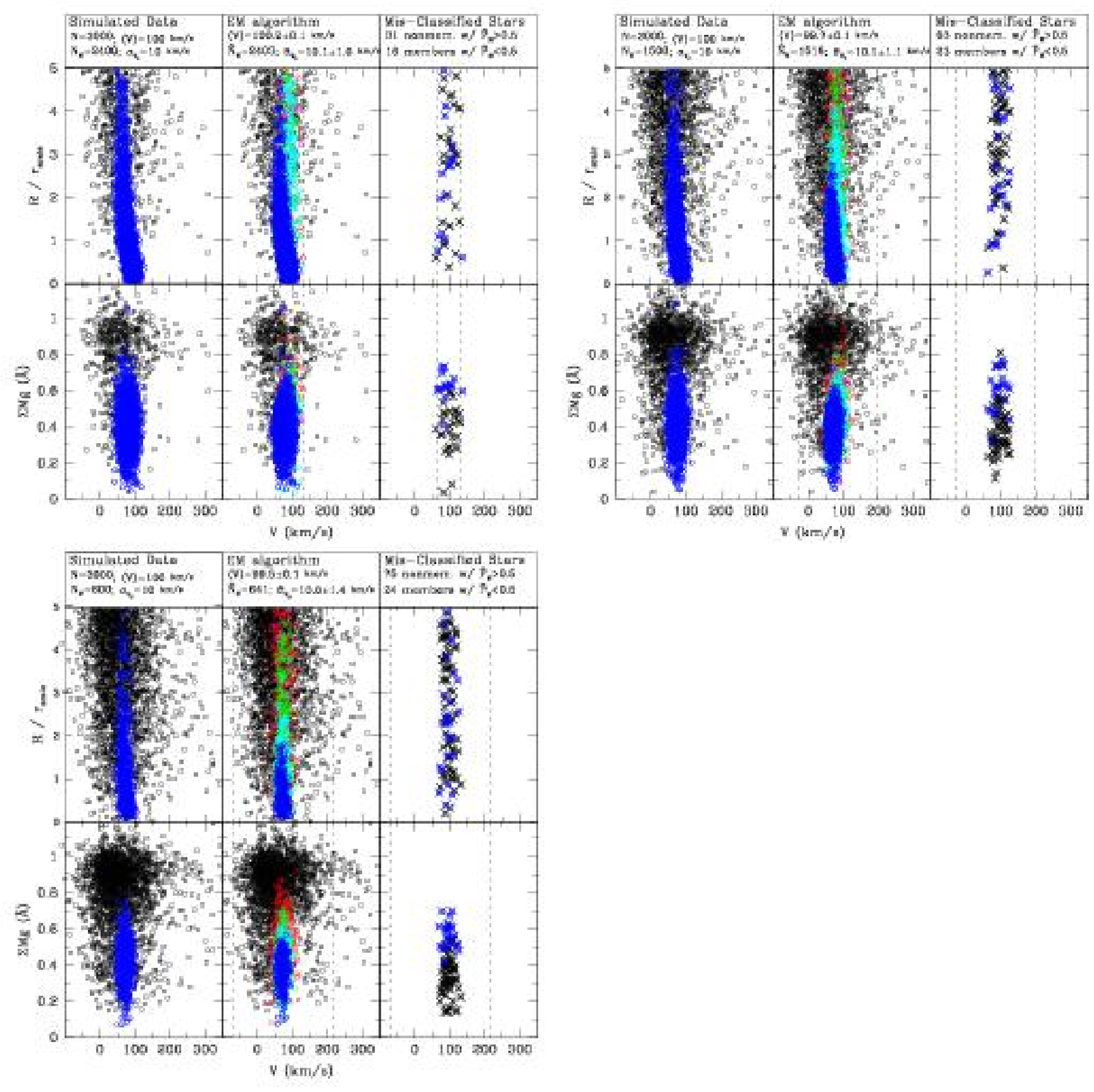}
  \caption{\scriptsize Same as Figure \ref{fig:sim_members_30_50}, but for simulated data sets with $N=3000$, $\langle V\rangle_{mem}=100$ km s$^{-1}$ and $\sigma_{V_0,mem}=10$ km s$^{-1}$.}
  \label{fig:sim_members_3000_100}
\end{figure*}
\clearpage
\begin{figure*}
  \plotone{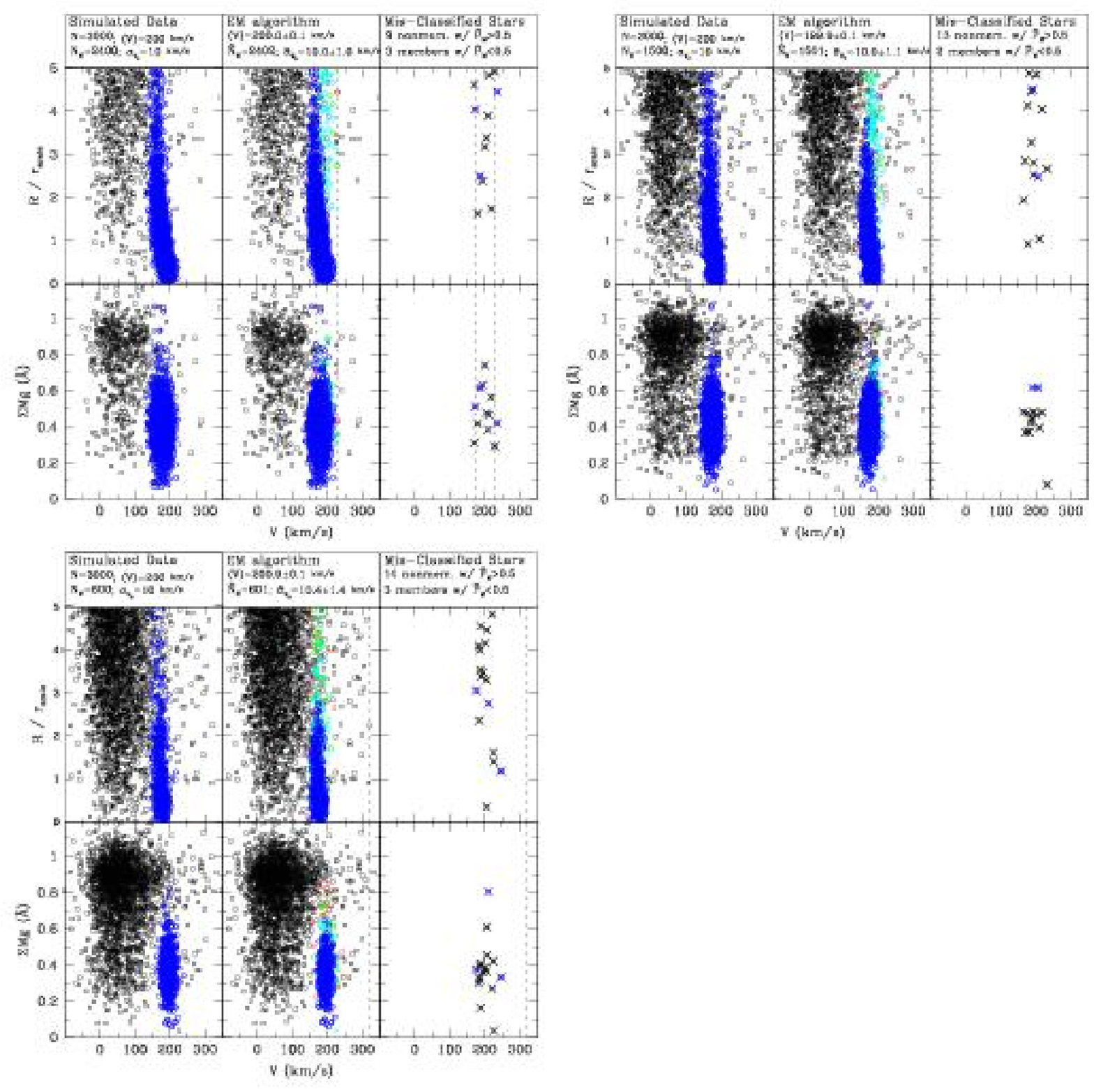}
  \caption{\scriptsize Same as Figure \ref{fig:sim_members_30_50}, but for simulated data sets with $N=3000$, $\langle V\rangle_{mem}=200$ km s$^{-1}$ and $\sigma_{V_0,mem}=10$ km s$^{-1}$.}
  \label{fig:sim_members_3000_200}
\end{figure*}
\clearpage
\begin{figure*}
  \plotone{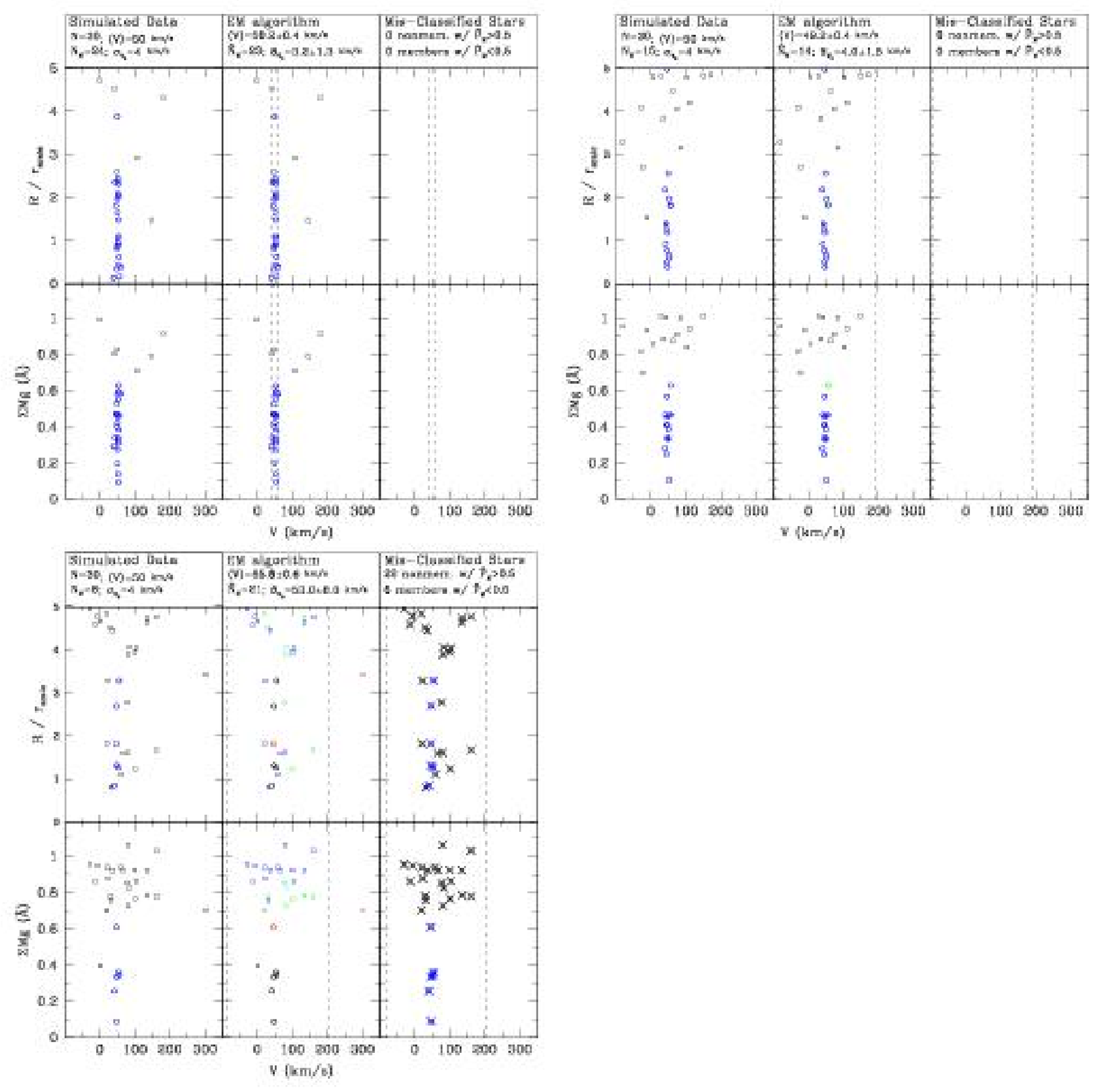}
  \caption{\scriptsize Same as Figure \ref{fig:sim_members_30_50}, but for simulated data sets with $N=30$, $\langle V\rangle_{mem}=50$ km s$^{-1}$ and $\sigma_{V_0,mem}=4$ km s$^{-1}$.}
  \label{fig:sim_smalldisp_members_30_50}
\end{figure*}
\clearpage
\begin{figure*}
  \plotone{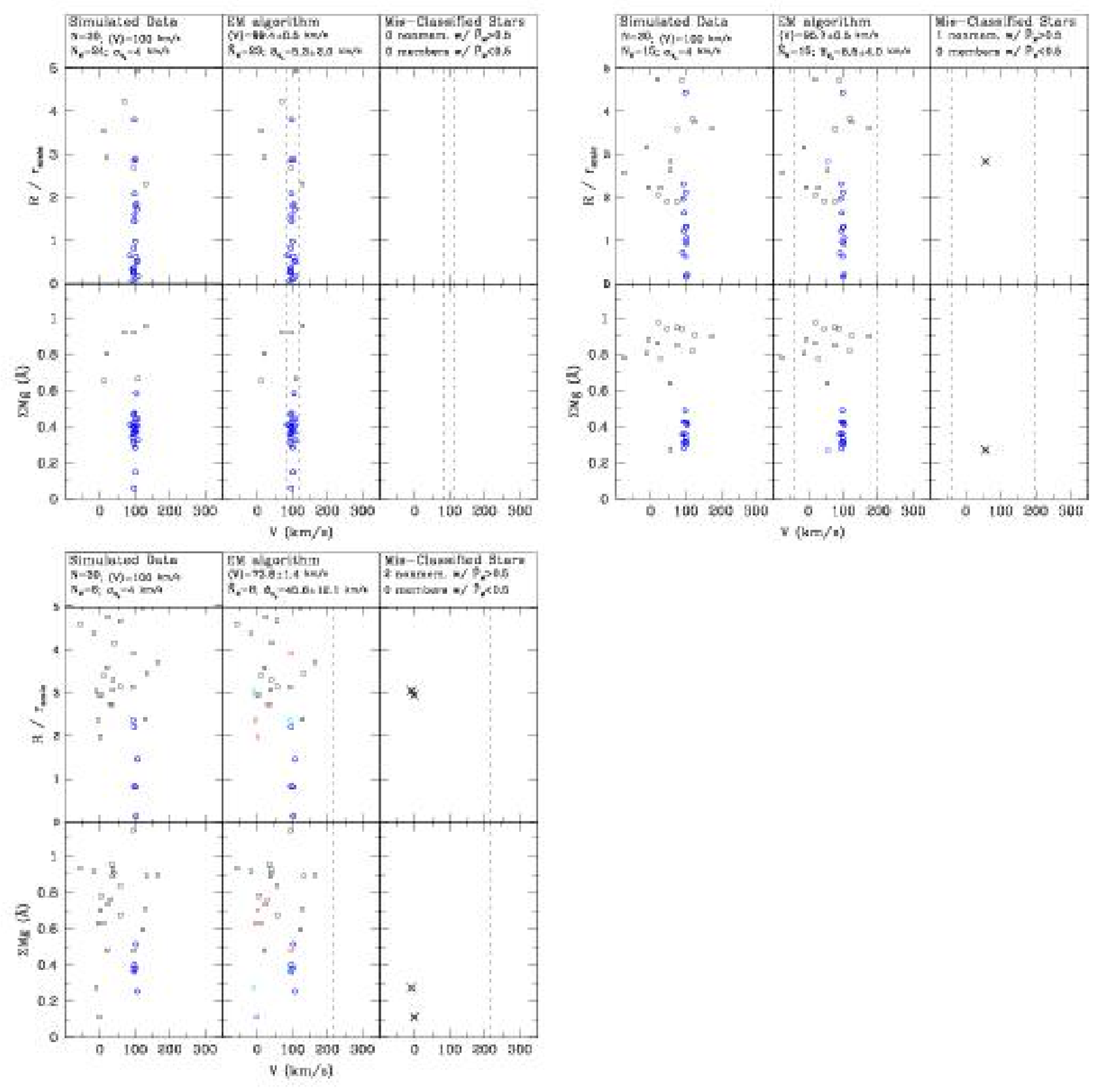}
  \caption{\scriptsize Same as Figure \ref{fig:sim_members_30_50}, but for simulated data sets with $N=30$, $\langle V\rangle_{mem}=100$ km s$^{-1}$ and $\sigma_{V_0,mem}=4$ km s$^{-1}$.}
  \label{fig:sim_smalldisp_members_30_100}
\end{figure*}
\clearpage
\begin{figure*}
  \plotone{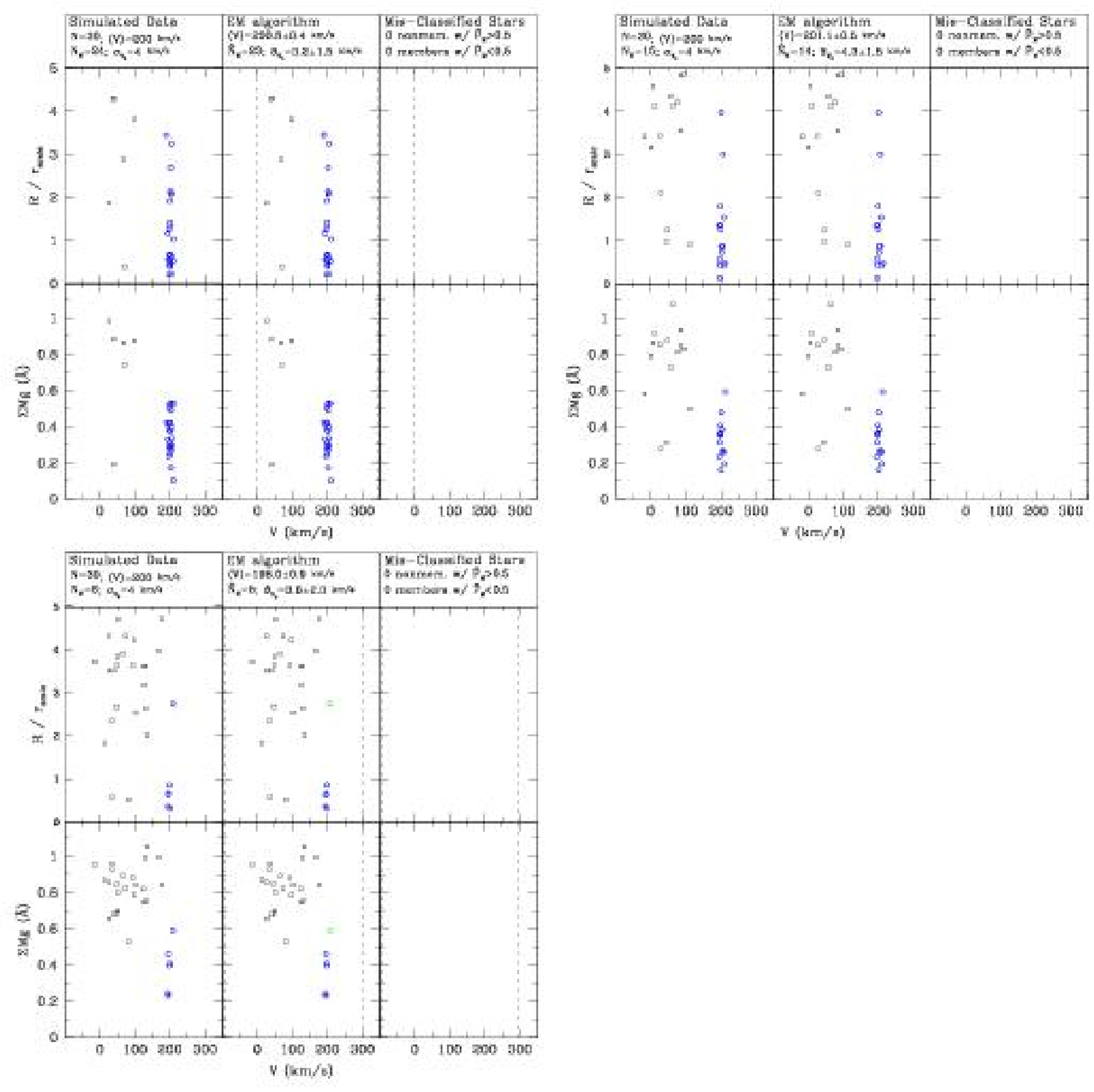}
  \caption{\scriptsize Same as Figure \ref{fig:sim_members_30_50}, but for simulated data sets with $N=30$, $\langle V\rangle_{mem}=200$ km s$^{-1}$ and $\sigma_{V_0,mem}=4$ km s$^{-1}$.}
\label{fig:sim_smalldisp_members_30_200}
\end{figure*}

\end{appendix}